\documentclass[journal=jacsat,manuscript=article]{achemso}
\usepackage{textcomp}
\DeclareUnicodeCharacter{2212}{\textminus}
\usepackage{chemformula} 
\usepackage[T1]{fontenc} 
\usepackage{lmodern}
\usepackage[utf8]{inputenc}
\usepackage{amsmath}
\usepackage{lineno}

\author{Lottie L. Murray$^{\dagger}$}
\author{Eric Herrmann$^{\dagger}$}
\author{Igor Evangelista}
\author{Sai Rahul Sitaram}
\author{Ke Ma}
\author{Anderson Janotti}
\author{Xi Wang}
\email{wangxi@udel.edu}
\author{Matthew F. Doty}
\affiliation[UDMSEG]{Dept.~of Materials Science and Engineering, Univ.~of Delaware, Newark, DE, 19716, US $^{\dagger}$,* These authors contributed equally to this work}
\email{doty@udel.edu}
\title{Deterministic Realization of Complex Local Strain Fields and Bandgap Profiles in Two-Dimensional Materials}

\begin{document}






\begin{abstract}
Emerging classical and quantum device concepts demand precise spatial control over the optoelectronic properties of two-dimensional (2D) materials, but deterministic engineering via local multiaxial strain distributions remains challenging. Using Ga$_2$Se$_2$, we demonstrate a material-agnostic platform in which nanostructure geometry deterministically prescribes in-plane strain profiles in suspended van der Waals membranes. We first use hyperspectral photoluminescence mapping and experimentally-constrained finite element analysis to quantify the experimental biaxial and uniaxial strain gauge factors that relate strain to the change in bandgap. We next show that a two-component analytical model can predict, with less than 12\% error, spatially-resolved bandgap shifts arising from multiaxial strain distributions in complex geometries, including the interactions between adjacent nanostructures. Finally, we demonstrate that this approach can be extended to other materials. The results demonstrate that nanostructure design provides a quantitative, deterministic framework for the realization of designed strain and bandgap distributions in 2D materials.

\end{abstract}

\vspace{1em}
\noindent\textbf{Keywords: local strain engineering, two-dimensional materials, optical properties, hyperspectral imaging, photoluminescence, gallium selenide, tungsten disulfide, van der Waals materials, finite element}

Strain engineering of two-dimensional (2D) materials provides a uniquely powerful handle for controlling their optoelectronic properties: the low dimensionality enables band structure modulation through mechanical deformation at strain magnitudes exceeding those achievable in bulk semiconductors \cite{e3}. Local, spatially varying strain gives rise to emergent phenomena including pseudo-magnetic fields in graphene\cite{e8,e9,e10,e11,e12}, strain-induced magnetic phase transitions in magnetic 2D materials \cite{e13,e14,e15,hassan}, exciton funneling in transition metal dichalcogenides \cite{e16,e17,e18,e19,e20,e21,excitons,exciton02,2dspes}, and the activation and tuning of single-photon emitters \cite{e22,e23,induced_strain,e25,induced_strain02,e27,e28,e29,e30}. This control over strain is critical for applications in nanophotonics\cite{strain_2D}, quantum information processing\cite{e7}, quantum networking\cite{network,network02}, quantum key distribution\cite{qkd,qkd02,2dspes}, and quantum sensing \cite{sensing,sensing02}. Reliably realizing these device concepts at scale requires moving beyond global strain approaches, such as substrate bending\cite{e4,e5, flex_strain}, that lack spatial resolution and geometric flexibility. 


Of the various methods to achieve local strain control \cite{e7}, patterned substrates have emerged as a promising approach for inducing large-magnitude, spatially-localized strain gradients \cite{e31}. In brief, 2D material flakes stamped directly onto nanostructures conform to the surface topology, developing spatially modulated strain environments through local out-of-plane deformation. Despite challenges associated with flake-nanostructure interactions \cite{e37,e38} and unintended wrinkling or bubble formation, this approach has enabled numerous strain-related discoveries \cite{e32,e33,e34,e35,e36}. However, the lack of a quantitative framework linking nanostructure design to predictable, spatially-resolved multiaxial strain distributions has limited the degree to which strain profiles can be rationally designed. We bridge this gap by demonstrating a quantitative framework for predicting the in-plane strain distributions induced in suspended semiconductor membranes by lithographically defined nanostructures. 

A central insight of this work is that 2D material flakes stamped onto nanostructured substrates develop significant in-plane tensile strain as the membrane stretches to conform to the underlying topology. While prior patterned substrate studies have primarily emphasized out-of-plane deformation, the resulting in-plane strain components are often underappreciated and not directly quantified. To establish a quantitative relationship between this in-plane strain and bandgap modulation, we combine hyperspectral photoluminescence (PL) mapping, atomic force microscopy (AFM), and finite element analysis (FEA) to extract spatially resolved multiaxial strain distributions and correlate them directly with optical response. We first develop and validate a model using isolated nanostructures of controlled geometry. We then demonstrate its predictive power across complex multi-structure geometries and additional material systems, establishing a predictive, geometry-driven framework for rational strain engineering in van der Waals materials.

\section{Quantifying Strain Gauge Factors via Isolated Nanostructures}

We focus primarily on the van der Waals semiconductor gallium selenide (Ga$_2$Se$_2$), which has attracted considerable interest for its bright bulk emission \cite{e48}, thickness- and strain-tunable optical properties \cite{strain_gase,e50,flex_strain,e52,degradation,ptmd,gase_structure,band_engineering}, and ability to host single-photon emitters \cite{Luo2023,e55,e56,e57,induced_strain02}. Prior investigations of strain-induced bandgap modification in Ga$_2$Se$_2$, as measured by PL, relied on transferring exfoliated flakes onto flexible substrates and bending the material to create wrinkles \cite{flex_strain}. While these studies established that the bandgap of Ga$_2$Se$_2$ responds to induced strain, they were limited in three important respects: i) wrinkles impose a purely uniaxial strain state with limited impact on local band structure \cite{strain_2D,strain_2D02}; ii) the approach is neither scalable nor controllable; and iii) the resulting small datasets precluded statistically robust conclusions. The platform introduced here overcomes all three limitations. Nanostructure geometry provides direct control over the local strain, enabling the systematic study of uniaxial, biaxial, and multiaxial strain environments and their distinct effects on the band structure \cite{strain_2D}. Hyperspectral PL mapping across arrays of structures generates datasets of sufficient scale for statistically-rigorous analysis. The experimental strain gauge factors are also validated against density functional theory (DFT) calculations. 

\begin{figure}[h]
  \includegraphics[width=1\linewidth]{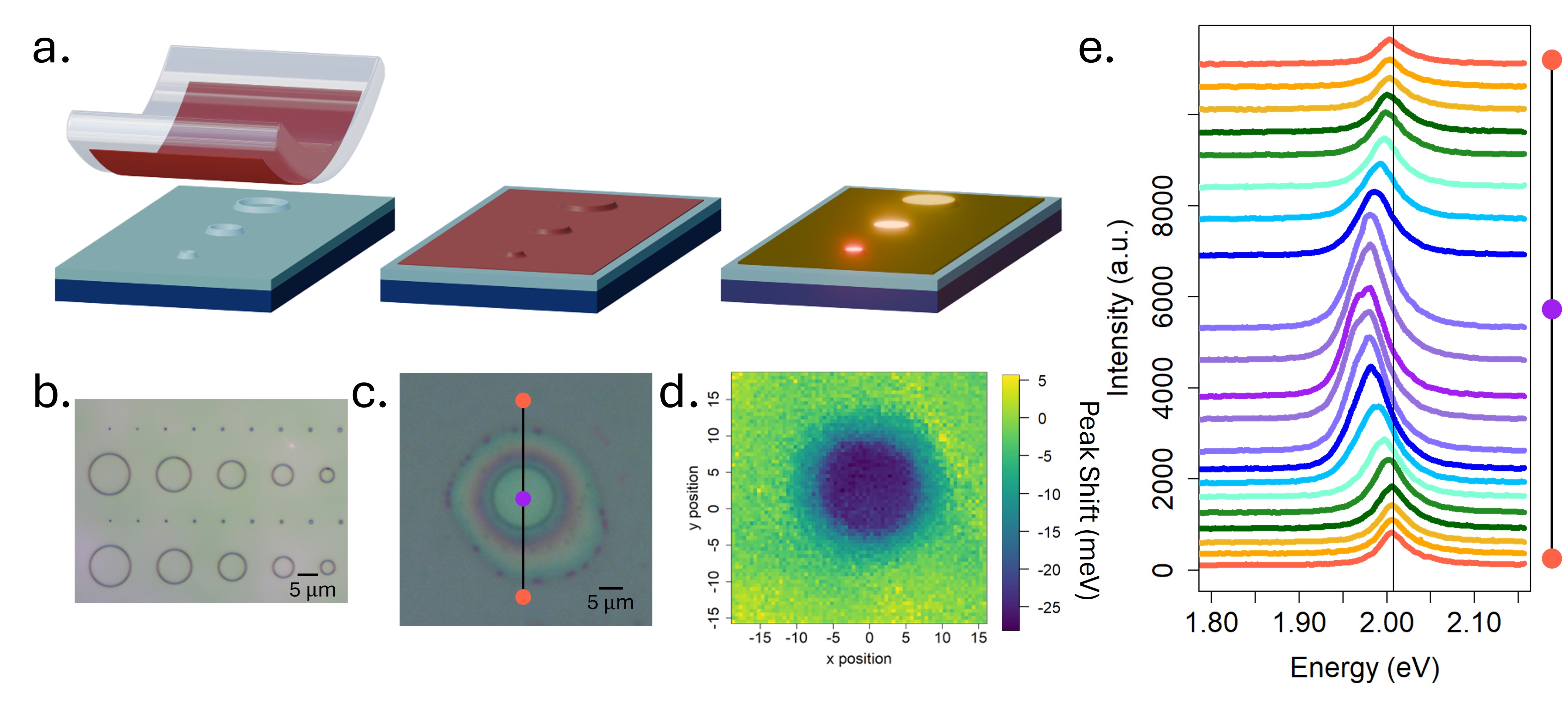}
  \caption{\textbf{Strain engineering in Ga$_2$Se$_2$ on isolated nanostructures.} (a) Schematic illustration of the transfer process placing Ga$_2$Se$_2$ flakes onto a patterned nanostructured substrate. (b) Optical image of the bare nanopatterned substrate showing ring structures and (c) of a Ga$_2$Se$_2$ flake suspended over the patterned substrate. (d) Spatially resolved PL peak shift map for a Ga$_2$Se$_2$ membrane suspended over a ring of $\sim$12 $\mu$m diameter, revealing a significant redshift localized to the suspended region. (e) Representative PL spectra as a function of position along the line indicated in (c), showing spatially-dependent redshift and enhanced emission intensity toward the center of the suspended region.}
  \label{single}
\end{figure}

Nanostructured features were fabricated on silicon-on-insulator (SOI, WaferPro) substrates using a dry-etching process that results in features protruding approximately 430 nm from the substrate surface \cite{Herrmann2024} (see Supporting Information). Large-area $\epsilon$-Ga$_2$Se$_2$ flakes were mechanically exfoliated and transferred onto these structures, forming suspended membranes over hollow regions, as illustrated in Figure \ref{single}(a–c). Outside the nanostructure perimeter the membrane conforms to the substrate. AFM measurements were performed on each structure to determine both the membrane thickness and the overall deformation profile (see Extended Date Figure 1). 
Room-temperature hyperspectral PL mapping was performed across 23 structures, yielding 84,525 spatially-resolved spectra. The resulting peak shift map (Figure \ref{single}(d)) and representative cross-sectional spectra (Figure~\ref{single}(e)) reveal enhanced emission intensity and bandgap shifts localized to the strained suspended regions, with the largest effects occurring directly above the hollow nanostructures.

We compute the local strain using a COMSOL Multiphysics FEA. A critical element of our approach is that the COMSOL strain calculation is fully constrained by experimental data. Specifically, the measured membrane thickness, Young's modulus, and Poisson's ratio for Ga$_2$Se$_2$ \cite{Chitara2018} are fixed inputs to the FEA. The only free parameter in the simulation is the boundary load that brings the 2D membrane into contact with the nanostructure and the flat substrate. As illustrated in Extended Data Figure 2, the magnitude of this boundary load is uniquely determined for each individual structure by minimizing the root-mean-square deviation (RMSE) between the computed deformation profile and that measured experimentally by AFM. Once the optimized boundary load is established, the full spatially resolved strain tensor is extracted from the simulation (see Extended Data Figure 3 for our estimation of error). From the tensor components ($\epsilon_{xx}$, $\epsilon_{yy}$, $\epsilon_{xy}$), the biaxial and uniaxial strain at each point are computed as:

\begin{equation}
    \epsilon_{bi}=\frac{\epsilon_{xx}+\epsilon_{yy}}{2}
    \label{hyd}
\end{equation}

and 

\begin{equation}
    \epsilon_{uni} = \sqrt{ \Biggl( \frac{\epsilon_{xx}-\epsilon_{yy}}{2} \Biggl)^2+\epsilon_{xy}^2}
    \label{dev}
\end{equation}

\noindent where $\epsilon_{uni} = 0$ when $\epsilon_{xx} = \epsilon_{yy}$ and $\epsilon_{xy}=0$ (see Supporting Information for derivation). These spatially resolved strain components are then correlated with the measured PL peak shifts at each point to extract experimental strain gauge factors. 

\begin{figure}[h]
\includegraphics[width=1\linewidth]{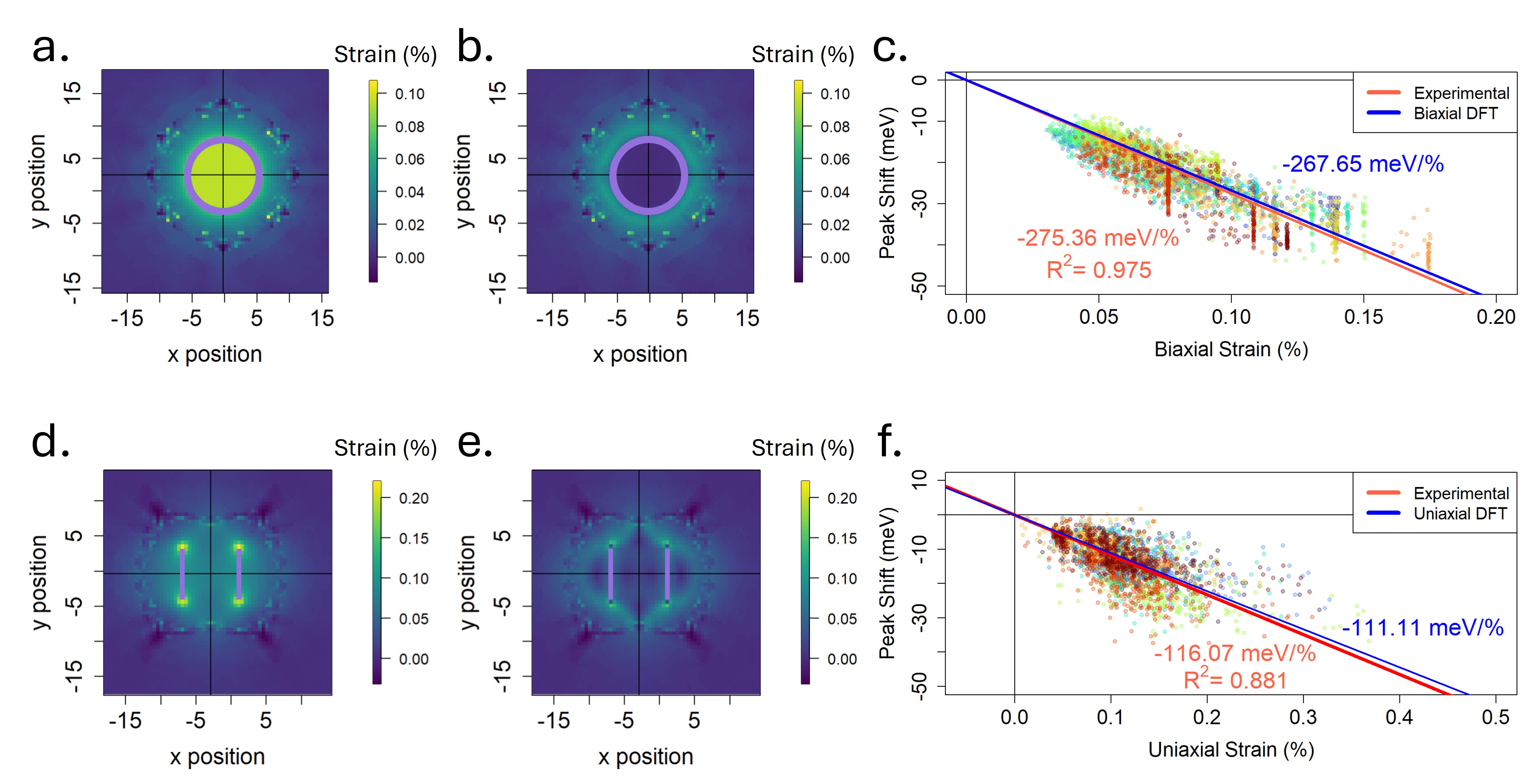}
  \caption{\textbf{Biaxial and uniaxial strain gauge factors for Ga$_2$Se$_2$.} (a,b) Biaxial and (d,e) uniaxial strain maps for a representative ring structure of 12 $\mu$m diameter (a,b) and ridge structure of 12 $\mu$m separation (d,e), each covered by a Ga$_2$Se$_2$ membrane. Purple lines indicate the location of the underlying ring or ridge nanostructures. (c,f) Experimentally measured PL peak shift as a function of (c) biaxial and (f) uniaxial strain, drawn from the suspended regions of 16 ring structures (4,523 spectra) and 7 ridge structures (2,803 spectra), respectively. Point colors distinguish individual samples. Red lines show linear fits to the experimental data; blue lines show gauge factors predicted by DFT.}
  \label{bi_uni}
\end{figure}

To isolate and quantify the effects of biaxial and uniaxial strain, we begin by studying two classes of isolated nanostructures: rings, which generate primarily biaxial strain in the suspended membrane, and parallel ridges, which generate primarily uniaxial strain. We studied 16 distinct ring structures with diameters ranging from 4$\mu$m to 14$\mu$m and 7 distinct ridge structures with separation distances ranging from 6 $\mu$m to 14 $\mu$m. We focus first on the suspended region over the hollow rings, where the geometry generates primarily biaxial strain. Figure \ref{bi_uni}(a,b) show the computed biaxial and uniaxial strain distributions, respectively, for a 12 $\mu$m diameter ring covered by a 250 nm thick Ga$_2$Se$_2$ membrane. The uniaxial strain within the suspended region is negligibly small (Figure \ref{bi_uni}(b)). In Figure \ref{bi_uni}(c), the experimentally measured PL peak shift is plotted as a function of computed biaxial strain across 4,523 spectra obtained from the 16 ring structures (flake thickness of 249 and 257 nm). The data follow a linear trend with high fidelity (R$^2 = 0.975$), yielding an experimental biaxial strain gauge factor of $\beta_{exp}$=-275.4 meV / \% strain. To confirm that the measured bandgap shifts arise from strain rather than other sources, we independently compute the expected bandgap shift as a function of biaxial strain in Ga$_2$Se$_2$ using Density Functional Theory (DFT, see Extended Data Figure 4 and Supporting Information). The experimentally-measured gauge factor is in excellent agreement with that computed by DFT ($\beta_{DFT}=-268$ meV / \% strain, blue line); the discrepancy of 2.8\% is well within the uncertainty of the material parameters used as simulation inputs. As shown in Extended Data Figure 5,  decreasing the ring diameter generates increasing biaxial strain, which paves the way toward predictive design of substrate features to achieve target strain values.

Ridge structures with varying separation distances and flake thicknesses from 75 nm to 108 nm were characterized following the same procedure. Figures \ref{bi_uni}(d,e) show the corresponding biaxial and uniaxial strain distributions for a representative ridge structure with a separation distance of 12 $\mu$m and a flake thickness of 107 nm. To isolate nearly pure uniaxial strain, we numerically screen the computed strain at every spatial point to identify locations where the ratio of the principal strain components ($\epsilon_1$ and $\epsilon_2$, see Supporting Information) is less than or equal to 0.15, below the Poisson ratio of 0.23 \cite{Chitara2018}, ensuring these locations are dominated by uniaxial deformation. Figure \ref{bi_uni}(f) plots the PL peak shift as a function of uniaxial strain for this screened subset of the data, comprising 2,803 spectra from 7 distinct ridge structures. A linear fit yields an experimental uniaxial strain gauge factor of $\alpha_{exp}$= -116.1 meV / \% strain (R$^2$=0.8812). The lower R$^2$ for the uniaxial data reflects the fact that ridges produce strain fields with less well-defined boundaries between uniaxial and mixed-strain regions, producing greater sample-to-sample variability. Despite this variability, the linear relationship between uniaxial strain and PL peak shift remains well-established across the full dataset. As shown in Extended Data Figure 6, DFT calculations give $\alpha_{DFT, z}$ = -111 meV / \% strain for uniaxial strain along the zigzag crystal axis and $\alpha_{DFT, a}$ = -108 meV/ \% strain along the armchair axis; the zigzag value is shown by the blue line in Figure \ref{bi_uni}(f) and differs from $\alpha_{exp}$ by only 4.4\%. The similarity of $\alpha_{DFT, a}$ and $\alpha_{DFT, z}$ is consistent with the absence of any experimentally measurable dependence of PL peak shift on strain orientation relative to the crystal axes, as described in the Supporting Information.

\section{Multiaxial Strain Prediction Across Full Spatial Profiles}
\begin{figure}[h]
  \includegraphics[width=1\linewidth]{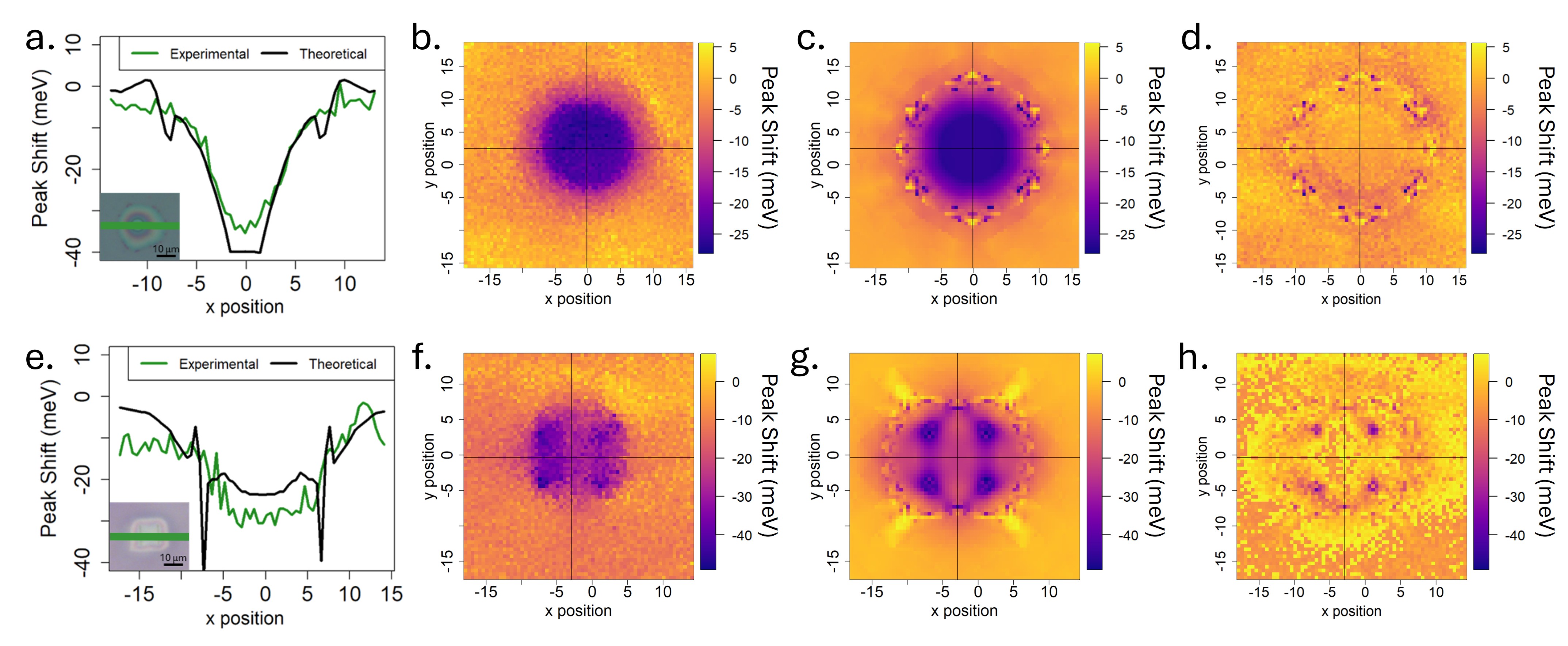}
  \caption{\textbf{Multiaxial PL peak shift prediction across full spatial profiles.} (a,e) Experimentally measured (green) and theoretically predicted (black) PL peak shift along representative cross-sections for (a) a 12 $\mu$m diameter ring with a 259 nm thick Ga$_2$Se$_2$ membrane and (e) an 8 $\mu$m separation ridge with a 75 nm thick flake. Two-dimensional maps of (b,f) experimentally measured peak shift, (c,g) peak shift predicted by Equation~\ref{PL peak shift} using experimentally determined strain gauge factors, and (d,h) the residual difference between experiment and prediction, for the same (b–d) ring and (f–h) ridge structures. Near-zero values across the difference maps confirm strong spatial agreement between experiment and model, with residual discrepancies confined to regions of peak strain concentration and substrate contact points.}
  \label{full}
\end{figure}

Having independently established the biaxial and uniaxial strain gauge factors for Ga$_2$Se$_2$, we now show that an analytic two-component model combining these two gauge factors predicts the PL peak shift at any spatial location. The expected change in bandgap energy at each point is calculated by

\begin{equation}
    \Delta E \approx \beta_{exp}\frac{\epsilon_1+\epsilon_2}{2}+\alpha_{exp}\frac{|\epsilon_1-\epsilon_2|}{2}
    \label{PL peak shift}
\end{equation}

\noindent where $\beta_{exp}$ and $\alpha_{exp}$ are the experimentally determined strain gauge factors and $\epsilon_{1}$ and $\epsilon_{2}$ are the local principal strain components extracted from the COMSOL simulations \cite{strain_calc,strain_calc02}. Figure \ref{full}(a,e) shows cross-sectional comparisons between the experimentally measured PL peak shift and the predictions of Equation \ref{PL peak shift} for a representative ring and ridge structure, demonstrating excellent agreement across the full spatial profile. Two-dimensional maps of the experimentally measured peak shift, the predicted peak shift from Equation \ref{PL peak shift}, and their difference are shown for a 12 $\mu$m diameter ring with a 259 nm thick Ga$_2$Se$_2$ membrane (Figure \ref{full}b-d) and for an 8 $\mu$m separation ridge with a 75 nm thick flake (Figure \ref{full}f-h). The near-zero difference maps in Figures \ref{full}(d,h) confirm that the two-component model captures the spatially resolved bandgap response across both structure types with high fidelity; residual discrepancies are confined to regions of peak strain concentration at ridge tips and substrate contact points where finite element meshing artifacts are expected.

Across all 23 structures, the normalized root mean square error (RMSE) between experimental PL peak shifts and the predictions of Equation \ref{PL peak shift} is 11.29\% (10.15\% for rings, 13.88\% for ridges; see Supporting Information). To identify the primary source of this error, we compute the normalized RMSE between the AFM-measured and COMSOL-predicted height profiles for each structure, finding a mean height RMSE of 7.22\% across all 23 structures (7.25\% for rings, 7.16\% for ridges). The close correspondence between the height RMSE and the peak shift RMSE indicates that the dominant source of prediction error is not model inadequacy but the inherent inability of finite element simulations to capture small variations in flake conformation that are experimentally unavoidable. Importantly, this implies that Equation \ref{PL peak shift} is capable of predicting spatially resolved PL peak shifts with less than 12\% error provided the local strain distribution is known with sufficient precision. Improved precision is readily achievable with improved AFM resolution or higher mesh-density simulations.

\subsection{Strain Cross-Talk in Interacting Nanostructures}
\begin{figure}[h]
  \includegraphics[width=1\linewidth]{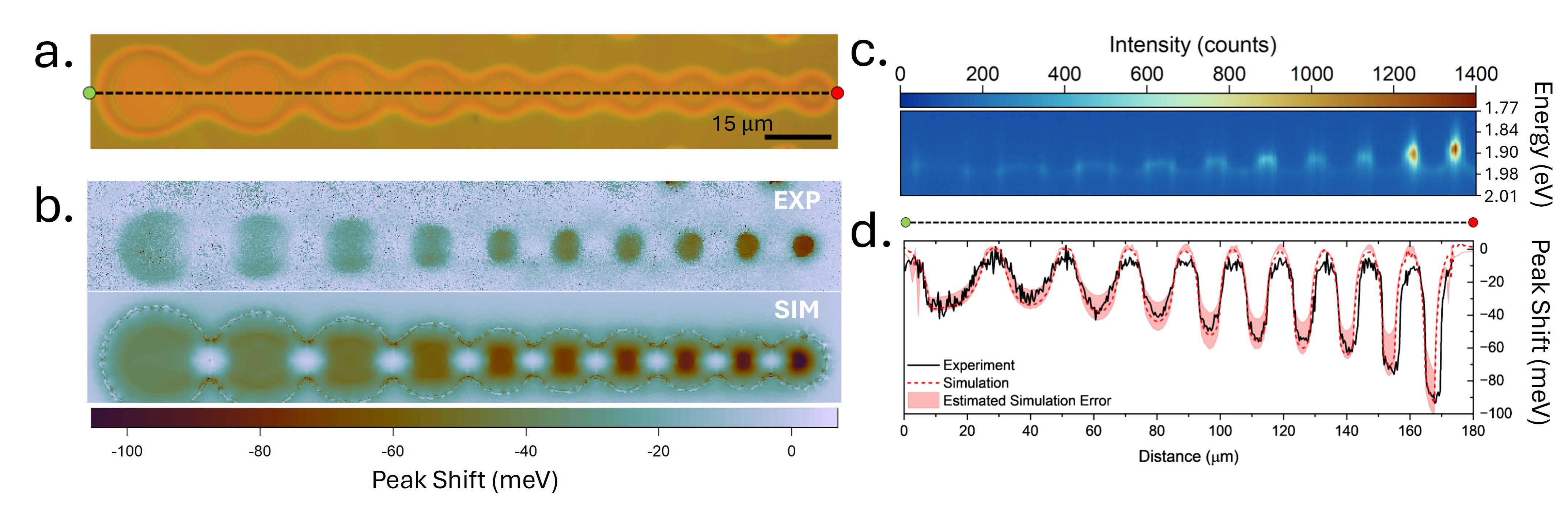}
  \caption{\textbf{Strain cross-talk between interacting nanostructures in Ga$_2$Se$_2$} (a) Optical microscope image of a Ga$_2$Se$_2$ flake transferred onto a row of rings with diameters of 15.0, 12.5, 10.0, 7.5, 5.0, 4.5, 4.0, 3.5, 3.0, and 2.5 $\mu$m from left to right, separated by a constant 10 $\mu$m gap. (b) Spatially resolved maps of the experimental (top) and simulated (bottom) PL peak shift, showing close correspondence between the optical response and the underlying strain field. (c) Heatmap of cross-sectional PL spectra taken along the dashed line in (a), from the green to the red circle, revealing progressive redshift and intensity enhancement with decreasing ring diameter. (d) Quantitative comparison of experimentally measured and FEA-predicted PL peak shift along the same cross-section, with shaded region indicating the estimated simulation uncertainty (see Supporting Information for derivation).}
  \label{ringrow}
\end{figure}

The isolated structures discussed above establish the quantitative foundation of our framework. However, practical device geometries will inevitably involve arrays of nanostructures in close proximity, where the strain fields of neighboring features interact. For example, we observed that when the spacing between neighboring rings is insufficient for the flake to fully adhere to the flat substrate surface between structures, the resulting deformation profile spans multiple features. Figure \ref{ringrow}(a) shows one such region, comprising a row of rings with diameters of 15.0, 12.5, 10.0, 7.5, 5.0, 4.5, 4.0, 3.5, 3.0, and 2.5 $\mu$m from left to right, separated by a constant 10 $\mu$m gap. This feature spacing prevents the flake from adhering to the substrate between structures, producing a complex, interconnected topography in which the strain field of each ring is influenced by its neighbors; a regime described as strain cross-talk.
 

As expected, PL mapping reveals a progressive redshift of the peak position with decreasing ring diameter, reaching a maximum redshift of 93 meV above the smallest ring (Figure \ref{ringrow}(b), top panel). However, we also observe spatial asymmetry in the PL peak shifts along the ring perimeters, with larger shifts occurring along the outer edges of the array compared to the centers of individual rings; this is a signature of the cross-talk strain field that develops when structures are closely spaced. Cross-sectional PL spectra taken along the dashed line in Figure \ref{ringrow}(a), plotted as a heatmap in Figure \ref{ringrow}(c), show that PL intensity increases with decreasing ring diameter, consistent with strain-enhanced radiative recombination. The bottom panel of Figure~\ref{ringrow}(b) shows the peak shift calculated from the simulated strain values using our two-component model. The excellent agreement between the predicted and experimentally-measured PL peak shifts confirms that the observed optical response is governed by the underlying strain distribution even in this complex geometry.

The simulated strain distribution also reveals several noteworthy features of the cross-talk geometry. Within each individual ring, the strain distribution follows a ``dumbbell'' shape, with higher strain values on the $\pm$ y edges of each circle. This asymmetry is absent in the isolated rings and attributable to the influence of neighboring rings on the local flake conformation. The maximum strain above each ring increases with decreasing ring diameter, reaching a maximum of 0.72\% above the smallest ring, consistent with the progressive PL redshift observed experimentally. The end rings of the array, which have only one neighboring structure rather than two, display a more symmetric strain distribution, further supporting the origin of the dumbbell asymmetry. Figure \ref{ringrow}(d) compares experimentally measured bandgap shifts with predictions using Equation \ref{PL peak shift} along the line profile indicated in Figure \ref{ringrow}(a), including the estimated simulation uncertainty. The excellent quantitative agreement demonstrates that the two-component model generalizes to this more complex strain environment, establishing FEA-guided design as a reliable route to tailored bandgap profiles across geometrically complex substrates.

\subsection{Tailoring Strain Distributions via Nanostructure Geometry}
Building on this predictive framework, we next demonstrate that deliberate variation of nanostructure geometry provides direct control over the spatial profile of strain induced in suspended Ga$_2$Se$_2$ membranes. As illustrated in Extended Data Figure 7, we fabricated a series of lobed nanostructures in which the strain gradient is engineered by radially varying the width of each lobe, controlling the degree of stretching as the flake conforms to the nanostructure and underlying substrate. These structures serve as a test case for the broader design principle that geometry can be systematically varied to produce target strain profiles, with FEA providing the quantitative link between a given geometry and its resulting strain distribution prior to fabrication. 

\begin{figure}[h]
  \includegraphics[width=1\linewidth]{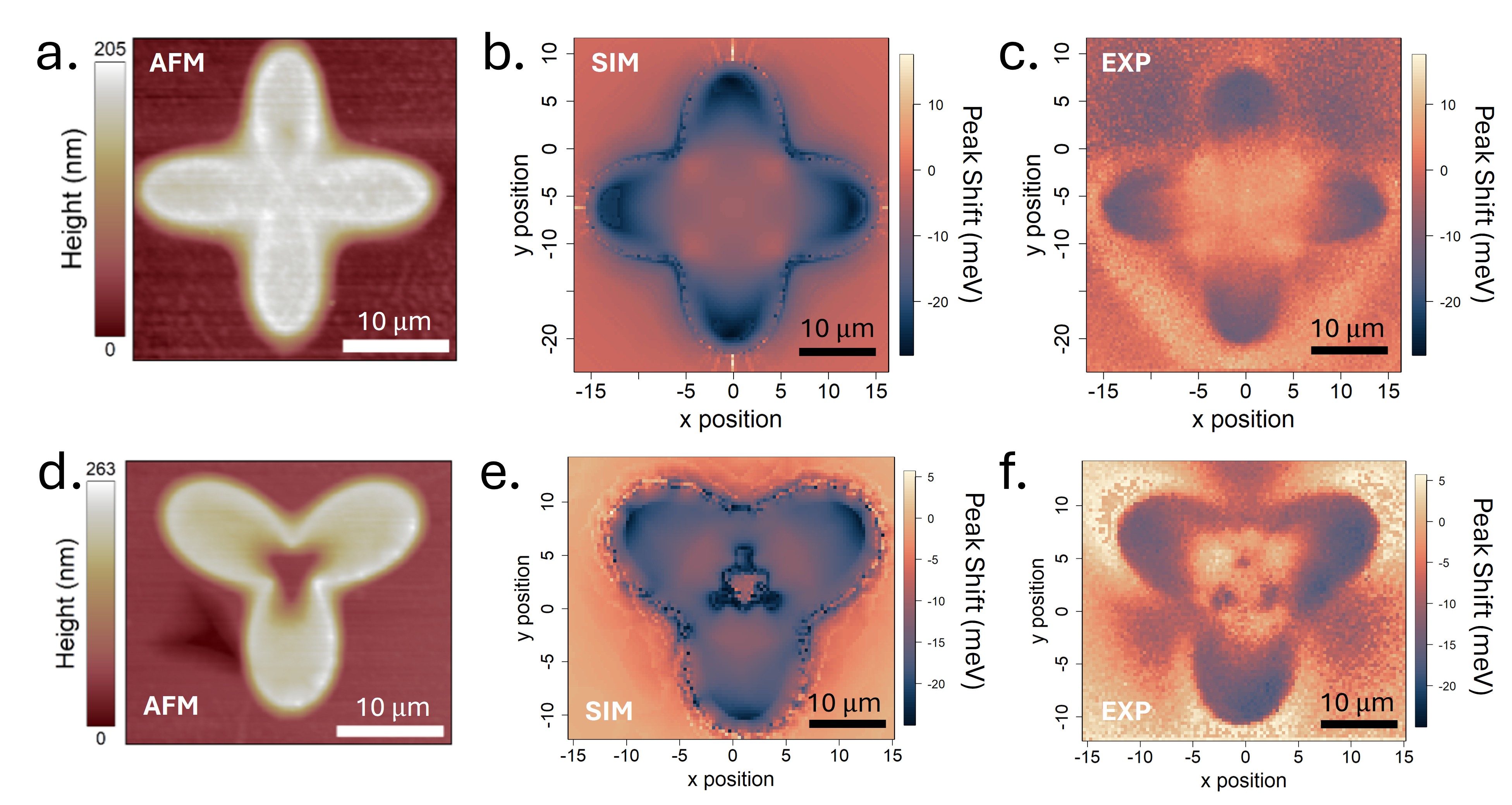}
  \caption{\textbf{Tailored strain gradient distributions in Ga$_2$Se$_2$ via lobed nanostructure geometry.} (a) AFM height map of a 140 nm thick Ga$_2$Se$_2$ flake transferred onto a four-lobed nanostructure. (b) Simulated and (c) experimentally measured PL peak shift maps for the four-lobed structure, showing a smooth strain gradient with the largest redshifts localized at the outer portions of the lobes and a low-strain region at the center. (d) AFM height map of the same flake transferred onto a three-lobed nanostructure. (e) Simulated and (f) experimentally measured PL peak shift maps for the three-lobed structure; the Ga$_2$Se$_2$ flake adheres to the substrate at the center while remaining suspended above the lobes, producing a distinct strain profile that is well reproduced by the FEA simulation.}
  \label{lobes}
\end{figure}
Figure \ref{lobes} shows data for a 140 nm-thick Ga$_2$Se$_2$ flake transferred onto a four-lobed and a three-lobed structure. An AFM map of the four-lobed structure is shown in Figure \ref{lobes}(a), with the corresponding predicted and experimentally-measured PL peak shift maps shown in Figures \ref{lobes}(b,c), respectively. The largest strains, and consequently largest PL redshifts, are localized at the outer portions of the lobes, while the center remains in a low-strain state. Between these extrema, the strain varies smoothly and continuously, confirming that radial variation of lobe width produces a well-controlled strain gradient. The analogous data for the three-lobed structure, shown in Figures \ref{lobes}(d–f), reveals that the flake adheres to the substrate at the center of the structure while remaining suspended above the lobes. We note that in the transition region where the flake moves from suspended to adhered, the simulated magnitudes exceed the corresponding PL redshift (see Supporting Information Figure S14). We attribute this to spatial averaging: the diffraction-limited PL experiment effectively integrates the response over an area of approximately 1 $\mu$m$^2$, which is large relative to the length scale over which the strain field varies. These results demonstrate that nanostructure geometry can be used to control both the strain magnitude and the strain gradient, providing a flexible and predictable design space for engineering complex bandgap profiles.


\subsection{Material Versatility}
\begin{figure}[h]
  \includegraphics[width=1\linewidth]{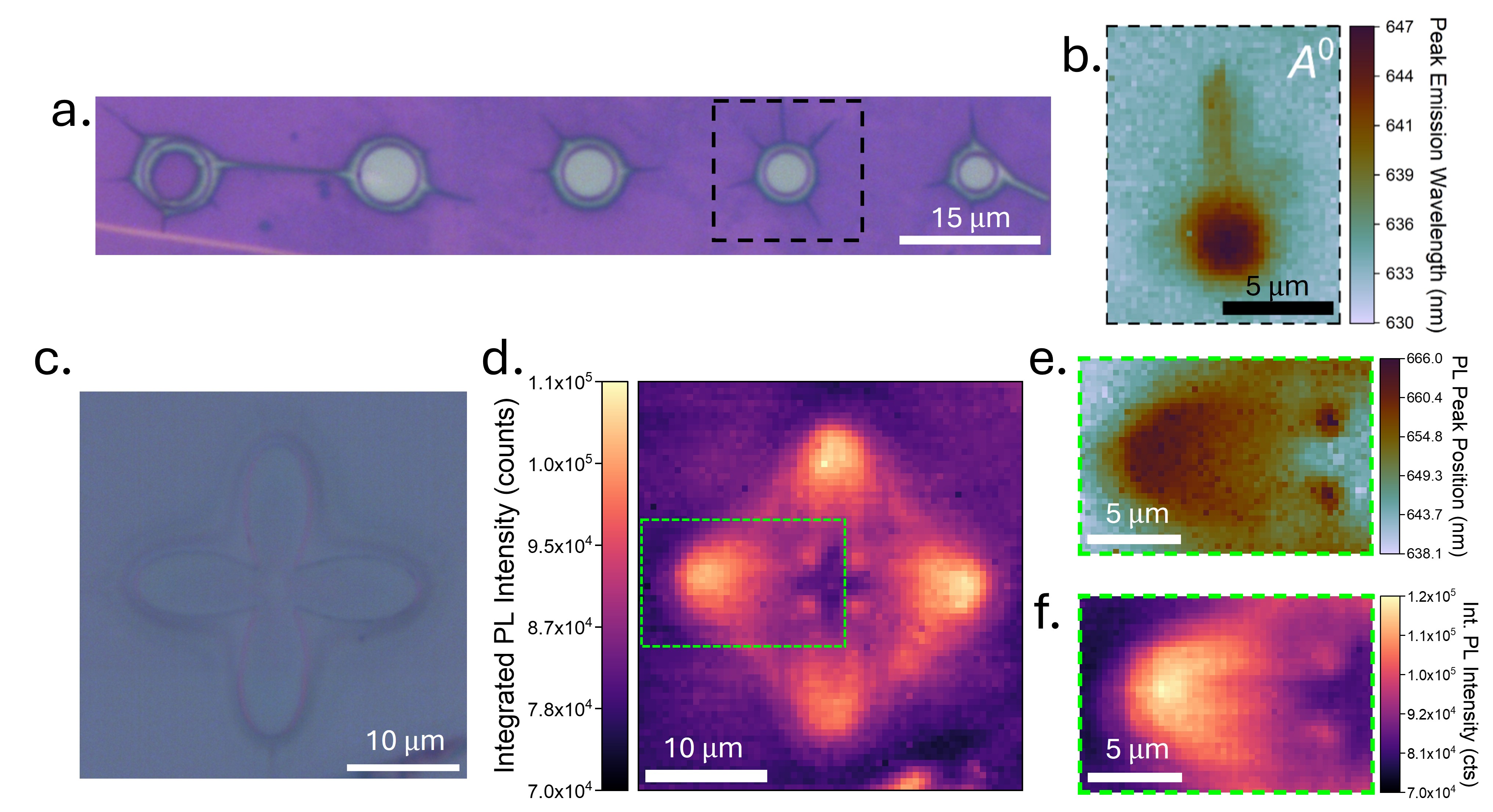}
  \caption{\textbf{Strain engineering of WS$_2$ over ring and lobed nanostructures.} (a) Optical microscope image of a 10~nm thick WS$_2$ flake transferred onto a series of ring nanostructures. The dashed box indicates the region for which a spatially-resolved map of the neutral exciton (A$^0$) peak emission is reported in (b), revealing a redshift of 46.1~meV localized to the suspended region above the 3.0~$\mu$m diameter ring. (c) Optical microscope image of a multilayered WS$_2$ flake transferred onto a four-lobed nanostructure. (d) Spatially-resolved map of integrated PL intensity showing enhanced emission localized to the suspended regions above the lobes. (e) High-resolution PL peak shift map and (f) integrated PL intensity map of the region highlighted by the green dashed box in (d), revealing spatially modulated optical response consistent with the underlying strain distribution. Scale bars in (e) and (f) are 5~$\mu$m.}
  \label{ws2}
\end{figure}

To further demonstrate the versatility of this platform, we extend our investigation to tungsten disulfide (WS$_2$) and conduct an analogous set of experiments. We quantitatively characterize the strain gauge factor by transferring a 10~nm-thick WS$_2$ flake onto a series of rings with diameters of 4.5, 4.0, 3.5, 3.0, and 2.5~$\mu$m (see Figure~\ref{ws2}(a)). The spatially resolved A$^0$ peak emission map (Figure~\ref{ws2}(b)) reveals a redshift of 46.1~meV localized to the suspended region above the ring. COMSOL FEA simulations constrained by AFM (see Supporting Information) indicate a biaxial strain magnitude of 0.35\% at the ring center, yielding an experimental biaxial strain gauge factor of 131.7~meV/\% that is in good agreement with calculated values reported for monolayer WS$_2$~\cite{strain_calc02}. We then transferred a multilayered WS$_2$ flake onto a four-lobed nanostructure, as shown in Figure~\ref{ws2}(c-f). The resulting gradients in PL intensity and peak shift parallel those reported above for Ga$_2$Se$_2$. These data  confirm that this nanostructured strain engineering platform generalizes to van der Waals semiconductor systems with distinct electronic structures and mechanical properties, requiring only material-specific gauge factor calibration.

\section{Conclusions}
We demonstrated a material-agnostic platform for deterministically realizing spatially-resolved multiaxial strain distributions in suspended van der Waals semiconductor membranes through deliberate nanostructure geometry design. We first combined FEA with hyperspectral PL mapping across 84,525 spectra and 23 distinct structures to extract experimental biaxial and uniaxial strain gauge factors for Ga$_2$Se$_2$ that agree with DFT calculations to within 3\% and 5\%, respectively. We next demonstrated that an analytical model predicts spatially-resolved bandgap shifts with less than 12\% error across isolated, interacting, and arbitrarily shaped nanostructure geometries. Finally, we used WS$_2$ to confirm that the platform and framework generalizes across van der Waals semiconductor systems, requiring only material specific gauge factor calibration. While this work uses bulk excitonic emission wavelength as the measure of local bandgap, the platform can immediately be applied to engineer strain distributions that create single-photon emitters, a critical capability for quantum photonic and quantum information applications. More broadly, the ability to prescribe spatially-varying strain profiles through substrate geometry alone provides a deterministic route to engineering local potentials and gradients across a wide range of strain-tunable phenomena in 2D materials, including magnetism, charge transport, and pseudomagnetic fields. These findings, in combination with recent efforts to dynamically engineer non-uniform strain~\cite{e64}, establish a quantitative foundation for the rational design of strain-engineered 2D material devices.

\section{Methods}
\subsection{Sample Preparation}
Patterned substrates were fabricated in the University of Delaware Nanofabrication Facility (UDNF) using methods adapted from a previously reported process \cite{Herrmann2024}, with full details provided in the Supporting Information. Nanostructures were defined on silicon-on-insulator (SOI) substrates using electron beam lithography (Raith EBPG 5200, 100 kV) with a bilayer resist stack, followed by chromium (Cr) hard mask deposition via electron beam evaporation (PVD Products). The nanostructures were etched into the silicon device layer using inductively coupled plasma etching, and the Cr hard mask was subsequently removed by wet chemical etching.

Bulk Ga$_2$Se$_2$ crystal (2D Semiconductors) was mechanically exfoliated into large-area flakes using polydimethylsiloxane (PDMS) stamps (Gel-Pak) and deterministically transferred onto the fabricated substrates via a water-assisted transfer procedure (see Supporting Information) \cite{Li2015}. Because of the optical degradation that occurs following exfoliation and transfer, a process we have previously characterized in detail \cite{degradation}, all samples were allowed to stabilize for 24 hours prior to PL mapping.
\subsection{Measuring PL}

A Horiba LabRAM HR Evolution bench top microscope was used for all PL collection. Measurements were taken at room temperature (298K) and were performed using an excitation wavelength of 532 nm with an intensity of 1.7 mW at the surface of the sample. Each optical measurement was performed using a 50X objective (1 $\mu$m spot size), dispersed with a 300 groove/mm grating to accumulate spectra spanning 575-725nm (1.71-2.16 eV), and measured with a Silicon CCD array detector with a 0.1 s integration time. For each structure, a map was taken over the entire strained area. The number of points in each map varied between 55x55 points to 80x80 points, depending on the size of the feature. The step size of the map was consistently 0.5 $\mu$m in both the x and y direction. 

All data analysis was done using RStudio running R version 4.3.2 using the packages `car',  `stats', `fields', `mclust', `viridis', `RColorBrewer', and `scico' \cite{car,stats,fields, mclust, viridis, brewer,scico}. Raw data was imported from text files extracted from Horbia's LabSpec 6 Spectroscopy Suite during collection. All PL data was normalized to unit area and the baseline was subtracted.

\subsection{Atomic Force Microscopy}
Atomic Force Microscopy (AFM) was conducted with an Anasys Nano IR2 in the Advanced Materials Characterization Laboratory (AMCL) at the University of Delaware. The AFM scans were first conducted on the edge of the flake to determine the flake thickness, which is used as an input for the COMSOL simulations. The thickness of flakes was determined by averaging the height profile across a 100 pixel area using the Gwyddion software package. Full AFM scans of every suspended flake provide the full deformation profile that was used to constrain the boundary load for each COMSOL simulation. The height of the bare nanostructures (i.e.~rings, ridges) was also confirmed via AFM.

\subsection{Quantifying Strain}

The finite element method in COMSOL Multiphysics was used to perform strain simulations and estimate the local strain components (see Supporting Information). The Ga$_2$Se$_2$ flakes were modeled as a membrane with biaxial mechanical properties, a Young's modulus of 82 GPa, and Poisson's ratio of 0.22 \cite{Chitara2018}. The flake thickness and nanostructure height obtained via AFM were also used as inputs to the simulation. To simplify the simulations, the deformation behavior of the Ga$_2$Se$_2$ flakes were approximated by applying a boundary load in the direction of the substrate. Deformation profiles were extracted for various boundary loads and compared to profiles obtained via AFM using a least-squares approach (see Supporting Information). Once the appropriate boundary load is determined, the simulation results are exported from COMSOL. These results are over a spatial grid 30 $\mu$m by 30 $\mu$m with 0.1 $\mu$m spacing. To directly compare to experimental results we apply a Gaussian smoothing filter (using the isoblur function in R with a sigma value of 5) and trimmed and centered the results to match the spatial coordinates of experimental maps. This process was performed using RStudio (version 4.3.2) using the packages `soundgen', `pracma', `imager', and `akima' \cite{soundgen,pracma,imager, akima}. 

\subsection{DFT calculations}
Structural parameters for DFT calculations were determined using the PBEsol exchange-correlation functional,\cite{Perdew2008} while the electronic properties were calculated with the Heyd-Scuseria-Ernzerhof (HSE) screen hybrid functional including spin-orbit coupling (SOC). All simulations were conducted using the VASP code with projector augmented plane wave (PAW) potentials, a plane-wave cutoff of 400 eV, and a 9×9×1 Monkhorst-Pack k-point mesh. Strain was applied along both the armchair (y) and zigzag (x) directions in the plane of the crystal. For each case, the lattice was adjusted to maintain zero stress in the transverse direction, and the out-of-plane lattice parameter was fully relaxed. The band structure was computed for strains ranging from 0\% to 5\%. Results show a linear decrease in band gap with increasing uniaxial strain: -0.108 eV/\% for strain along the armchair direction and -0.111 eV/\% along the zigzag direction. Importantly, the material retains its direct band gap across the full strain range, with the conduction and valence band edges remaining at the $\Gamma$ point. 

\section{Conflict of Interest}
EH and XW are listed as inventors on a pending patent application by the University of Delaware.

\begin{acknowledgement}
LLM and MFD gratefully acknowledge support from the National Science Foundation (Grant 2217786) and the II-VI / Coherent Foundation through their Block-Gift Program. EH and XW acknowledge the support from the U.S. National Science Foundation under Grant DMR-2128534. EH acknowledges funding by the AG Microsystems MEMS Research Fund. The authors also acknowledge the use of equipment in the Surface Analysis Faculty at UD supported by the National Science Foundation (Major Research Instrumentation Award Number: CHE-1828325).
\end{acknowledgement}

\begin{suppinfo}
\noindent The Supporting Information includes detailed explanation of sample preparation, AFM, PL hyperspectral imaging, DFT calculations, COMSOL Multiphysics simulations, data analysis, and tungsten disulfide strain engineering.  
\end{suppinfo}


\bibliography {ref}

\newpage
\section{Extended Data}
\begin{figure}[h]
  \includegraphics[width=0.9\linewidth]{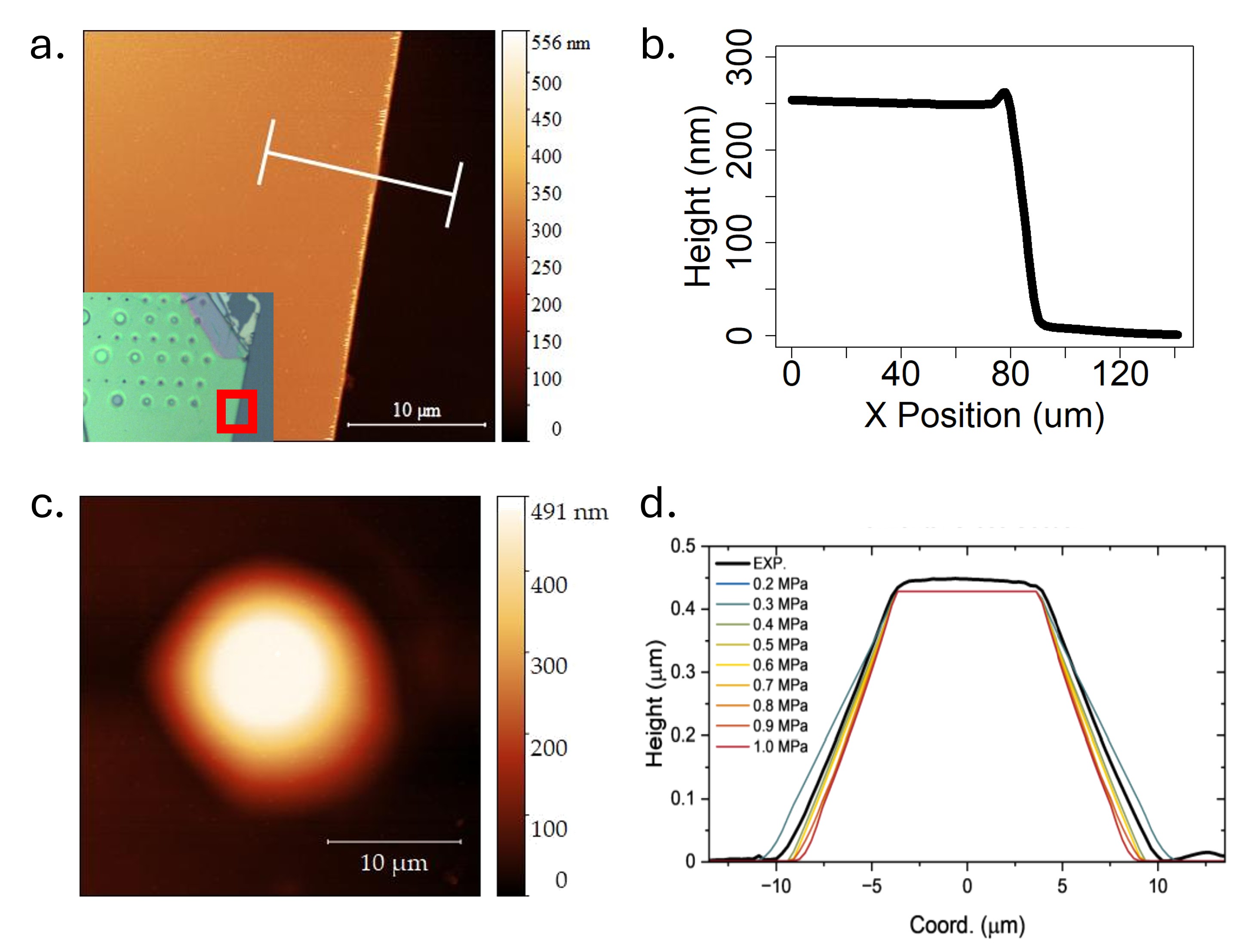}
  \caption*{\textbf{Extended Data Fig.~1$|$ AFM characterization of Ga$_2$Se$_2$ flake thickness and membrane deformation.}  (a) AFM scan across the edge of a transferred Ga$_2$Se$_2$ flake. (b) Height profile extracted from the scan in (a), from which the flake thickness is determined by averaging over a 100 pixel area. (c) AFM scan of the Ga$_2$Se$_2$ membrane suspended over a ring nanostructure. (d) Cross-sectional height profiles from the AFM data (black) and COMSOL simulations performed over a range of boundary loads (colors), illustrating the sensitivity of the simulated deformation profile to the applied load and the quality of the best-fit result.}
  \label{afm_strain}
\end{figure}

\begin{figure}
  \includegraphics[width=1\linewidth]{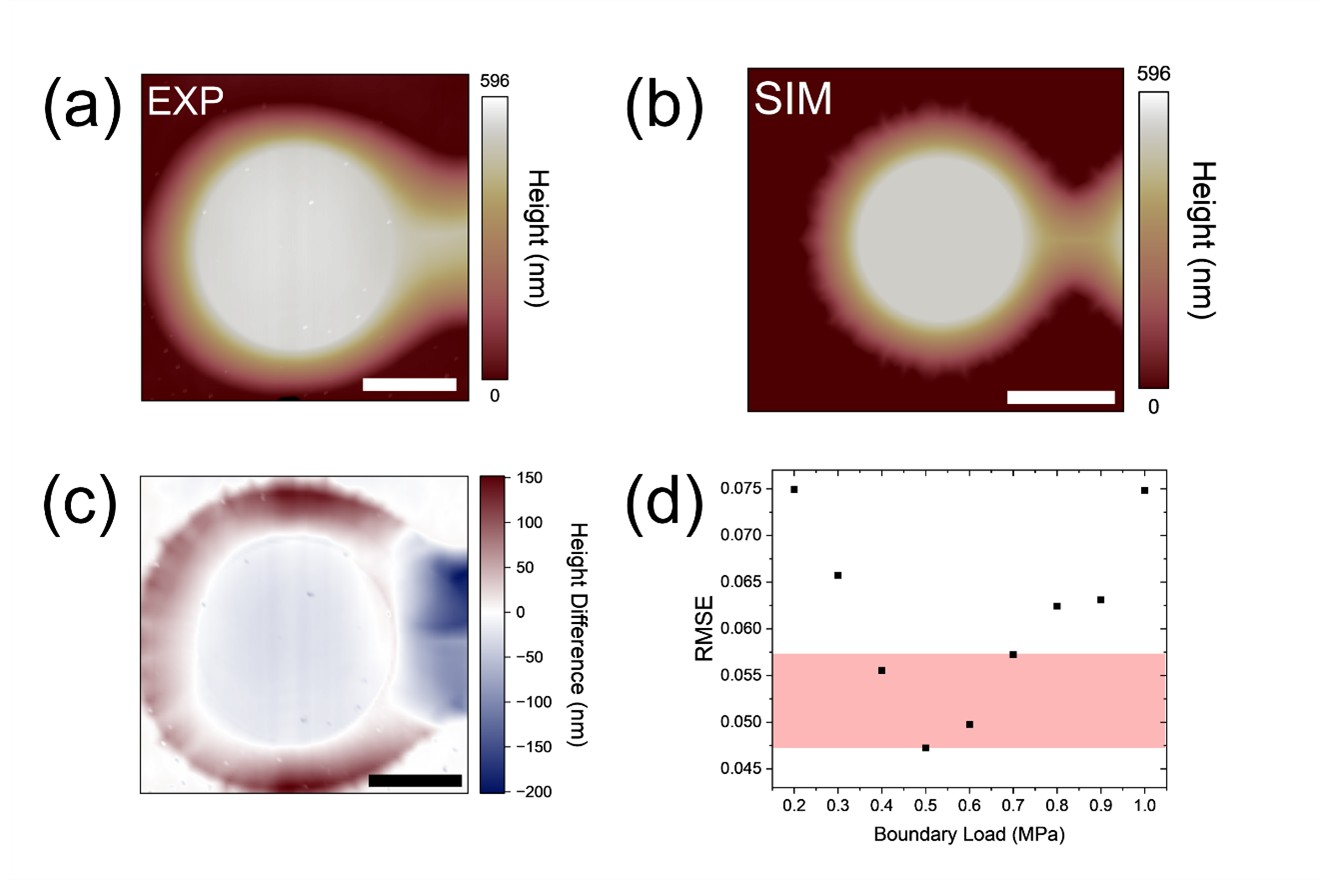}
  \caption*{\textbf{Extended Data Fig.~2$|$ Calibration of the finite element simulations.} (a) Experimental AFM height map of a Ga$_2$Se$_2$ flake deformed over the 15 $\mu$m diameter ring shown. (b) Simulated deformation profile for a boundary load of 0.5 MPa. (c) Spatial distribution of height differences between the aligned datasets in (a) and (b), illustrating the residual disagreement between experiment and simulation at the optimal load. (d) Root mean square error (RMSE) between the AFM map and simulated deformation profiles across all tested boundary loads; the red shaded region represents a 20\% deviation from the minimum RMSE and defines the boundary load confidence interval used to estimate strain uncertainty. All scale bars are 10 $\mu$m.}
  \label{si_cal}
\end{figure}

\begin{figure}
  \includegraphics[width=1\linewidth]{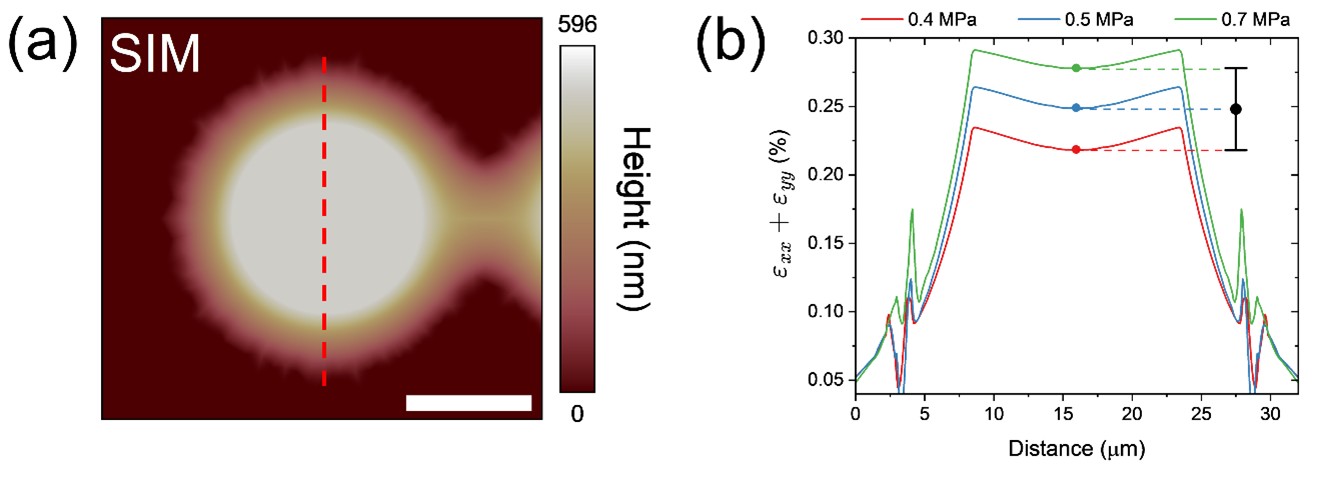}
  \caption*{\textbf{Extended Data Fig.~3$|$Estimated error in the simulated strain distribution.} (a) Simulated deformation profile of the Ga$_2$Se$_2$ membrane for a boundary load of 0.5 MPa. Scale bar is 10 $\mu$m. (b) Simulated cross-sectional strain distributions taken along the red dashed line in (a) for boundary loads of 0.4, 0.5, and 0.7 MPa, corresponding to the edges and center of the boundary load confidence interval. The estimated strain error, indicated by the black error bar, is constructed from the spread in strain values at the center of the suspended region across these three boundary loads, yielding a representative uncertainty of $\pm$12\% in reported strain magnitudes.}
  \label{si_err}
\end{figure}

\begin{figure}
  \includegraphics[width=0.75\linewidth]{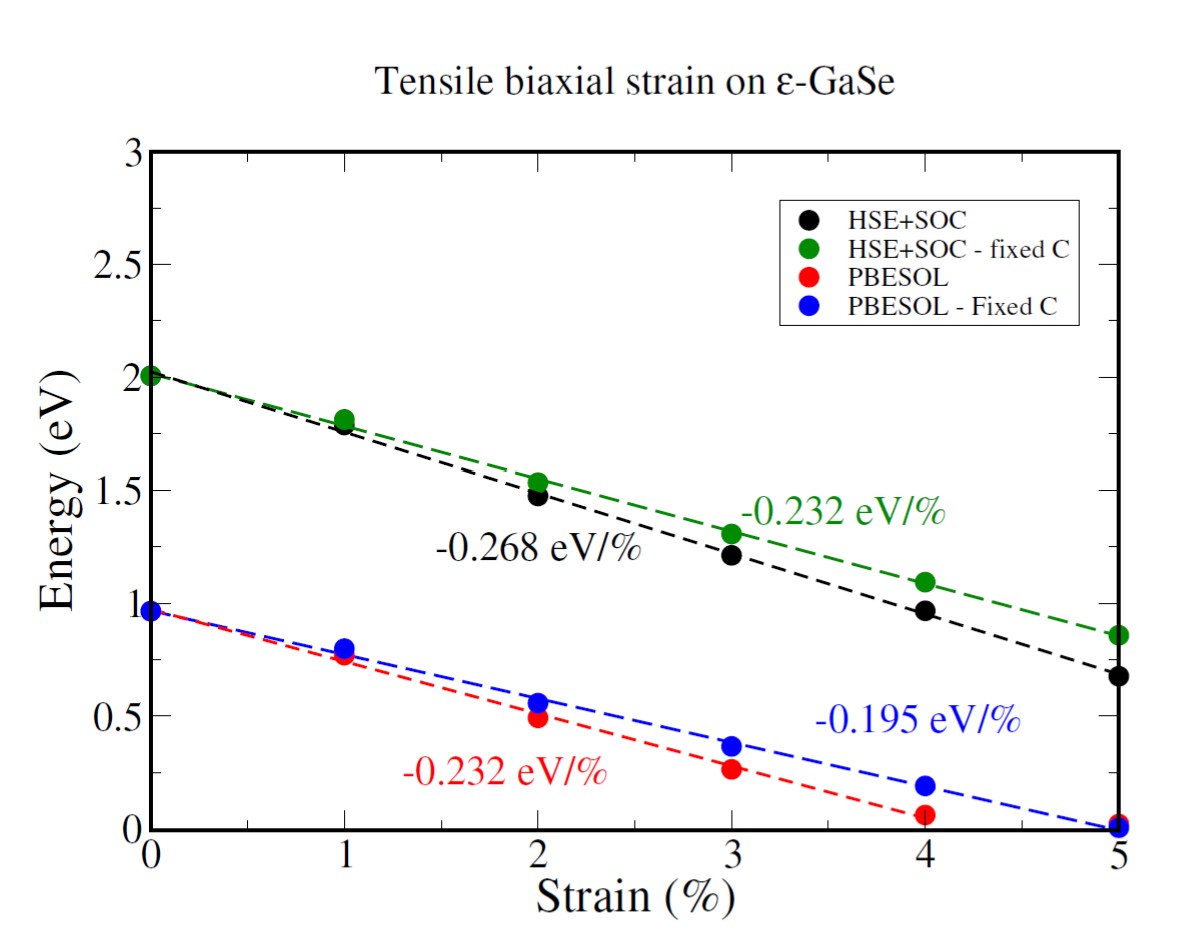}
  \caption*{\textbf{Extended Data Fig.~4$|$ Bandgap evolution of Ga$_2$Se$_2$ with applied biaxial strain computed using different DFT functionals.} Bandgap energy as a function of applied biaxial tensile strain computed using the PBEsol functional with (red) and without (blue) out of plane lattice relaxation, and the HSE$+$SOC functional with (black) and without (green) out of plane relaxation. The HSE$+$SOC functional with relaxation yields a strain gauge factor of $-0.268$ eV$/\%$ strain, which is taken as the theoretical biaxial gauge factor for comparison with experiment.}
  \label{SI_dft_biaxial}
\end{figure}

\begin{figure}[h]
  \includegraphics[width=1\linewidth]{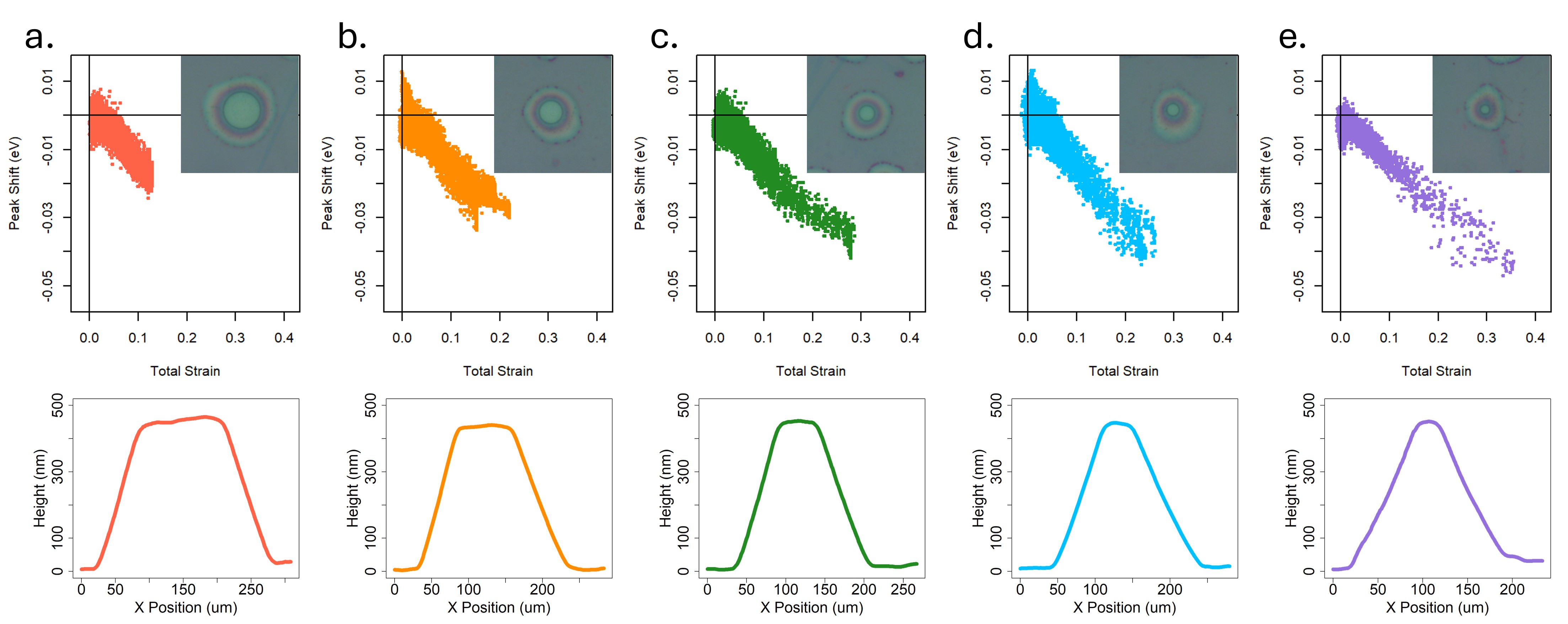}
  \caption*{\textbf{Extended Data Fig.~5$|$ Strain dependent PL peak shift for rings of varying diameter.} (a-e) PL peak shift as a function of total strain ($\epsilon_{bi}+\epsilon_{uni}$) for rings of approximately 14, 11, 6.5, 5.5, and 4 $\mu$m diameter, respectively. Insets show an optical images of each ring structure. Lower panels show AFM cross-sectional height profiles for each corresponding structure. The maximum total strain and corresponding PL redshift increase systematically with decreasing ring diameter, consistent with greater membrane stretching over smaller suspended regions.}
  \label{diameters}
\end{figure}

\begin{figure}
  \includegraphics[width=0.75\linewidth]{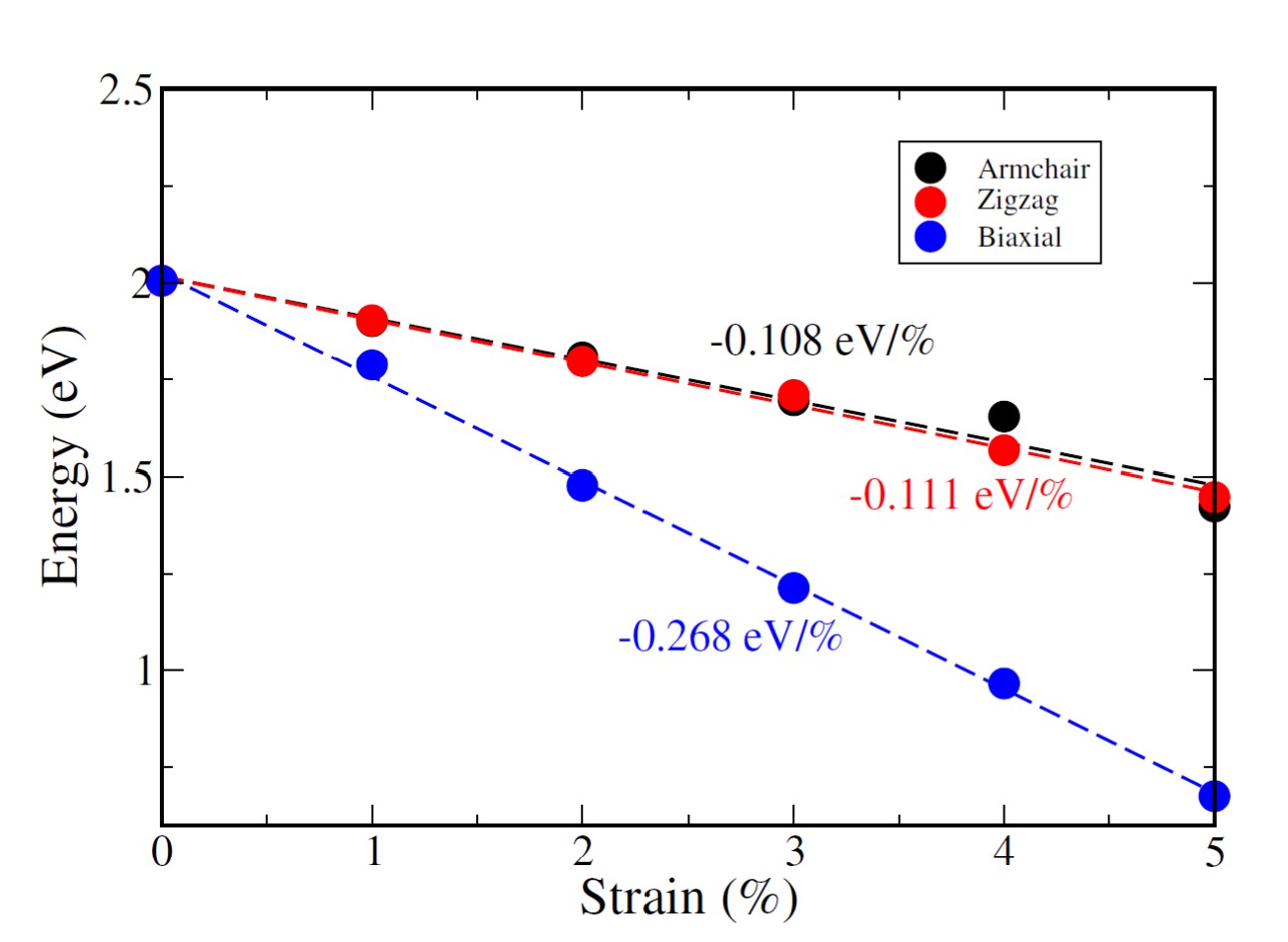}
  \caption*{\textbf{Extended Data Fig.~6$|$ Bandgap evolution of Ga2Se2 with applied uniaxial and biaxial strain.} Bandgap energy as a function of applied tensile strain along the armchair (black) and zigzag (red) crystal directions, and under biaxial strain (blue), computed using the HSE+SOC functional. The bandgap decreases linearly with strain in all cases, with gauge factors of -0.108 eV/\% (armchair), -0.111 eV/\% (zigzag), and -0.268 eV/\% (biaxial). The near-identical armchair and zigzag values are consistent with the absence of any experimentally measurable dependence of PL peak shift on strain orientation relative to the crystal axes.}
  \label{SI_dft_tensile}
\end{figure}

\begin{figure}[h]
  \includegraphics[width=1\linewidth]{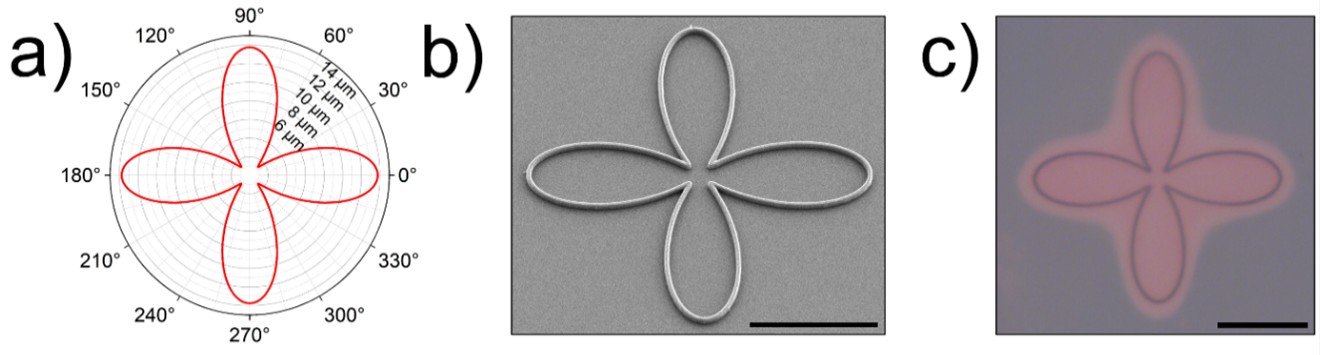}
  \caption*{\textbf{Extended Data Fig.~7 $|$ Fabrication and optical characterization of the four-lobed nanostructure.}
  (a) Polar coordinate equation r$ = 7.5 + 6.25\cos(4\theta)$ used to define the four-lobed geometry, with radial contours indicating feature dimensions. (b) Tilted scanning electron microscope image of the fabricated nanostructure etched to a height of approximately 200 nm. (c) Optical microscope image of a 140 nm thick Ga$_2$Se$_2$ flake transferred onto the nanostructure, confirming successful membrane transfer over the lobed geometry. All scale bars are 10 $\mu$m.}
  \label{4lobed}
\end{figure}
\end{document}


\section{Sample Preparation}
All nanostructures were fabricated on commercially available silicon-on-insulator (SOI) substrates using methods adapted from a previously reported process \cite{Herrmann2024}. The substrates were first cleaned with an acetone and isopropyl alcohol (IPA) wash. A bilayer resist stack consisting of lift-off (LOR3A) and electron beam (AR-P 6200.04) resists with a combined thickness of ~250 nm was then spun onto the bare SOI surface. The resist was patterned using electron beam lithography (Raith EBPG 5200, 100kV) and subsequently developed in AR 600-546 and IPA for the lift-off and electron beam resists, respectively. Rings and ridges of various diameters and separations were defined, each with an in-plane thickness of approximately 100 nm. A 25 nm thick chromium (Cr) hard mask was then deposited at 0.2 Å/s by electron beam evaporation (PVD Products). Following liftoff, the nanostructures were etched into the SOI substrate using an inductively coupled plasma etching system with CHF$_3$ (30 sccm) and C$_4$F$_8$ (40 sccm)  gases. The etch time was set to 10 minutes, resulting in a nanostructure height of approximately 420 nm. The remaining Cr hard mask was removed by immersing the samples in commercially available Cr etchant held at 40°C for one hour. 

Following the fabrication of the nanostructures, large-area $\epsilon-$Ga$_2$Se$_2$ flakes of relatively uniform thickness were exfoliated from a bulk crystal (2D Semiconductors) using polydimethylsiloxane (PDMS) strips (Gel-Pak). Exfoliated flakes were then deterministically transferred onto the patterned substrates through a water-assisted procedure \cite{Li2015}. A bead of water is first picked up at the location of the flake and then aligned above the target nanostructure. The flake is then lowered and brought into contact with the substrate at room temperature. With the flake and substrate in contact, the substrate is gradually heated to 95°C to enable the release of the flake from the PDMS and evaporation of the water. The sample is then allowed to cool to room temperature before performing photoluminescence mapping. 

We conduct atomic force microscopy (AFM) measurements on each structure to obtain a precise measure of the membrane thickness (Extended Data Figure 1(a, b)) and a spatially-resolved profile of the deformation of the suspended membrane (Extended Data Figure 1(c, d)). The AFM scans were first conducted on the edge of the flake to determine the flake thickness. The thickness of flakes was determined by averaging the height profile across a 100 pixel area using the Gwyddion software package. Full AFM scans of every suspended flake provide the full deformation profile that was used to constrain the boundary load for each COMSOL simulation. The height of the bare nanostructures (i.e.~rings, ridges) was also confirmed via AFM. The black line in Extended Data Figure 1(d) shows a cross section of the experimentally-measured deformation of the membrane suspended over a ring structure. The colored lines show COMSOL simulations for a range of boundary load values; the best fit is obtained for a boundary load of 0.4 MPa. 


\section{Finite Element Simulations}
We employ the finite element method in COMSOL Multiphysics to perform all strain simulations presented in the main text. In all simulations, Ga$_2$Se$_2$ is modeled as a membrane with isotropic mechanical properties, Young’s modulus of 82 GPa, and Poisson’s ratio of 0.22 \cite{Chitara2018}. The thickness of the Ga$_2$Se$_2$ flake and height of the nanostructures are defined using experimentally measured values. The presence of the nanostructure is approximated by projecting its in-plane geometry onto the Ga$_2$Se$_2$ membrane surface. The projected region is constrained to in-plane motion to mimic frictionless contact between the Ga$_2$Se$_2$ flake and the apex of the nanostructure. 

The experimental transfer procedure involves complicated interactions between the Ga$_2$Se$_2$ flake and nanostructured surface, including deformation of the PDMS stamp and capillary forces from the interaction with water. To simplify the simulations, we approximate the deformation behavior of the Ga$_2$Se$_2$ flake by applying a boundary load to the upper surface of the Ga$_2$Se$_2$ in the direction of the substrate. Contact between the Ga$_2$Se$_2$ and flat substrate surface is defined such that the Ga$_2$Se$_2$ domain adheres to the substrate once contact is made. To limit the number of fitting parameters, the simulations exclude frictional forces and decohesion of the Ga$_2$Se$_2$ from the substrate. The only free parameter in the fit is the boundary load, which is individually calibrated for each structure by varying the magnitude of the force per unit area and comparing the resulting simulated deformation profiles to the measured AFM height profiles using a least-squares approach. 

\subsection{Calibration Procedure}
\begin{figure}
  \includegraphics[width=1\linewidth]{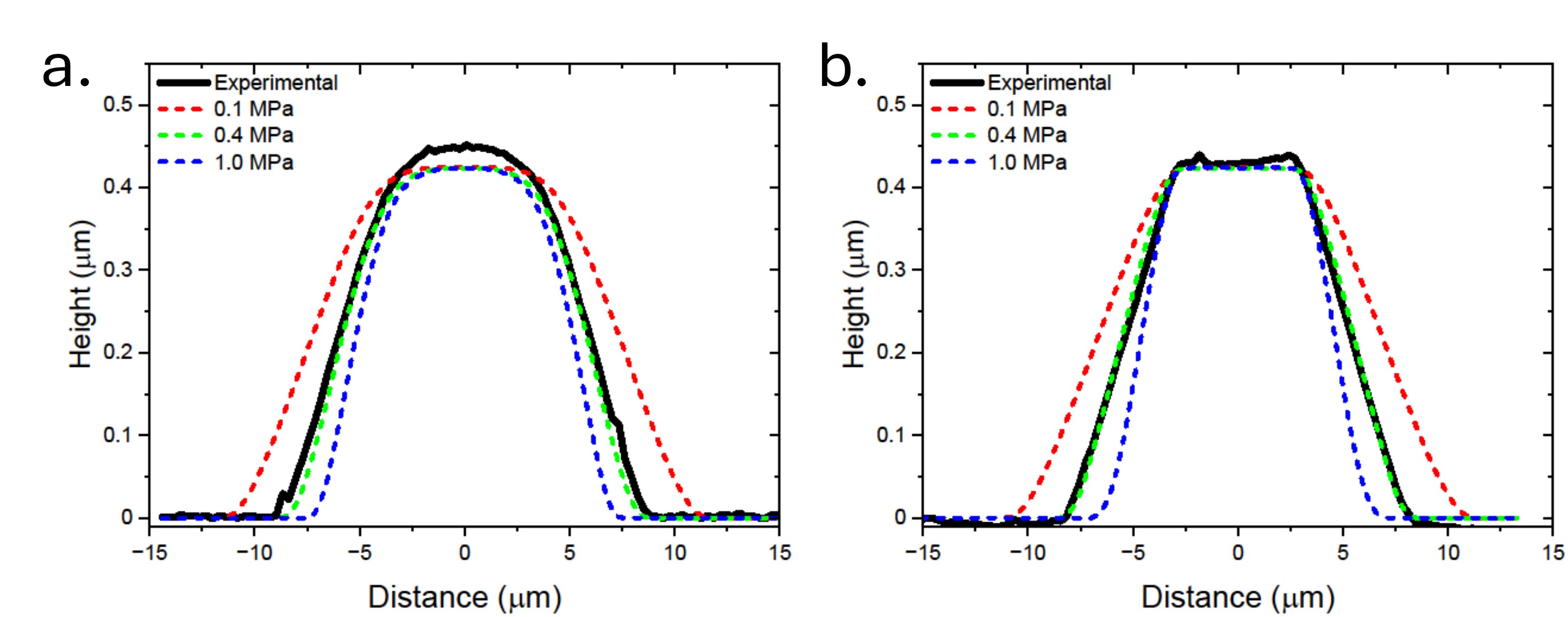}
  \caption{Comparisons of experimental and simulation height profiles taken along the (a) horizontal and (b) vertical cross-sections for a Ga$_2$Se$_2$ flake transferred onto a representative ridge nanostructure.}
  \label{si_ridge_calibration}
\end{figure}

The calibration procedure was performed on all nanostructures and an optimized boundary load was used for each individual structure. In Figure S\ref{si_ridge_calibration}, we compare simulated and AFM cross-sectional height profiles of Ga$_2$Se$_2$ transferred onto representative ridge nanostructures under various boundary loads. 

\begin{figure}
  \includegraphics[width=0.55\linewidth]{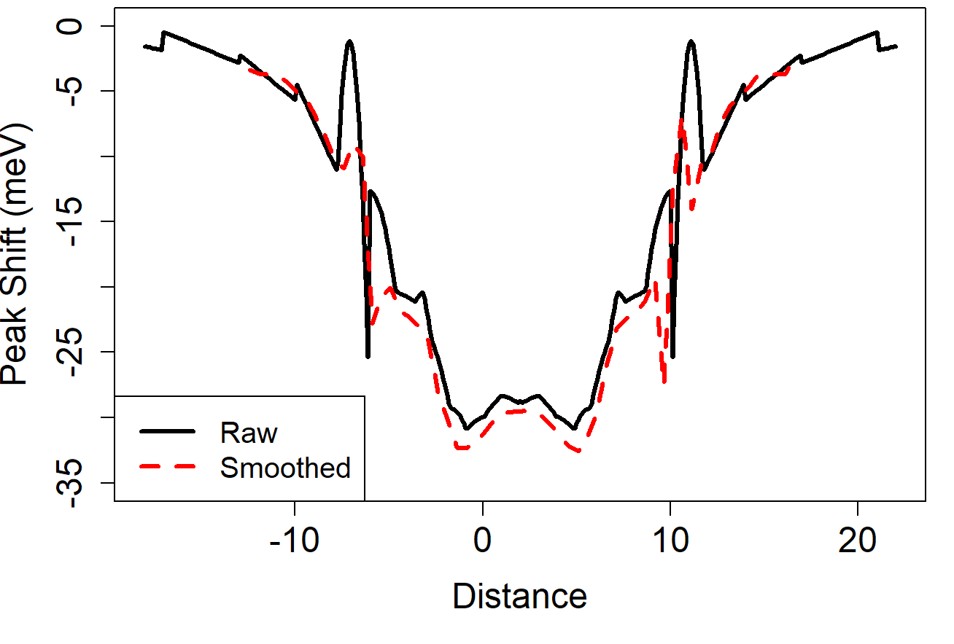}
  \caption{Comparison between raw (black) and smoothed (dashed red) simulated PL shift data taken along a cross-section through a Ga$_2$Se$_2$ flake suspended on a representative ridge nanostructure.}
  \label{si_shift}
\end{figure}

The strain components are extracted from the deformed Ga$_2$Se$_2$ simulation domain over a square grid of 30 µm width and point spacing of 0.1 µm. The finite nature of the simulation mesh elements introduces sharp transitions in strain magnitude and direction that are not captured by experimental photoluminescence maps. For direct comparison with experimental results, we apply a Gaussian smoothing filter of 2 µm diameter to the extracted simulation data to approximate the averaging effects of the laser spot size on the acquired data. We present an example of this smoothing process on the simulation data in Figure S\ref{si_shift}, where we show simulated PL shift data both before and after smoothing.

In Extended Data Figure 2(a), we show the AFM map used in the calibration of the boundary load. This AFM map corresponds to the 96 nm thick GaSe flake transferred over the largest diameter (15 µm) ring that was first presented in Figure 4 of the main text. In Extended Data Figure 2(b), we show an example deformation profile obtained from a single boundary load simulation. To compare the simulated deformation profile to that obtained via AFM, the two datasets are first spatially aligned and the simulation dataset is interpolated onto the AFM data grid. To quantify the agreement between these two datasets, we calculate the root mean square error (RMSE) over the entire grid using the equation

\begin{equation}
    \text{RMSE} = \sqrt{\frac{\sum^{N}_{i=1}(z_{SIM,i}-z_{AFM,i})^2}{N}} 
    \label{rmse}
\end{equation}

where $z_{SIM}$ and $z_{AFM}$ are the simulated and AFM-measured height values at grid point $i$, respectively, and $N$ is the total number of grid points. For visualization purposes, we show in Extended Data Figure 2(c) a spatial distribution of the height disagreement corresponding to the quantity $z_{SIM}-z_{AFM}$ across all grid points for the datasets shown in Extended Data Figure 2(a,b). The RMSE thus provides a single measure that quantifies the agreement between the AFM data and each boundary load simulation. We note here that a lower RMSE value corresponds to better agreement between datasets. To calibrate the simulations to the AFM data in Extended Data Figure 2(a), we perform simulations for various boundary loads ranging from 0.2-1.0 MPa in steps of 0.1 MPa. The RMSE is then calculated for each independent simulation using the above RMSE equation and the resultant RMSE values are plotted as a function of boundary load in Extended Data Figure 2(d).  


\subsection{Estimated Error}
The limitations inherent to the membrane simulation model prevent a precise description of Ga$_2$Se$_2$ deformation, even when the optimized boundary load is applied. This limitation is illustrated by the distribution of height differences depicted in Extended Data Figure 2(c). To estimate the error in simulated strain, we consider a deviation of approximately 20\% from the minimum RMSE value obtained at a boundary load of 0.5 MPa for the suspended flake shown in Figure 4 in the main text. Under this approximation, the boundary load that represents the “true” strain is estimated to fall within the range of 0.4 to 0.7 MPa. We will refer to this range as our boundary load confidence interval (BLCI). We then use the strain values simulated at the edges of the BLCI to define an error range for the reported strain magnitudes. In Extended Data Figure 3(a) we show the simulated deformation profile of the calibrated Ga$_2$Se$_2$ region for a boundary load of 0.5 MPa. In Extended Data Figure 3(b), we plot the simulated biaxial strain magnitudes $\epsilon_{xx}+\epsilon_{yy}$ for boundary loads of 0.4, 0.5, and 0.7 MPa along a cross-sectional line through the deformed membrane, as depicted by the red dashed line in Extended Data Figure 3(a). We compare simulated strain magnitudes at the center of the suspended region and obtain values of approximately 0.22\%, 0.25\%, and 0.28\%, for the 0.4 MPa, 0.5 MPa, and 0.7 MPa boundary loads, respectively. Thus, the true strain at the center of the suspended region is estimated to be within the range of $\epsilon_{xx}+\epsilon_{yy}=0.25\pm0.03\%$. This deviation corresponds to an approximate error in the simulated strain of ±12\%. We consider this error to be representative of all strain values reported in this work. 

We now extend the error analysis to our estimation of the bandgap shift from simulated strain. The region considered for calibration of the boundary load in this Section is a small region of the much larger strained GaSe flake previously shown in Figure 4(a) of the main text. Thus, the region shown in Extended Data Figure 2 represents only a small region of the membrane used during the calibration procedure. To approximate the error in the bandgap shift presented at a boundary load of 0.5 MPa, we construct an error range from the bandgap shift simulated using boundary loads of 0.4 and 0.7 MPa, the edges of the BLCI. This range is highlighted by the red shaded region in Figure 4(d) of the main text.

Although our model captures the dominant effects of strain on the PL peak behavior in Ga$_2$Se$_2$, we note the limitations of this model. First, we describe the three-dimensional geometry of Ga$_2$Se$_2$ as a two-dimensional membrane, which treats the thin film as having zero out-of-plane stiffness and ignores interfacial slip between layers. Further, frictional effects and decohesion of the Ga$_2$Se$_2$ membrane from the substrate are ignored to reduce complexity of the simulation. Including these physics in the simulations may improve the model and reduce error by better prescribing strain relaxation dynamics as the boundary load is removed from the membrane. 


\subsection{COMSOL Comparison to AFM}
\begin{figure}[h]
  \includegraphics[width=1\linewidth]{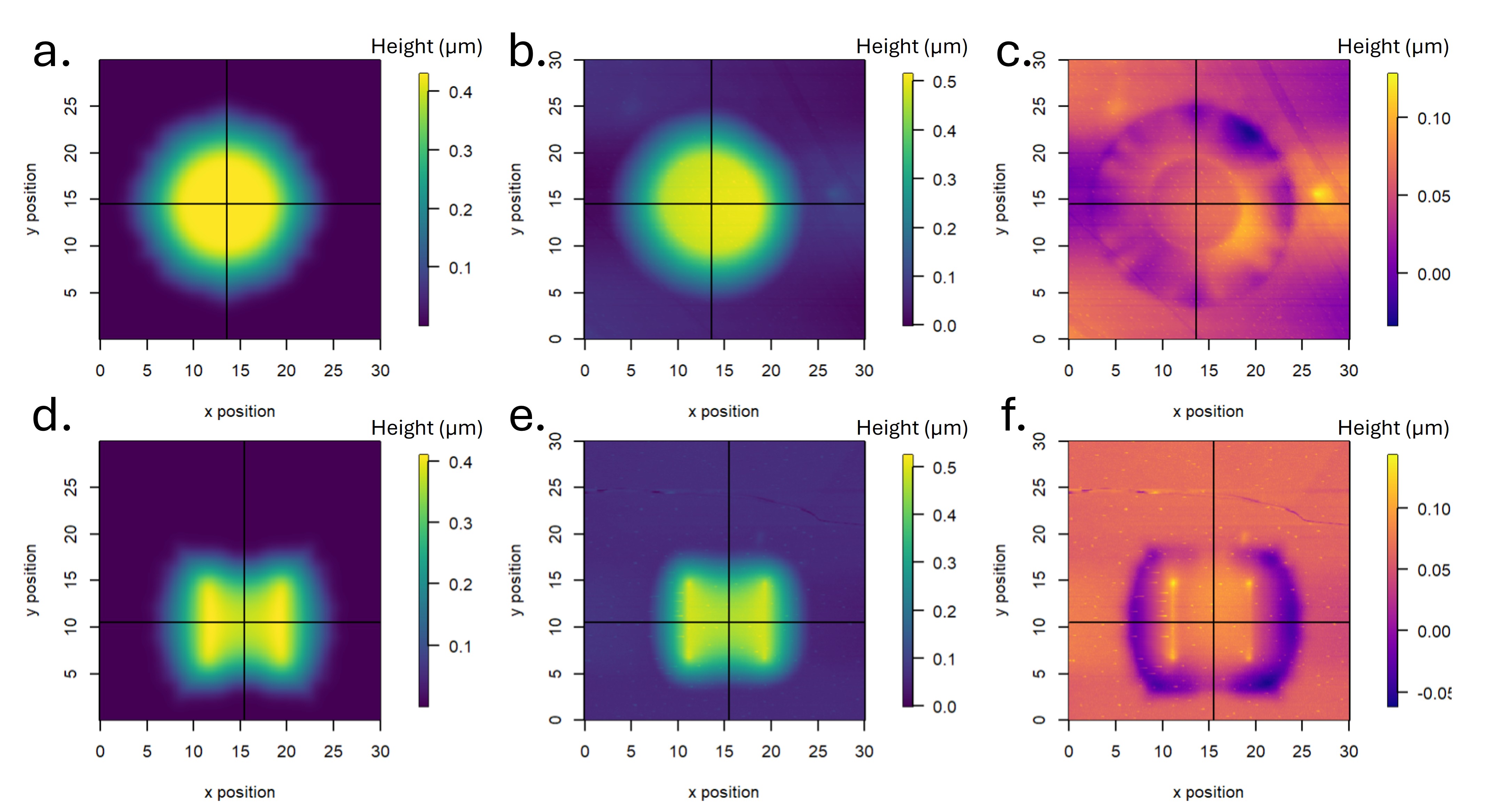}
  \caption{(a,d) The height profile extracted from the COMSOL simulations for an example ring and ridge structure. (b,e) The AFM height profile for the same example ring and ridge structures. (c,f) The difference between the experimental and predicted heights.}
  \label{afm_rmse}
\end{figure}

To probe one potential source of error in our model, we looked at the difference between the experimentally determined height profile (measured via AFM) and the extracted height from the COMSOL simulations. See Figure S\ref{afm_rmse} for one example. We then computed the normalized RMSE values across all structures, which was 7.22\%. 

\section{Density Functional Theory Calculations}

Density functional theory (DFT) calculations \cite{Hohenberg1964,Kohn1965} were performed using the projector augmented-wave (PAW) method \cite{Blochl1994} as implemented in the Vienna \textit{ab initio} Simulation Package (VASP).\cite{Kresse1996,Kresse1996b,Kresse1999} The $\epsilon$-Ga$_2$Se$_2$ model was constructed using Ga and Se PAW potentials with valence electronic configurations $4s^24p^1$ and $4s^24p^4$, respectively. The electron wave functions were expanded in a plane-wave basis with a kinetic-energy cutoff of 400 eV, and the Brillouin zone was sampled using a $9\times9\times1$ Monkhorst--Pack $k$-point grid. All structures were fully optimized until the convergence thresholds for total energy and force reached $10^{-7}$ eV and $10^{-4}$ eV/\AA, respectively.

Structural relaxations were performed using the PBEsol exchange-correlation functional.\cite{Perdew2008} PBEsol was chosen because it was specifically developed to improve the description of equilibrium lattice constants and structural properties of solids relative to conventional PBE, which commonly overestimates lattice constants. This choice is appropriate here because the role of semilocal DFT is to provide relaxed structural parameters and strain-dependent internal coordinates, not the final electronic band gap. The final electronic structures and strain-dependent band gaps were computed using the screened hybrid Heyd--Scuseria--Ernzerhof (HSE) functional including spin-orbit coupling (SOC).\cite{Heyd2003,Heyd2006}

The HSE exact-exchange mixing parameter was set to $\alpha=27\%$, which gives a relaxed band gap consistent with the experimentally relevant optical gap of $\epsilon$-Ga$_2$Se$_2$ within the uncertainty associated with thickness, sample environment, and photoluminescence peak extraction. The use of HSE+SOC is important because semilocal PBE-type functionals underestimate the absolute band gap, whereas hybrid functionals provide a more reliable description of the band-edge separation. We emphasize, however, that the primary DFT quantity used in this work is not the absolute value of the band gap but the strain gauge factor, i.e., the slope of the band gap with respect to applied strain. This slope provides the relevant comparison to the experimentally extracted PL peak shifts.

\begin{figure}
  \includegraphics[width=0.75\linewidth]{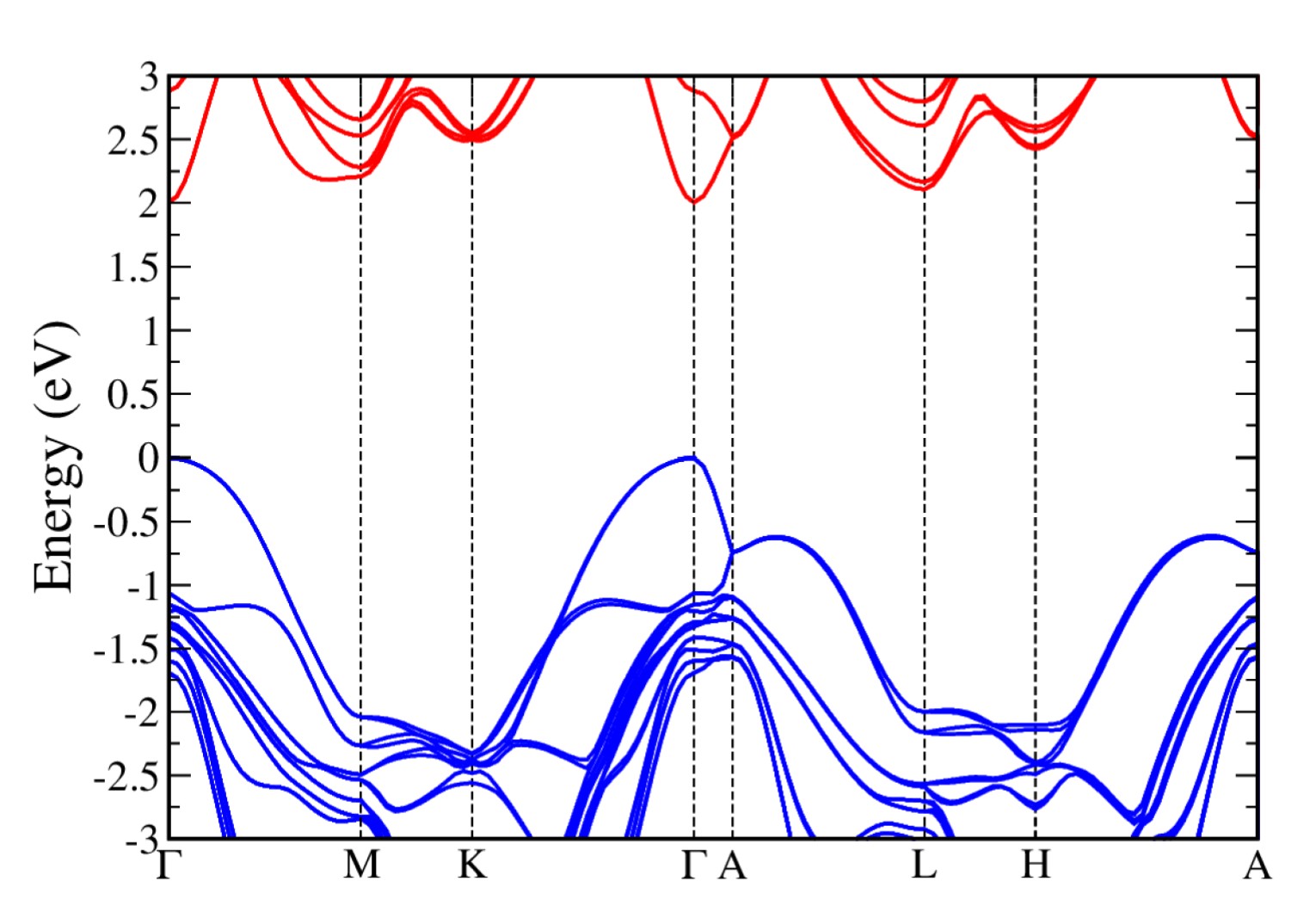}
  \caption{Band structure of $\epsilon$-Ga$_2$Se$_2$ calculated using the HSE functional including spin-orbit coupling (SOC). The calculated band gap is 2.0 eV and is direct at the $\Gamma$ point.}
  \label{SI_dft_bandstructure}
\end{figure}

For the relaxed unstrained structure, HSE+SOC predicts a direct band gap of approximately 2.0 eV at the $\Gamma$ point, as shown in Figure S\ref{SI_dft_bandstructure}. This value is in good agreement with available experimental data for layered GaSe/Ga$_2$Se$_2$.\cite{Luo2023} The calculated band structure therefore validates the use of HSE+SOC for extracting the strain-dependent band-gap response. The optimized structural parameters are consistent with previously reported values for layered GaSe-family compounds within the expected variation associated with thickness, stacking, and exchange-correlation functional.\cite{Chitara2018,Usman2022,Luxa2016} Because the experimentally measured PL originates from relatively thick, bulk-like flakes, comparison to bulk or multilayer GaSe/Ga$_2$Se$_2$ literature values is more appropriate than comparison to monolayer limits.

To compute the strain gauge factors, biaxial and uniaxial tensile strain were applied to the relaxed structure. For biaxial strain, both in-plane lattice vectors were increased by the same prescribed strain, while the out-of-plane lattice parameter and all internal atomic positions were relaxed. For uniaxial strain along the armchair direction, the armchair lattice vector was fixed at the prescribed strain and the transverse in-plane lattice vector, out-of-plane lattice parameter, and internal atomic coordinates were relaxed. The analogous procedure was used for uniaxial strain along the zigzag direction. This relaxation protocol reflects the experimental situation in which the suspended membranes are strained in plane but are not constrained to maintain a fixed out-of-plane lattice spacing.

Extended Data Figure 4 of the main manuscript shows the calculated biaxial-strain dependence of the band gap using PBEsol and HSE+SOC, with and without relaxation of the out-of-plane lattice parameter. As expected, PBEsol underestimates the absolute band gap. HSE+SOC gives the correct unstrained gap scale and, when out-of-plane relaxation is included, yields a biaxial strain gauge factor of
\[
\frac{dE_g}{d\epsilon_{\rm biaxial}}=-0.268~{\rm eV}/\%.
\]
This is the DFT biaxial gauge factor used for comparison with experiment.

For uniaxial strain, the HSE+SOC band gap also decreases approximately linearly with tensile strain, as shown in Extended Data Figure 6 of the main manuscript. The calculated slopes are
\[
\frac{dE_g}{d\epsilon_{\rm armchair}}=-0.108~{\rm eV}/\%,
\]
and
\[
\frac{dE_g}{d\epsilon_{\rm zigzag}}=-0.111~{\rm eV}/\%.
\]
The near equality of these two values indicates that the strain-induced band-gap shift is nearly isotropic with respect to the in-plane crystallographic direction, consistent with the experimental observation that the PL peak shift does not show a measurable dependence on the orientation of the local strain relative to the crystal axes.

The experimentally extracted local strains are generally much smaller than the maximum strain considered in the DFT calculations. The DFT strain range was extended up to 5\% not because the experimental membranes are assumed to reach this strain, but to establish the robustness of the linear band-gap response and to verify that no change in the nature of the band edge occurs over a broader tensile-strain range. The strain gauge factors used in the manuscript are the linear slopes of the band gap with respect to strain and are therefore directly applicable to the smaller experimental strain regime. In all cases considered, Ga$_2$Se$_2$ remains a direct-gap semiconductor with both the valence-band maximum and conduction-band minimum at $\Gamma$. No strain-induced direct-to-indirect transition or band-edge valley shift is observed, in contrast to many transition-metal dichalcogenides.\cite{Zhang2013}

\begin{figure}
  \includegraphics[width=0.75\linewidth]{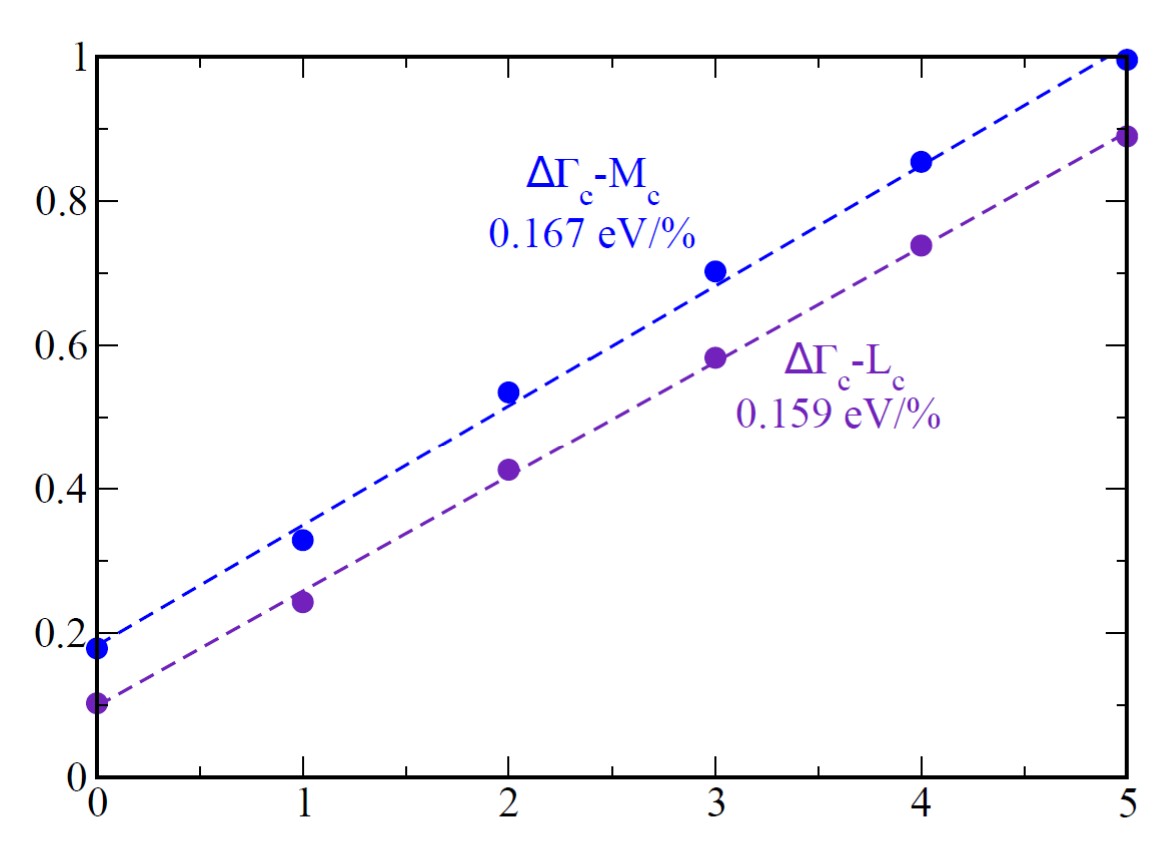}
  \caption{Energy separation between the conduction-band minimum at $\Gamma$ and the competing conduction-band states at $L$ (blue) and $M$ (purple) as a function of tensile strain. The increasing separation shows that strain stabilizes the direct band gap at $\Gamma$ in $\epsilon$-Ga$_2$Se$_2$.}
  \label{SI_dft_conduction}
\end{figure}

The calculated evolution of the conduction-band valleys also supports the observed enhancement of PL intensity in strained regions. As shown in Figure S\ref{SI_dft_conduction}, the energy separation between the conduction-band minimum at $\Gamma$ and the conduction-band states at the $M$ and $L$ points increases with tensile strain. Thus, tensile strain makes the material more strongly direct-gap-like by increasing the energetic separation between the $\Gamma$ valley and the competing off-$\Gamma$ conduction-band states.

The physical origin of the band-gap reduction can be understood from the strain-induced modification of orbital hybridization at the band edges. The valence- and conduction-band-edge states of layered Ga$_2$Se$_2$ near $\Gamma$ are derived primarily from Ga and Se orbitals with both in-plane and out-of-plane bonding character. Tensile strain increases the Ga--Se bond lengths and modifies the Ga--Se--Ga and Se--Ga--Se bond angles, thereby reducing orbital overlap and changing the bonding-antibonding splitting that determines the band gap. Biaxial strain produces a larger gauge factor than uniaxial strain because it simultaneously perturbs orbital overlap along both in-plane directions, whereas uniaxial strain allows partial relaxation in the transverse in-plane and out-of-plane directions. A fully quantitative decomposition of the band-gap shift into individual orbital-overlap or hopping-matrix-element contributions would require an additional Wannier-function, projected-orbital, or crystal-orbital Hamilton population analysis. Such an analysis is beyond the scope of the present work. For the purposes of this study, the central theoretical quantity is the directly computed HSE+SOC strain gauge factor, which quantitatively agrees with the experimentally extracted PL strain response.

\section{Photoluminescence (PL) Measurements}

\begin{figure}
  \includegraphics[width=1\linewidth]{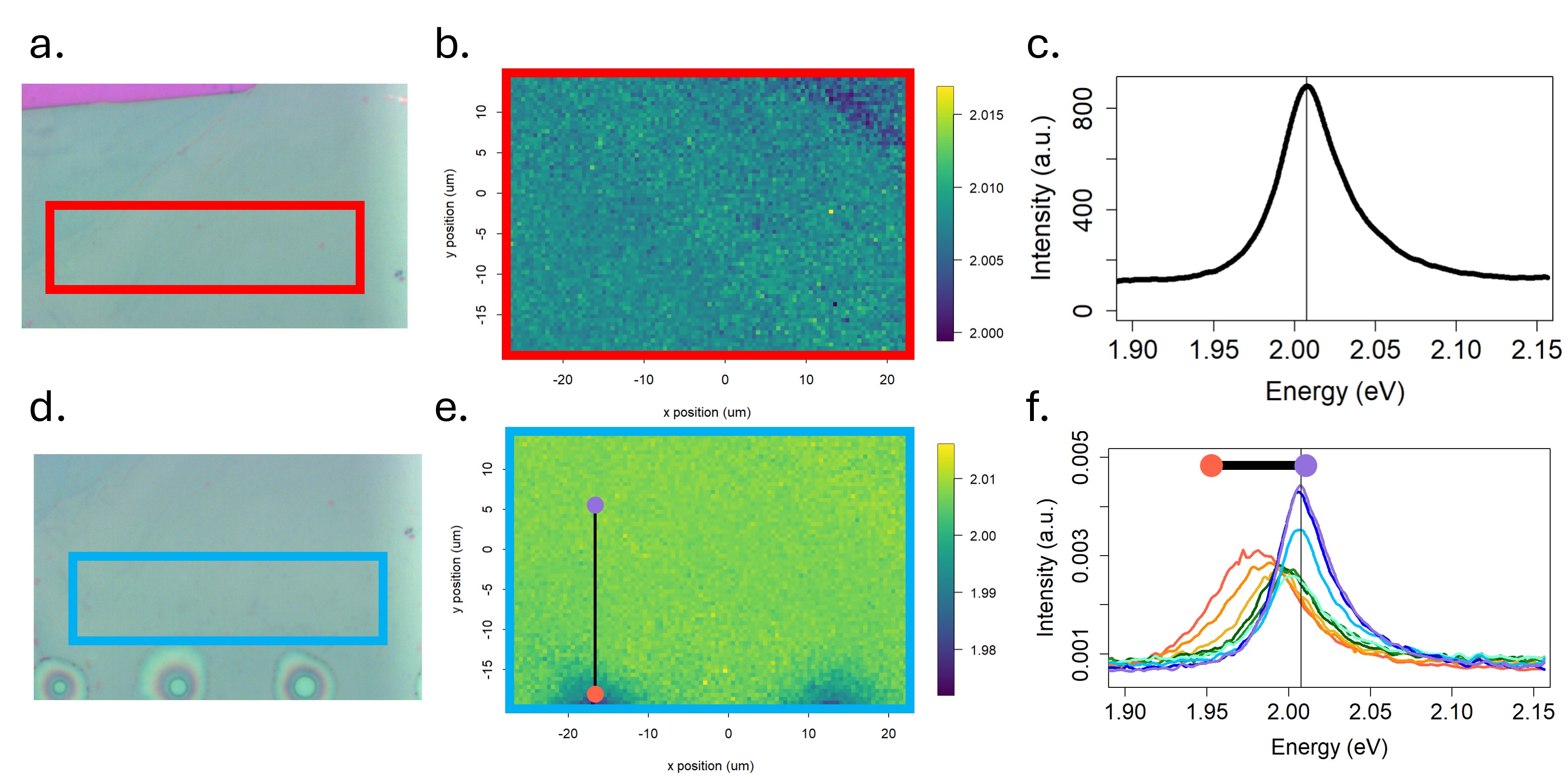}
  \caption{(a,d) Show the optical image of the measured area on a flat region of the flake that is suspended over the patterned substrate, with the resulting peak max energy shown in (b,e) for each spatial location. (c) Average PL signal across the entire area shown in (b). (f) Cross sectional view of the peak shift in (e) as the distance from the structure increases.}
  \label{flat}
\end{figure}

To determine the experimental PL peak shift, the baseline (unstrained) PL energy was first obtained from areas far away from any nanostructures, but within the same flake. The rectangles in Figures S\ref{flat} (a) and (d) provide two examples of optical images of `flat' areas that were probed. Figures S\ref{flat}(b) and (e) show the PL center energy measured at each point within the corresponding rectangles. Figure S\ref{flat}(c) shows the average of all points in Figure S\ref{flat}(b), which provides a clear measure of the unstrained PL peak energy. Figures S\ref{flat}(e) and (f) illustrate the importance of obtaining the baseline (unstrained) PL peak energy from regions that are far away from any nanostructures: a clear shift in PL energy is observed as a function of position due to the rings that are near, but not within, the lower boundary of the rectangle shown in Figure S\ref{flat}(d). The unstrained PL energy was determined independently for all unique flakes. These values of the unstrained PL peak energy were then used to calculate the peak shift across the suspended regions. 

\begin{figure}
  \includegraphics[width=0.75\linewidth]{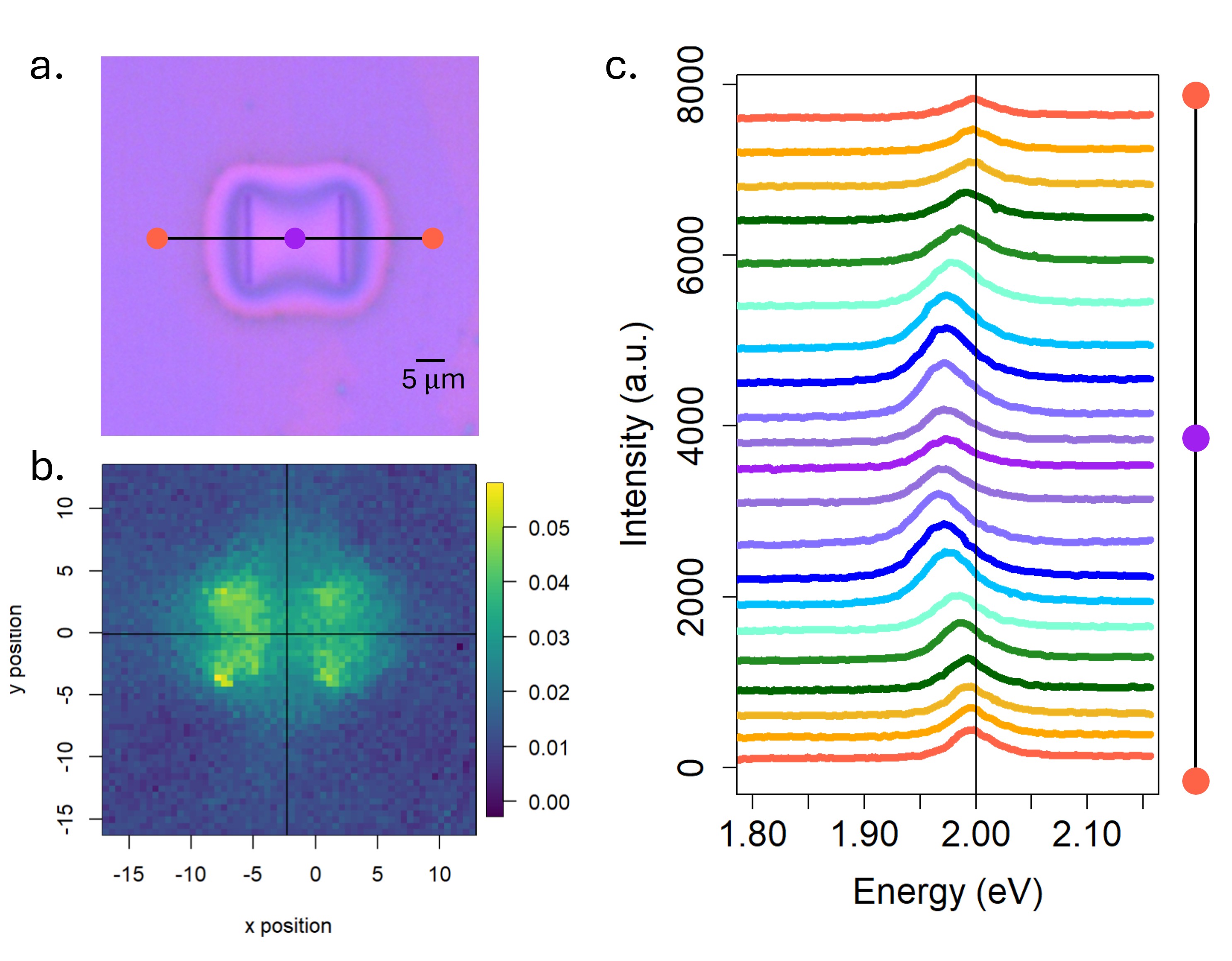}
  \caption{(a) Optical image of suspended flake over a ridge structure. (b) x-y peak shift map for this ridge structure. (c) Stacked spectra showing the peak shift along the line indicated in (a). }
  \label{SI_ridge_expPL}
\end{figure}

\section{Strain Components}
The 2D tensor that describes the local strain at any point is given by \cite{strain_calc,strain_calc02}:

\begin{equation}
    \epsilon = 
    \begin{pmatrix}
        \epsilon_{xx} & \epsilon_{xy} \\
        \epsilon_{xy} & \epsilon_{yy} \\
    \end{pmatrix}
    \label{strain_tensor}
\end{equation}

which can be broken into two components:

\begin{equation}
    \epsilon = 
    \begin{pmatrix}
        \epsilon_{xx} & \epsilon_{xy} \\
        \epsilon_{xy} & \epsilon_{yy} \\
    \end{pmatrix}
    =
    \begin{pmatrix}
        \frac{\epsilon_{xx}+\epsilon_{yy}}{2} & 0 \\
        0 & \frac{\epsilon_{xx}+\epsilon_{yy}}{2} \\
    \end{pmatrix}
    +
    \begin{pmatrix}
        \frac{\epsilon_{xx}-\epsilon_{yy}}{2} & \epsilon_{xy} \\
        \epsilon_{xy} & \frac{\epsilon_{yy}-\epsilon_{xx}}{2} \\
    \end{pmatrix}
    =
    \epsilon_{hyd}+\epsilon_{dev},
    \label{strain_tensor_decomposition}
\end{equation}

\noindent where $\epsilon_{hyd}$ is the hydrostatic component of the strain and $\epsilon_{dev}$ is the deviatoric component. It is straightforward to see that the hydrostatic component simply represents the application of biaxial strain, with equal strain along the $x$ and $y$ directions and no shear ($\epsilon_{xy}=0$). Mathematically, when $\epsilon_{xx}=\epsilon_{yy}$ and $\epsilon_{xy}=0$ the deviatoric component of the strain goes to zero and only the hydrostatic (biaxial) component remains. In other words the biaxial strain is given by

\begin{equation}
    \epsilon_{bi}=\frac{\epsilon_{xx}+\epsilon_{yy}}{2} = \frac{\epsilon_{1} + \epsilon_{2}}{2}
    \label{hyd}
\end{equation}

\noindent where $\epsilon_{1}$ and $\epsilon_{2}$ are the eigenvalues of the strain tensor (Equation \ref{strain_tensor})

\begin{equation}
    \epsilon_{1,2} = 
    \frac{\epsilon_{xx}+\epsilon_{yy}}{2}
    \pm
    \sqrt{ \Biggl( \frac{\epsilon_{xx}-\epsilon_{yy}}{2} \Biggl)^2+\epsilon_{xy}^2}.
    \label{principal_strain}
\end{equation}

\noindent known as the principal components. These principal components represent the maximum and minimum strain at a given spatial location. 

When we perform the biaxial strain gauge analysis reported in the main manuscript, we first numerically screen the computed strain to identify points at which the deviatoric component of strain is zero. We then use Equation \ref{hyd} to compute the local biaxial strain and plot the experimentally measured PL peak shift as a function of the biaxial strain. The fit to that data gives us the biaxial strain gauge factor $\beta_{exp}$.

We can see that both the maximum and minimum strains include the biaxial component. This makes sense because the biaxial strain is isotropic in the plane. If we subtract the isotropic biaxial component from each principal strain, the remainder describes any ``excess" strain along one dimension, i.e.~unixaial strain. In other words, the uniaxial strain is described by   

\begin{equation}
    \epsilon_{uni} = \sqrt{ \Biggl( \frac{\epsilon_{xx}-\epsilon_{yy}}{2} \Biggl)^2+\epsilon_{xy}^2} = \frac{\epsilon_{1} - \epsilon_{2}}{2}.
    \label{dev}
\end{equation}
 
Whenever strain stretches an atomic lattice along one direction, the lattice contracts along the transverse directions. The ratio of the contraction to the stretch is Poisson's ratio, which for Ga$_2$Se$_2$ is 0.23. When we perform the uniaxial strain gauge analysis reported in the main manuscript, we first numerically screen the computed strain at every point to identify those for which the ratio of $\epsilon_{1}$ to $\epsilon_{2}$ is less than 0.15. This threshold is below Poisson's ratio, which means that the strain at such a point is uniaxial. We compute the magnitude of the uniaxial strain at that point using Equation \ref{dev} and then plot the measured PL peak shift as a function of the uniaxial strain amplitude to determine the uniaxial strain gauge factor $\alpha_{exp}$.


Finally, we use the results of Equations \ref{hyd} and \ref{dev} to compose the model for bandgap shift as a function of local biaxial and uniaxial strain
\begin{equation}
    \Delta E \approx \beta_{exp}\frac{\epsilon_1+\epsilon_2}{2}+\alpha_{exp}\frac{|\epsilon_1-\epsilon_2|}{2}
    \label{PL peak shift}
\end{equation}
\section{Extending the Framework}
The strain gauge factors and analytical model developed using isolated ring and ridge nanostructures establish a quantitative foundation that is only useful if it generalizes beyond the geometries used to construct it. In this work we demonstrate two complementary extensions of this framework. First, we show that the model accurately predicts spatially resolved PL peak shifts induced by lobed nanostructures; geometries that produce continuously varying, multiaxial strain distributions fundamentally different in character from the nearly pure biaxial and uniaxial fields of rings and ridges. Second, we extend the platform to WS$_2$, demonstrating that the same nanostructure based approach produces quantifiable strain-induced optical shifts in a material with distinct electronic structure and mechanical properties. Together, these results establish both the geometric flexibility and material versatility of the framework.

\subsection{Lobed Structures}
A series of lobe-shaped nanostructures are designed to induce spatially-varying strain distributions across the suspended region of transferred Ga$_2$Se$_2$ flakes. In the simulations, these lobed nanostructures are defined using parametric curves and an offset geometry modifier is used to define the region that is constrained to in-plane motion. The four-lobed structure presented in the main text is defined using the polar coordinate equation $r=7.5+6.25\cos{(4 \theta)}$ plotted in Extended Data Figure 7(a). In Extended Data Figure 7(b,c), we show a tilted scanning electron microscope image of the four-lobed structure etched to a height of $\approx 200$ nm and an optical microscope image of a 140 nm thick GaSe flake transferred onto the structure, respectively.

\begin{figure}[h]
  \includegraphics[width=1\linewidth]{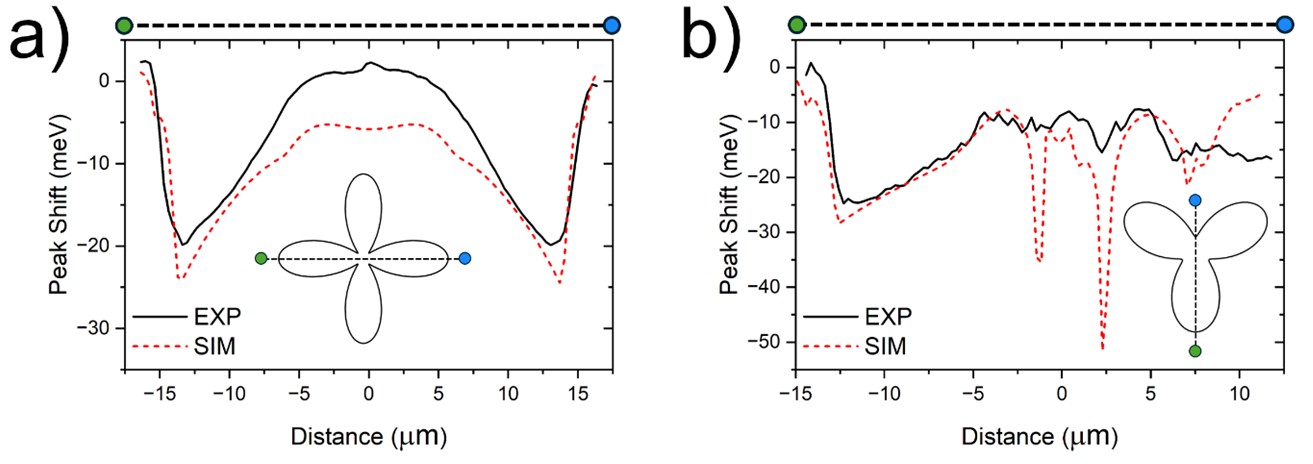}
  \caption{PL Shift induced by lobed nanostructures. (a) Experimental and simulation cross-sectional PL shift for the four-lobe structure taken along the dashed line in the inset.  (b) Similar comparison for the three-lobed structure.  }
  \label{3lobed}
\end{figure}

As discussed in the main text, the strain magnitude is highest at the outer regions of the lobes and decreases toward the center of the nanostructure. Similar behavior is observed for the three-lobe structure presented in the main text, but with the central region adhered to the underlying substrate. In Figure S\ref{3lobed}(a,b), we compare experimentally measured and simulated strain induced PL shift along cross-sections taken through the center of the four- and three-lobe structures, respectively. Simulations indicate that the transition from suspended to adhered regions in the three-lobe structure is associated with a relatively substantial increase in strain magnitude – behavior that is not captured in the experimental photoluminescence maps. This discrepancy is likely attributable to averaging effects caused by the finite spot size or the omission of decohesion effects in the simulations. Incorporating decohesion physics into the simulations may provide the necessary strain relaxation dynamics that better capture the behavior observed experimentally. 

\subsection{Tungsten Disulfide (WS$_2$) strain engineering}
\begin{figure}[h]
  \includegraphics[width=0.9\linewidth]{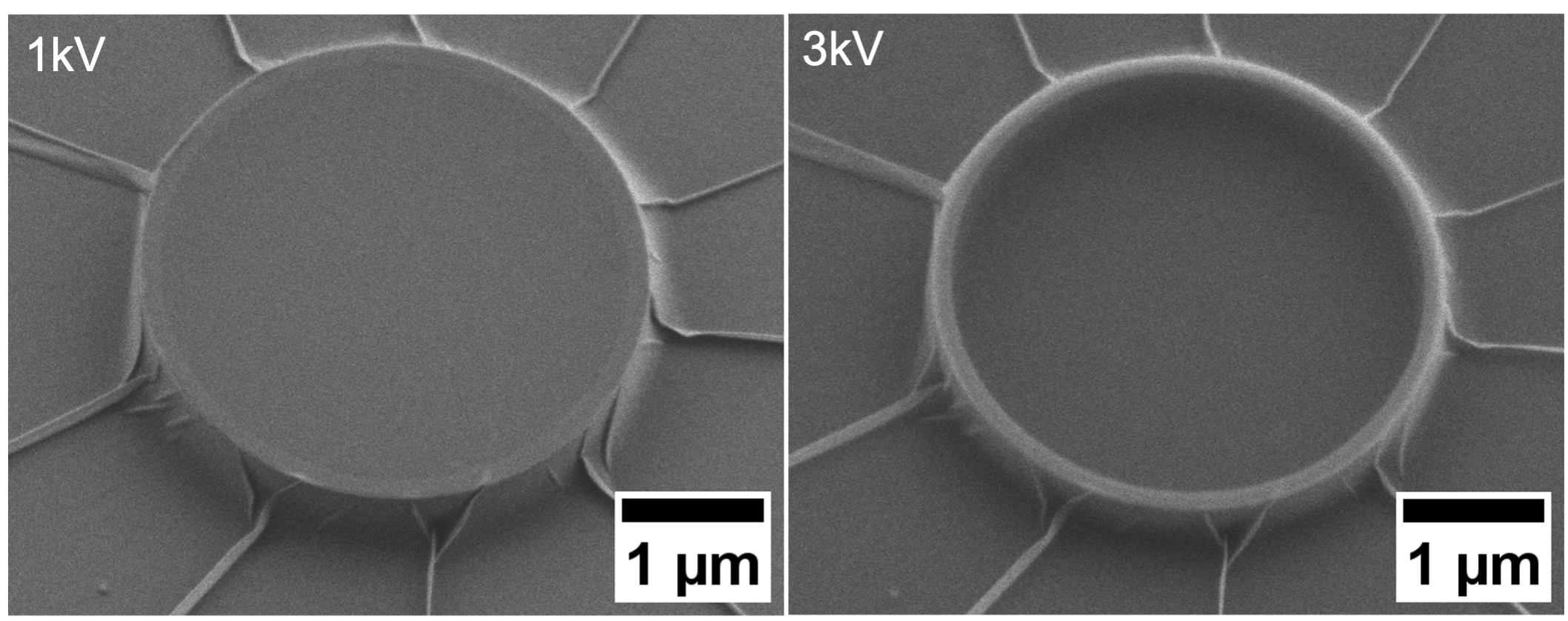}
  \caption{SEM images of suspended/strained monolayer WS$_2$. The two SEM images illustrate the same region with 1kV (left) and 3kV (right) electron acceleration voltage to show the continuity of the strained WS$_2$ and the underlying SiO$_2$ ring nanostructure, respectively.}
  \label{SEM WS2 Ring}
\end{figure}

As described in the main text, we performed a study of WS$_2$ to demonstrate that this strain-engineering platform and framework can be generalized to all 2D materials. Figure S\ref{SEM WS2 Ring} shows an SEM image of a monolayer of WS$_2$ transferred onto a ring structure. These images demonstrate the conformal adherence of the monolayer to the ring structure, and also demonstrate the absence of deformations, including sagging, in the suspended "drumhead" region. We note that the wrinkles and tight sidewall adhesion outside the ring are caused by the vacuum in the SEM and are not typically present when structures are measured at atmospheric pressure.

\begin{figure}[h]
  \includegraphics[width=1\linewidth]{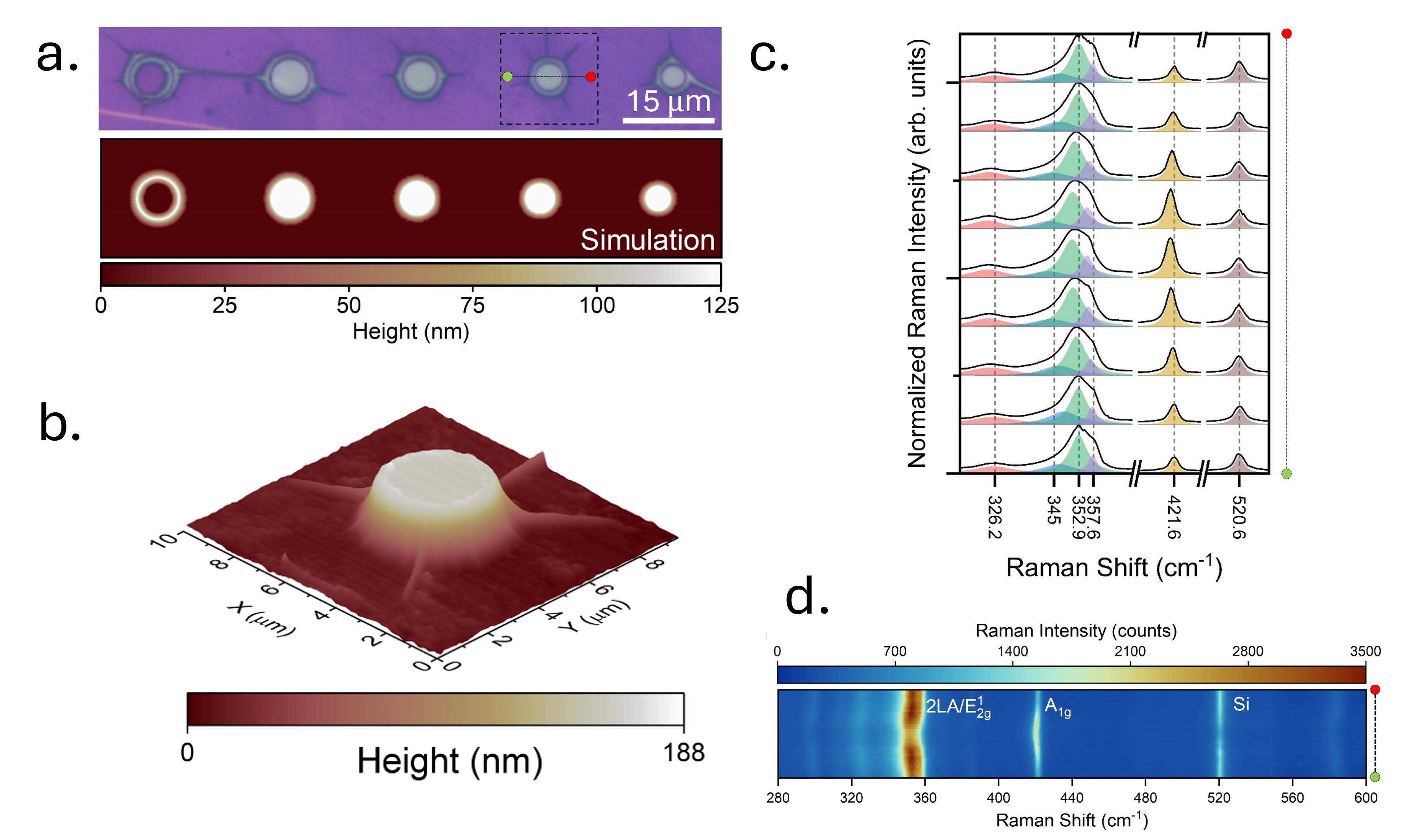}
  \caption{Strain engineering of WS$_2$. (a) Top: optical microscope image of a 10 nm thick WS$_2$ flake transferred onto a series of ring nanostructures of diameter 4.5, 4.0, 3.5, 3.0, and 2.5 $\mu$m, from left to right. Bottom: simulated deformation profile of the region shown above. (b) Three-dimensional AFM height map of the strained region highlighted in (a), showing a flat suspended membrane above the 3.0 $\mu$m diameter ring with slight wrinkle formation outside the ring perimeter. (c) Normalized Raman scattering spectra taken along the dashed line indicated in (a), from the red dot (top) to the green dot (bottom), showing spatially dependent shifts in the WS$_2$ Raman modes. (d) Cross-sectional Raman spectra taken along the dashed line in (a) from the green circle to the red circle.}
  \label{si_ws2}
\end{figure}

The top panel of Figure S\ref{si_ws2}(a) shows an optical microscope image of a 10 nm thick flake transferred onto five neighboring rings of diameters 4.5, 4.0, 3.5, 3.0, and 2.5 $\mu$m, from left to right. The bottom panel shows the simulated deformation profile across the ring pattern for a boundary load of 0.5 MPa, the same calibrated force per unit area that was determined for GaSe. All rings remain suspended above the ring except above the 4.5 $\mu$m ring, where the force applied is sufficient to push the flake down until contact with the flat substrate surface. Figure S\ref{si_ws2}(b) shows a three dimensional AFM height map of the strained region highlighted in (a), showing a flat suspended membrane above the 3.0 $\mu$m diameter ring with slight wrinkle formation outside the ring perimeter. In Figure S\ref{si_ws2}(c), we plot a series of Raman spectra taken over the cross-sectional dashed line shown in Figure S\ref{si_ws2}(a) and perform fitting using a series of Lorentzians. The vertical dashed lines mark the unstrained Raman shift values for all modes. Apart from the Si substrate Raman peak, all Raman modes undergo a redshift across the strained region of the flake. The Raman modes reach maximum shifts of 2.12 cm$^{-1}$, 2.02 cm$^{-1}$, and 1.28 cm$^{-1}$ for the 2LA, E$^{1}_{2g}$, and A$_{1g}$, respectively. These Raman mode shifts are accompanied by a $\approx$42.7 meV redshift of the neutral exciton photoluminescence peak, as discussed in the main text. 

\begin{figure}[h]
  \includegraphics[width=0.83\linewidth]{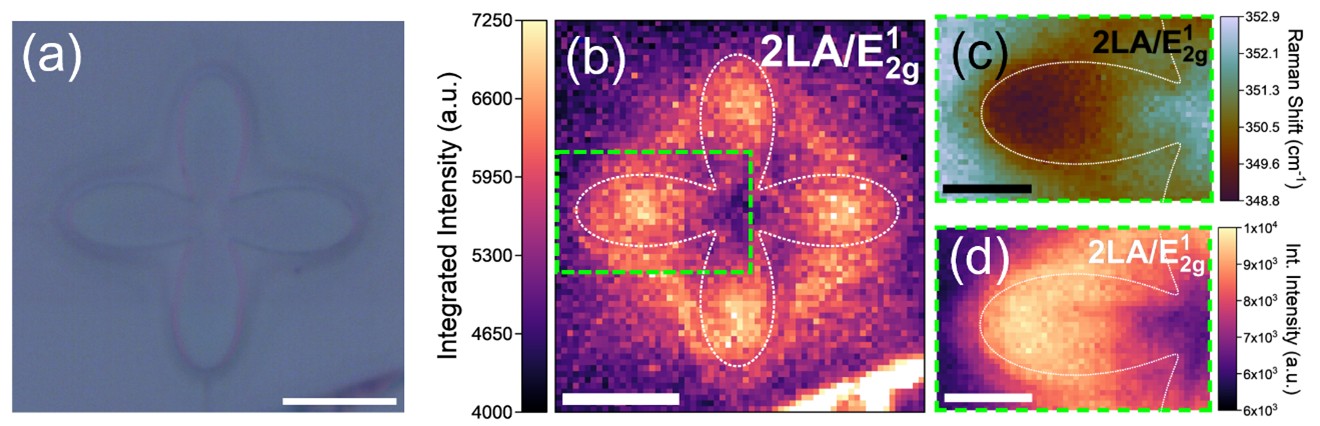}
  \caption{Raman shift of the WS$_2$ E$_{2g}^{1}$ mode induced by lobed nanostructures. (a) Optical microscope image of multilayered WS$_2$ on a four-lobed structure. Scale bar 10 $\mu$m (b) Integrated Raman scattering intensity for the E$_{2g}^{1}$ mode. We note that the color scale was adjusted to remove influence from a nearby flake in the lower right corner of the plot.  (c) High resolution Raman shift map of the outlined region in (b). (d) High resolution map of the integrated Raman scattering intensity for the E$_{2g}^{1}$ mode. The scale bars in (c) and (d) are 5 $\mu$m.}
  \label{si_ws2_raman_lobe}
\end{figure}

Because simulations of the WS$_2$ flake predict a symmetric biaxial strain distribution in the suspended region above the ring, the measured shifts in the Raman modes across this region are expected to be approximately uniform. We also note that there is some finite strain in regions of the flake that are in contact with the substrate. For these reasons, there is ambiguity in whether the observed shifts originate from in-plane strain as a result of the strain engineering process or because the WS$_2$ is not in contact with the substrate above the ring. To provide evidence that substrate-induced effects are not the predominant cause of the observed Raman mode shift \cite{e10,e11} behavior, we performed similar Raman scattering analysis for a multilayer WS$_2$ flake transferred onto a four-lobed structure as shown in Figure S\ref{si_ws2_raman_lobe}(a). In Figure S\ref{si_ws2_raman_lobe}(a), we show the integrated 2LA/E$_{2g}^{1}$ mode intensity over a range of 337-365 cm$^{-1}$ across the entire strained region and find that the intensity increases in regions of high strain. Here we note that the 2LA and E$_{2g}^{1}$ modes cannot be distinguished due to the lower acquisition time required for spatial mapping. Therefore, we fit the 2LA/E$_{2g}^{1}$ peaks using a single Lorentzian as an approximation. In Figure S\ref{si_ws2_raman_lobe}(c,d), we show higher resolution scans of the 2LA/E$_{2g}^{1}$ mode location and integrated intensity across a single lobe of the structure, respectively. Both the 2LA/E$_{2g}^{1}$ mode location and scattering intensity vary in accordance with the strain distribution while regions of the flake that are under minimal strain remain unaffected. 

We further show the PL distribution behavior in Figure~6 in the main text and find similar behavior to that observed for the 2LA/E$_{2g}^{1}$ Raman mode. The observation of spatially-varying Raman shift and PL emission within the suspended region provides evidence that the observed behavior is predominantly influenced by strain rather than substrate-induced effects. However, further investigation is necessary to rule out substrate-induced effects entirely or determine the extent to which they are present. 


\section{All Individual Structures: Experimental versus Model}
\begin{figure}[h]
  \includegraphics[width=0.83\linewidth]{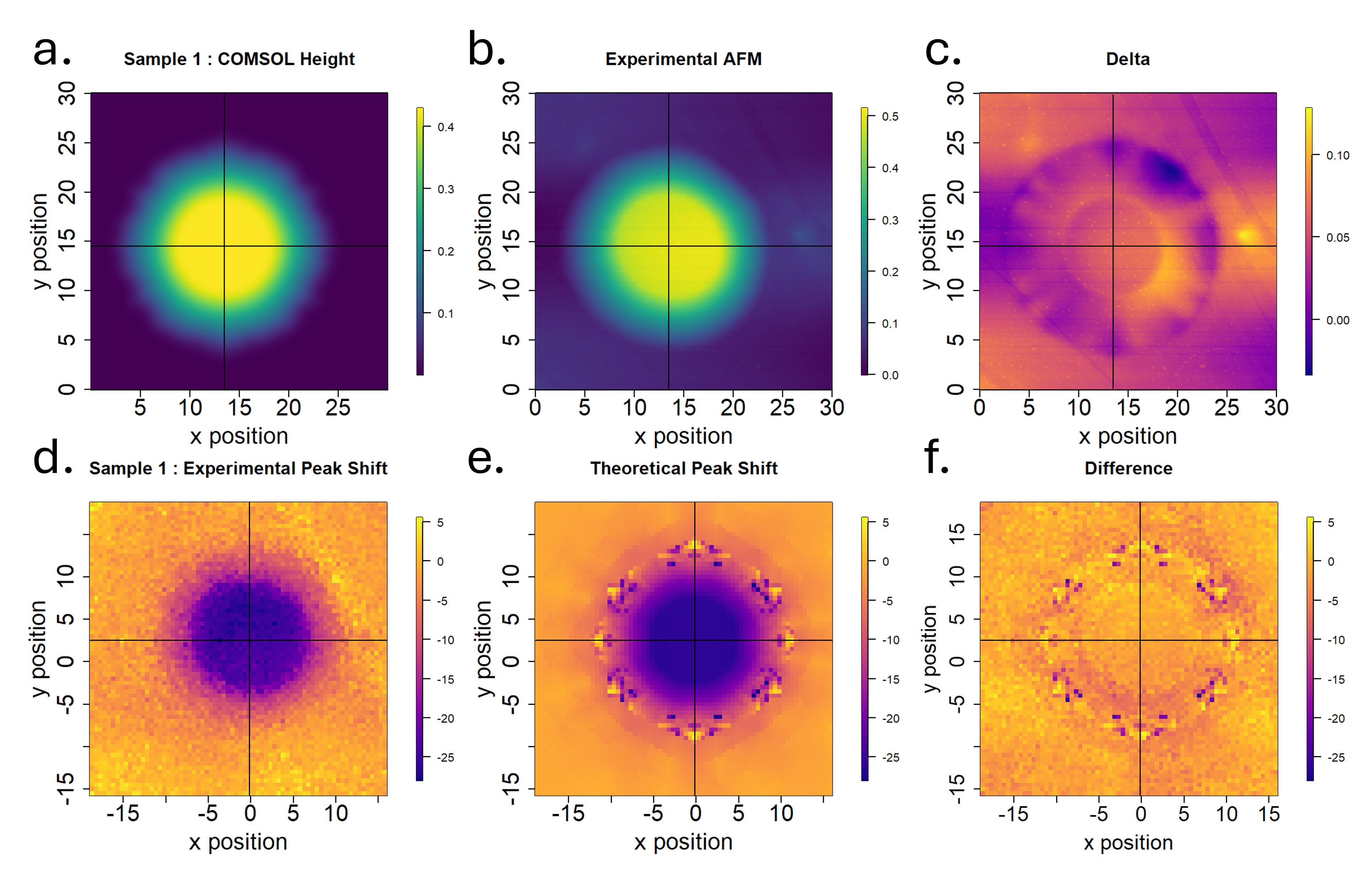}
  \caption{Sample 1. The height profile (a) simulated from COMSOL and (b) measured  by AFM, with (c) showing the difference between the experimental and predicted heights. (d) The experimental peak shift, (e) the calculated peak shift using the experimentally determined strain gauge factors, and (f) the difference between the two.}
  \label{sample_01}
\end{figure}

In Figure S\ref{sample_01}-S\ref{sample_23} we provide two-dimensional plots of the experimental and simulated results for each sample measured. In each case panels (a-c) show the height profiles from (a) the COMSOL simulations and (b) the AFM scan, with (c) showing the difference between the two. In each case panels (d-f) show (d) the experimentally measured PL peak shift, (e) the theoretical PL peak shift computed using Equation \ref{PL peak shift} and the experimentally determined strain gauge factors, and (f) the difference between the two. The main differences occur at points of high strain (e.g.~tips of the ridges) or where we see numerical artifacts in the COMSOL simulations (e.g.~where the flake touches the substrate).

\begin{figure}[h]
  \includegraphics[width=0.83\linewidth]{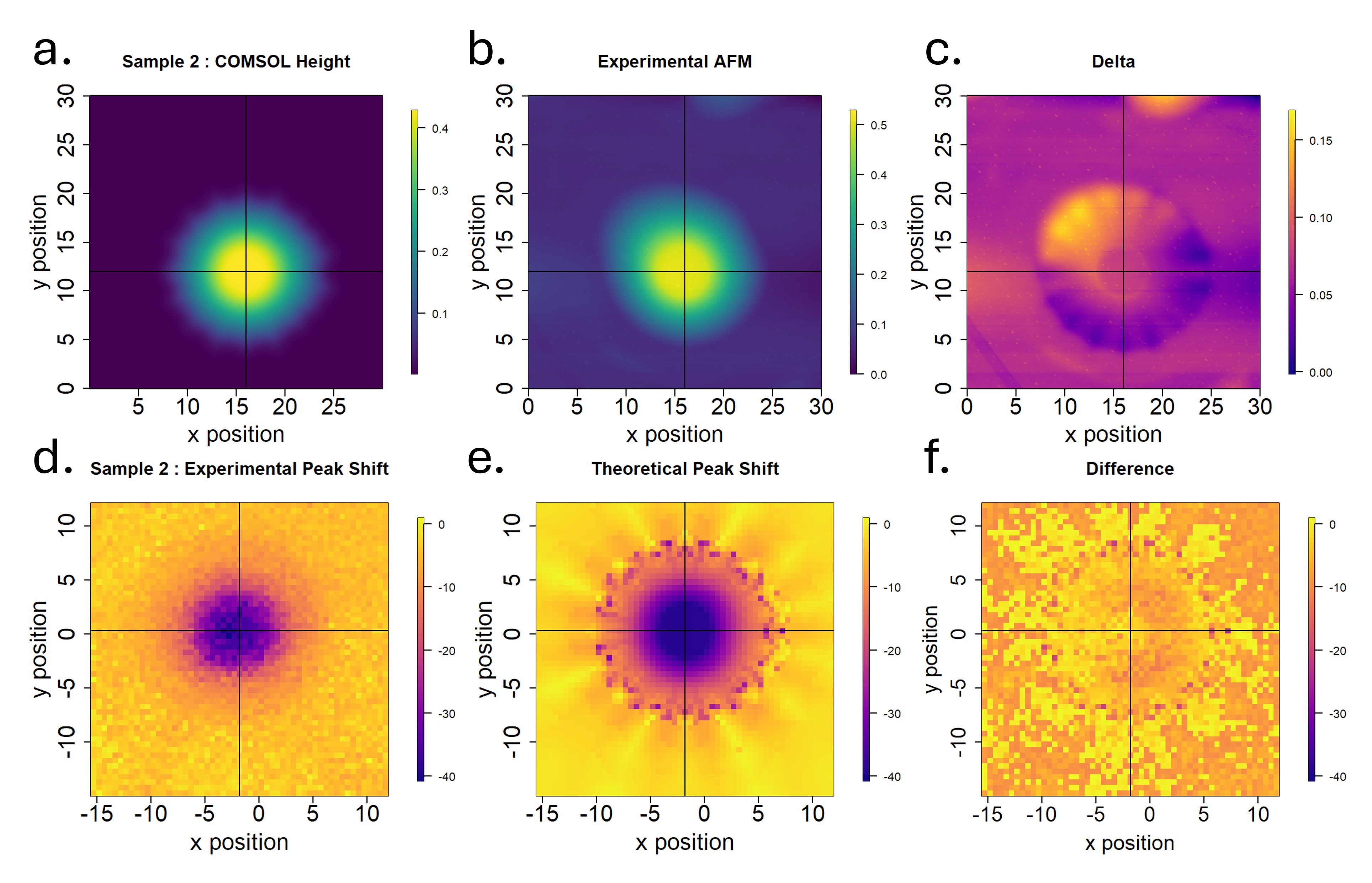}
  \caption{Sample 2. The height profile (a) simulated from COMSOL and (b) measured  by AFM, with (c) showing the difference between the experimental and predicted heights. (d) The experimental peak shift, (e) the calculated peak shift using the experimentally determined strain gauge factors, and (f) the difference between the two.}
  \label{sample_02}
\end{figure}
\begin{figure}[h]
  \includegraphics[width=0.83\linewidth]{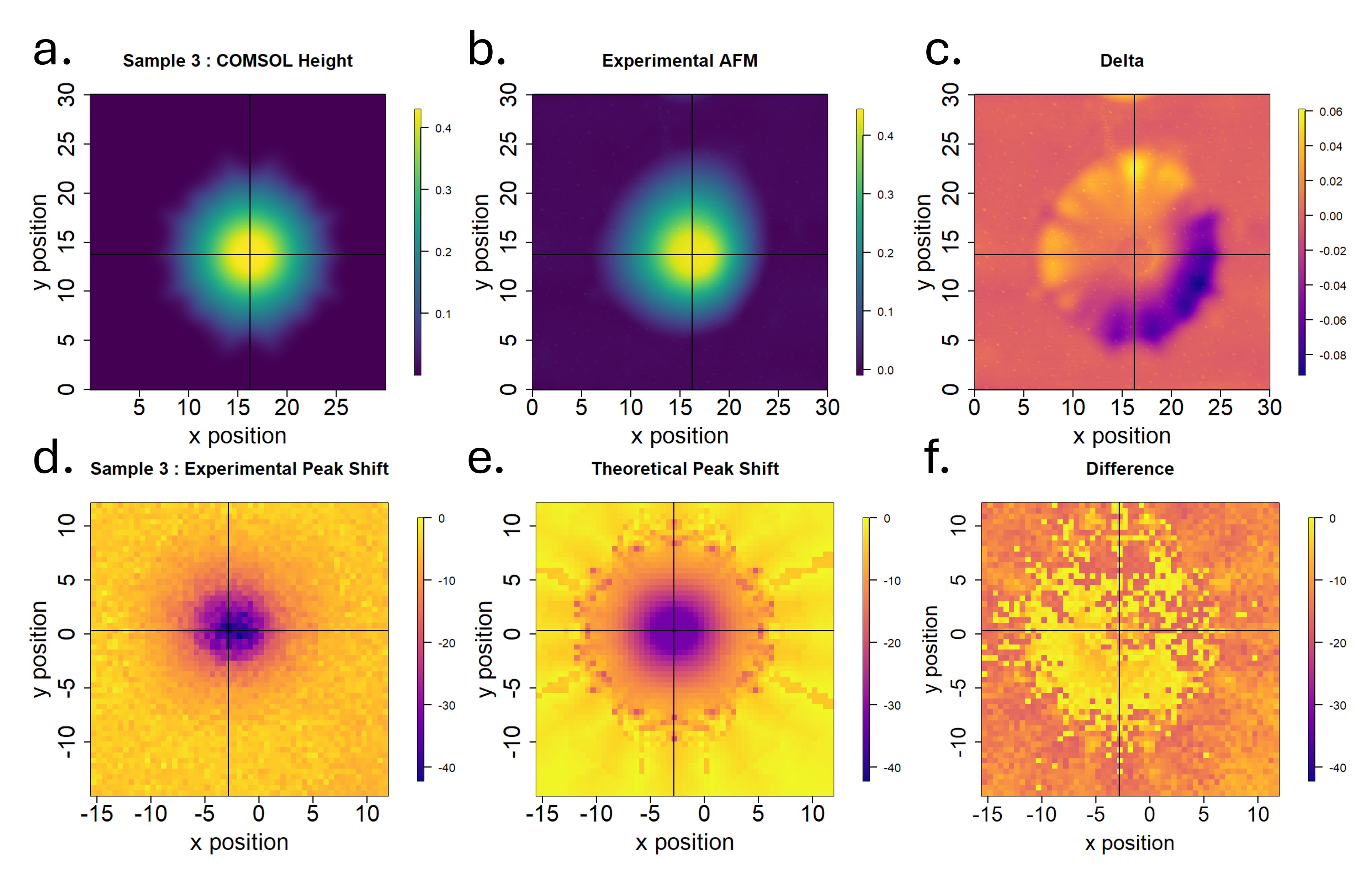}
  \caption{Sample 3. The height profile (a) simulated from COMSOL and (b) measured  by AFM, with (c) showing the difference between the experimental and predicted heights. (d) The experimental peak shift, (e) the calculated peak shift using the experimentally determined strain gauge factors, and (f) the difference between the two.}
  \label{sample_03}
\end{figure}
\begin{figure}[h]
  \includegraphics[width=0.83\linewidth]{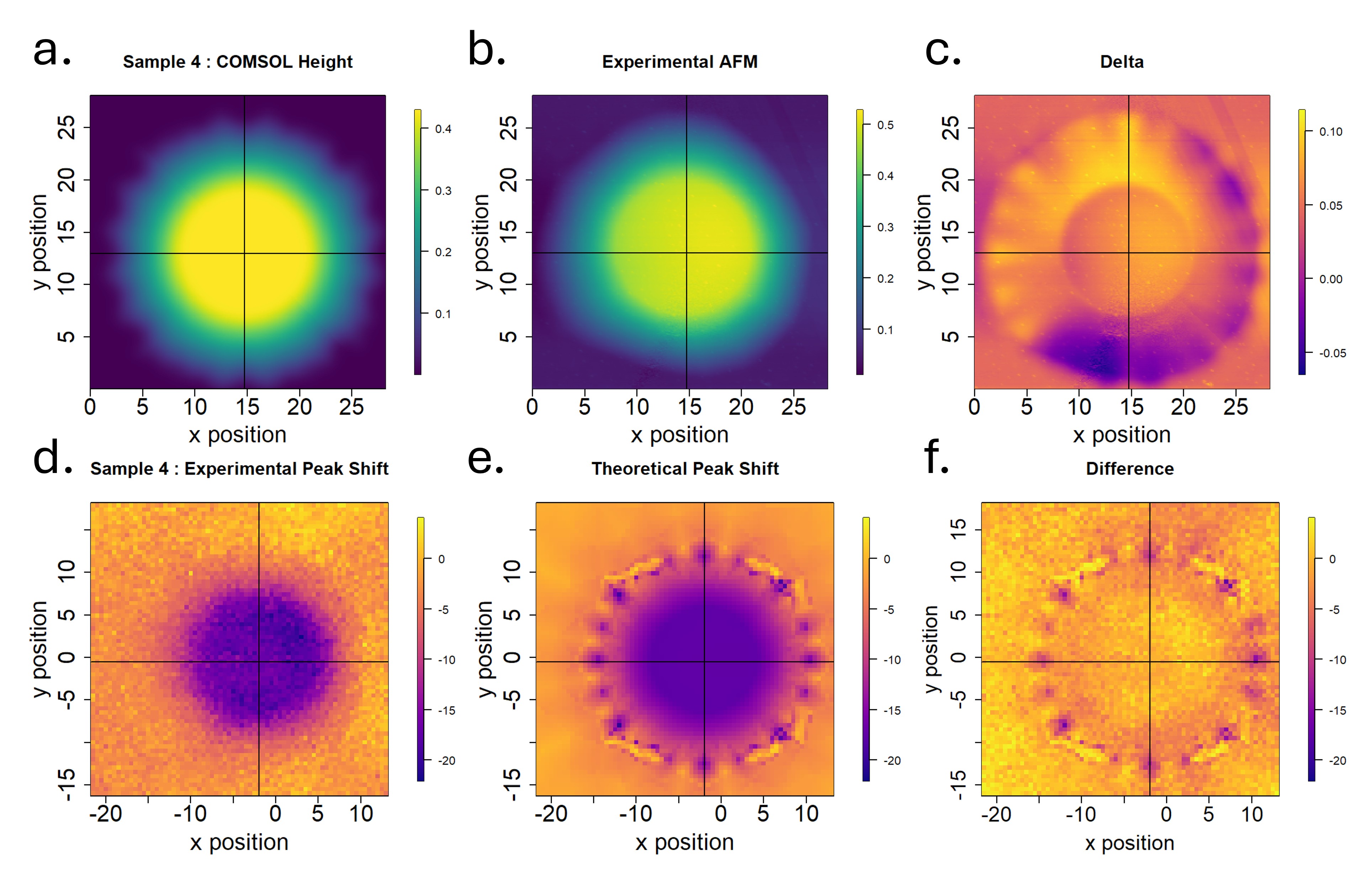}
  \caption{Sample 4. The height profile (a) simulated from COMSOL and (b) measured  by AFM, with (c) showing the difference between the experimental and predicted heights. (d) The experimental peak shift, (e) the calculated peak shift using the experimentally determined strain gauge factors, and (f) the difference between the two.}
  \label{sample_04}
\end{figure}
\begin{figure}[h]
  \includegraphics[width=0.83\linewidth]{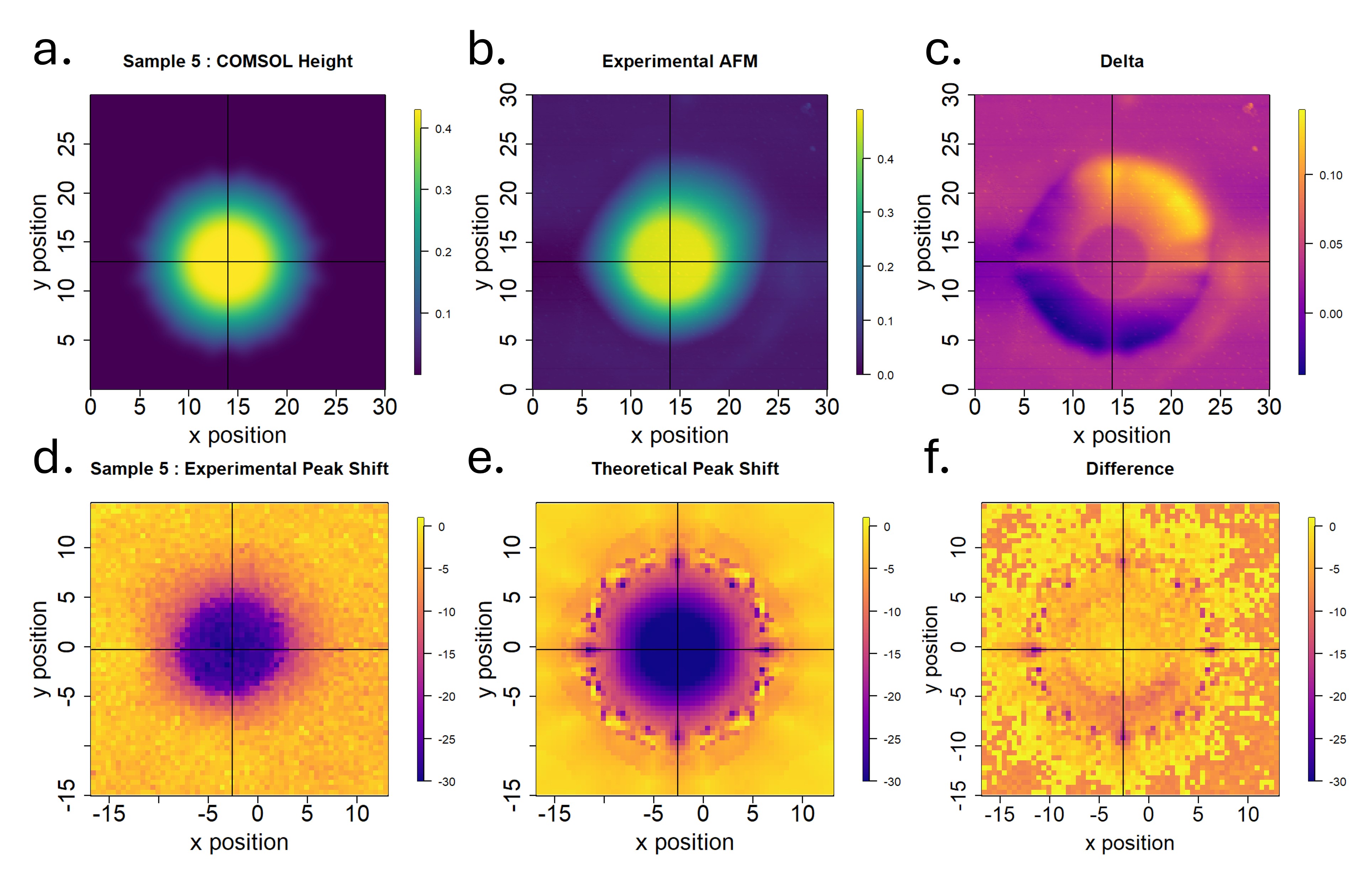}
  \caption{Sample 5. The height profile (a) simulated from COMSOL and (b) measured  by AFM, with (c) showing the difference between the experimental and predicted heights. (d) The experimental peak shift, (e) the calculated peak shift using the experimentally determined strain gauge factors, and (f) the difference between the two.}
  \label{sample_05}
\end{figure}
\begin{figure}[h]
  \includegraphics[width=0.83\linewidth]{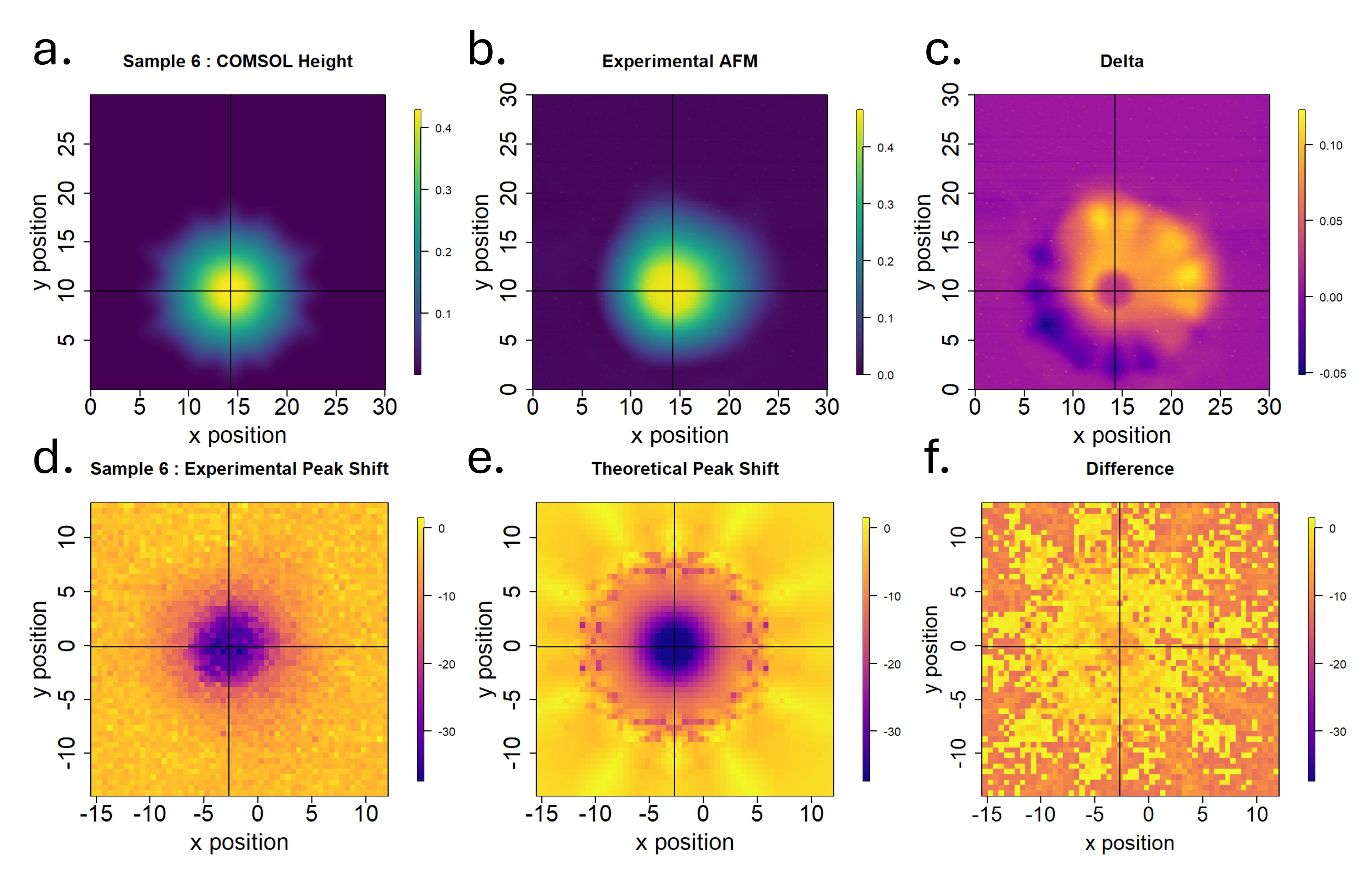}
  \caption{Sample 6. The height profile (a) simulated from COMSOL and (b) measured  by AFM, with (c) showing the difference between the experimental and predicted heights. (d) The experimental peak shift, (e) the calculated peak shift using the experimentally determined strain gauge factors, and (f) the difference between the two.}
  \label{sample_06}
\end{figure}
\begin{figure}[h]
  \includegraphics[width=0.83\linewidth]{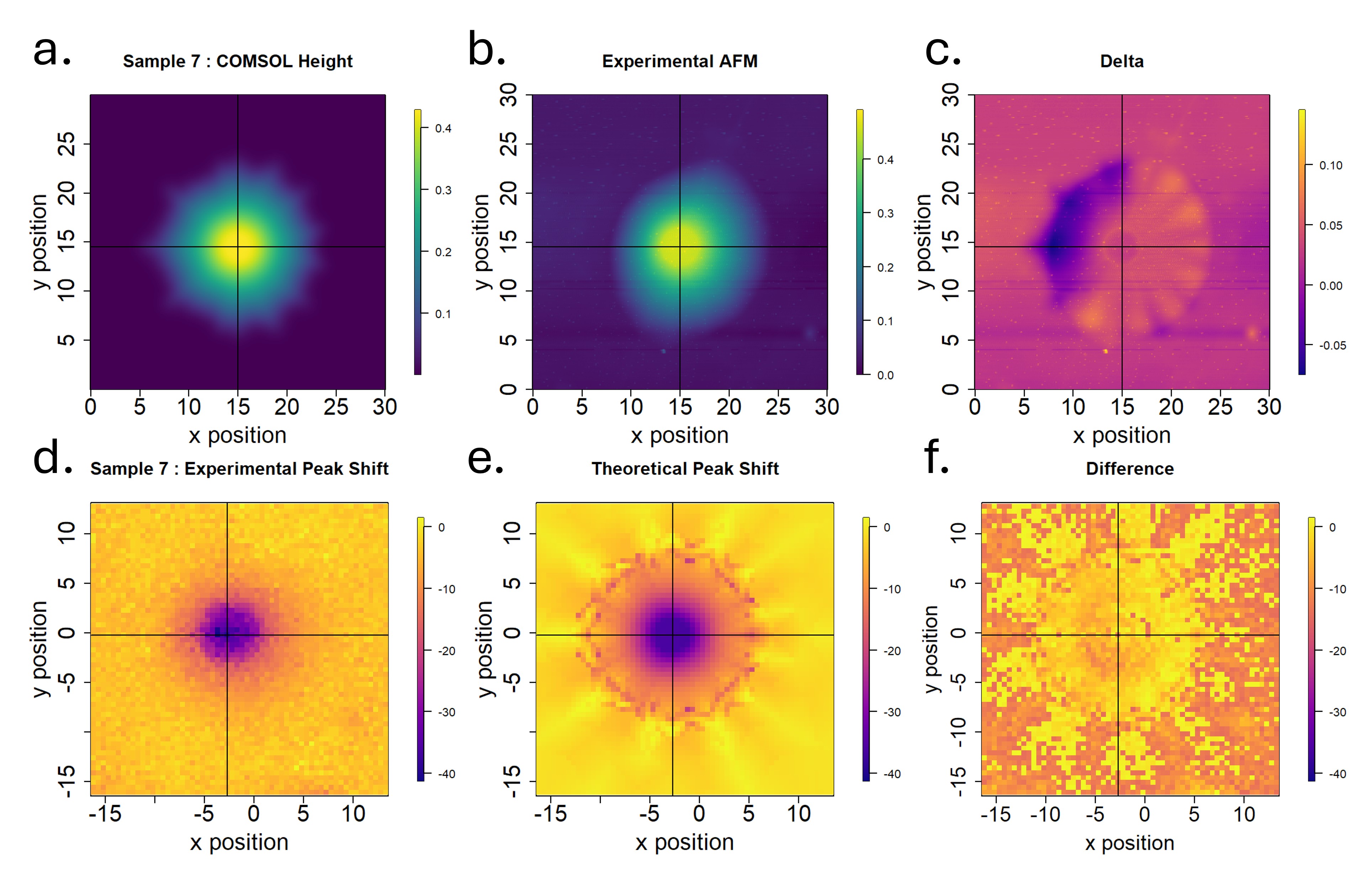}
  \caption{Sample 7. The height profile (a) simulated from COMSOL and (b) measured  by AFM, with (c) showing the difference between the experimental and predicted heights. (d) The experimental peak shift, (e) the calculated peak shift using the experimentally determined strain gauge factors, and (f) the difference between the two.}
  \label{sample_07}
\end{figure}
\begin{figure}[h]
  \includegraphics[width=0.83\linewidth]{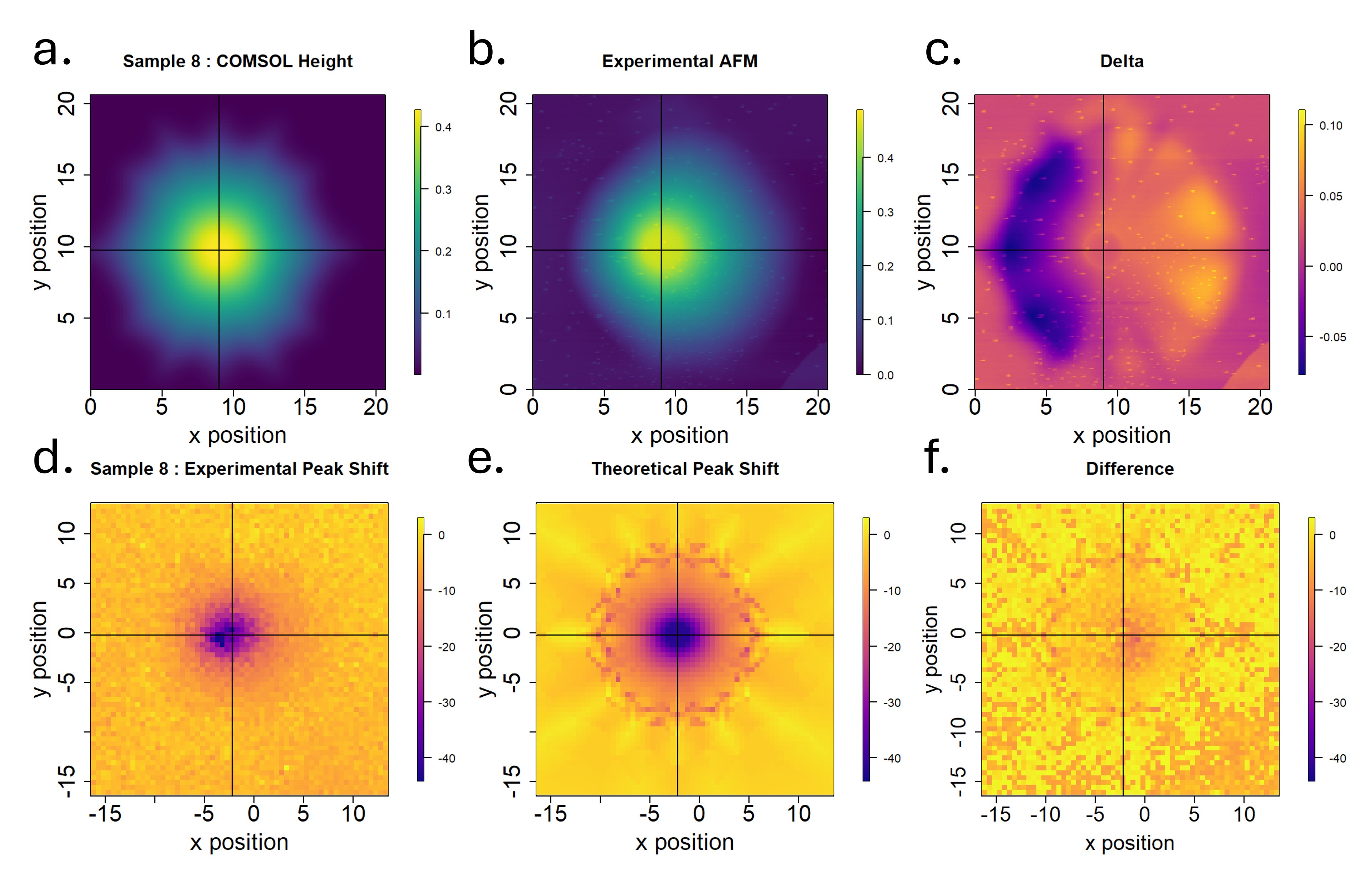}
  \caption{Sample 8. The height profile (a) simulated from COMSOL and (b) measured  by AFM, with (c) showing the difference between the experimental and predicted heights. (d) The experimental peak shift, (e) the calculated peak shift using the experimentally determined strain gauge factors, and (f) the difference between the two.}
  \label{sample_08}
\end{figure}
\begin{figure}[h]
  \includegraphics[width=0.83\linewidth]{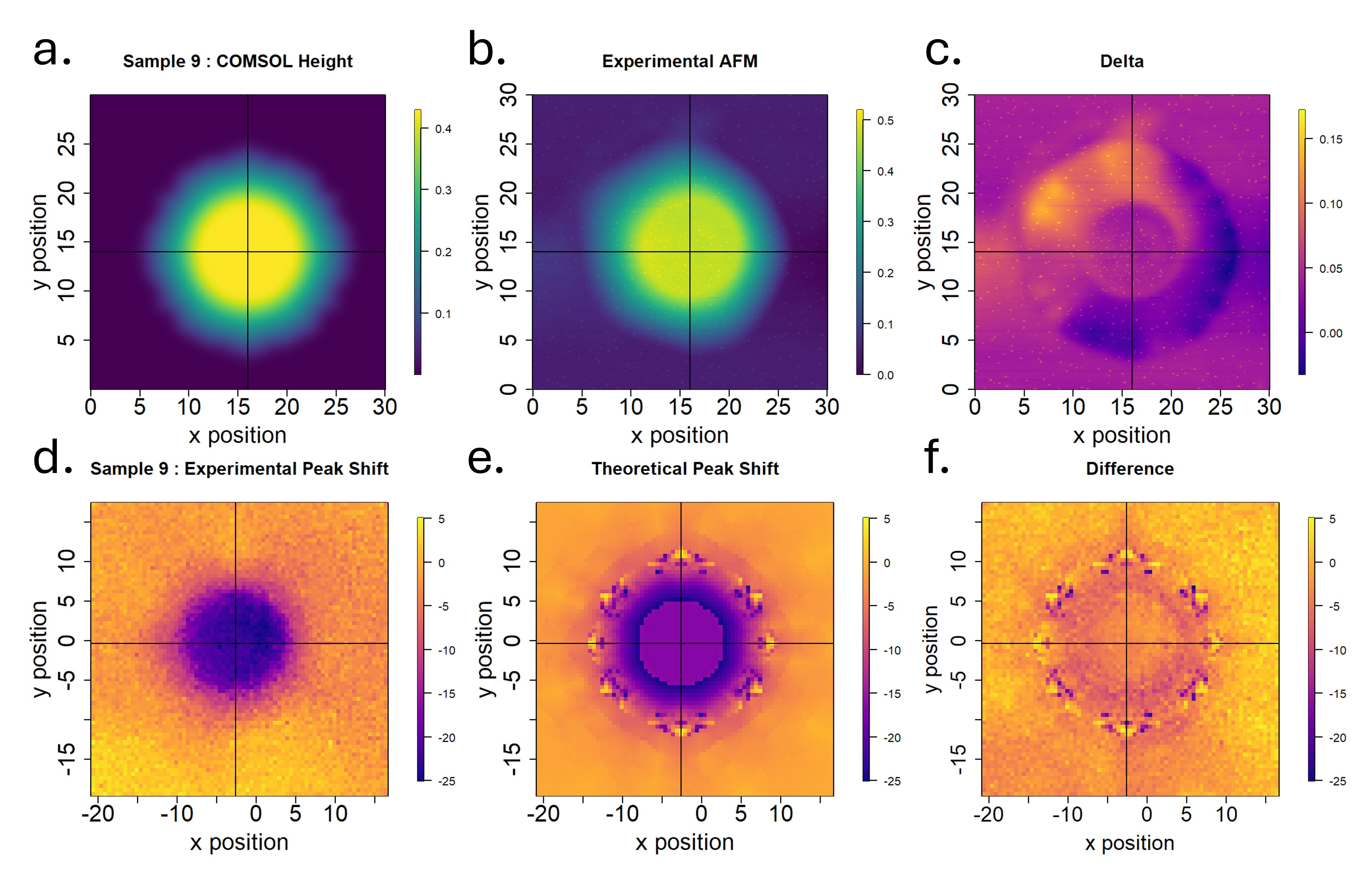}
  \caption{Sample 9. The height profile (a) simulated from COMSOL and (b) measured  by AFM, with (c) showing the difference between the experimental and predicted heights. (d) The experimental peak shift, (e) the calculated peak shift using the experimentally determined strain gauge factors, and (f) the difference between the two.}
  \label{sample_09}
\end{figure}
\begin{figure}[h]
  \includegraphics[width=0.83\linewidth]{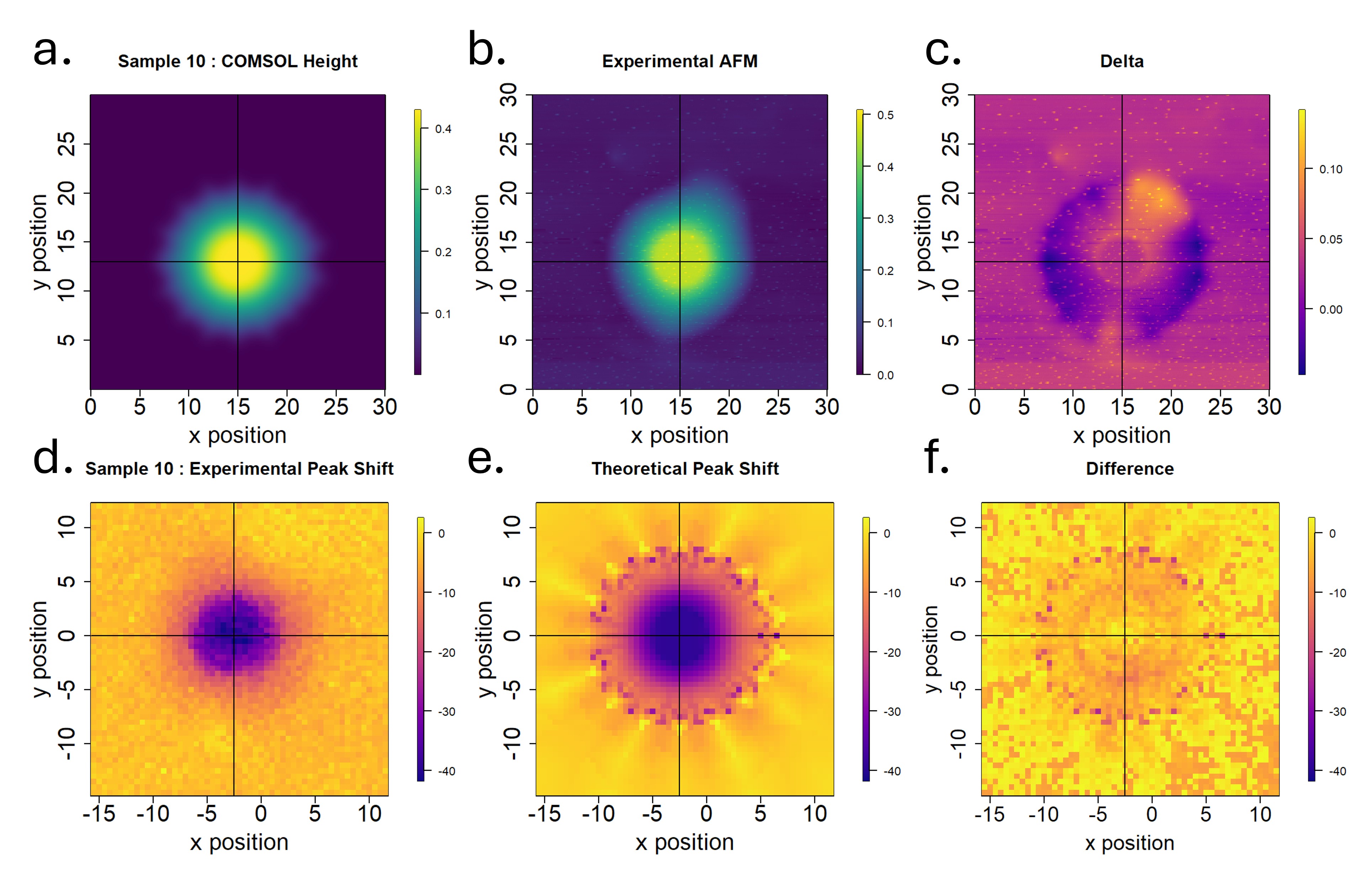}
  \caption{Sample 10. The height profile (a) simulated from COMSOL and (b) measured  by AFM, with (c) showing the difference between the experimental and predicted heights. (d) The experimental peak shift, (e) the calculated peak shift using the experimentally determined strain gauge factors, and (f) the difference between the two.}
  \label{sample_10}
\end{figure}

\begin{figure}[h]
  \includegraphics[width=0.83\linewidth]{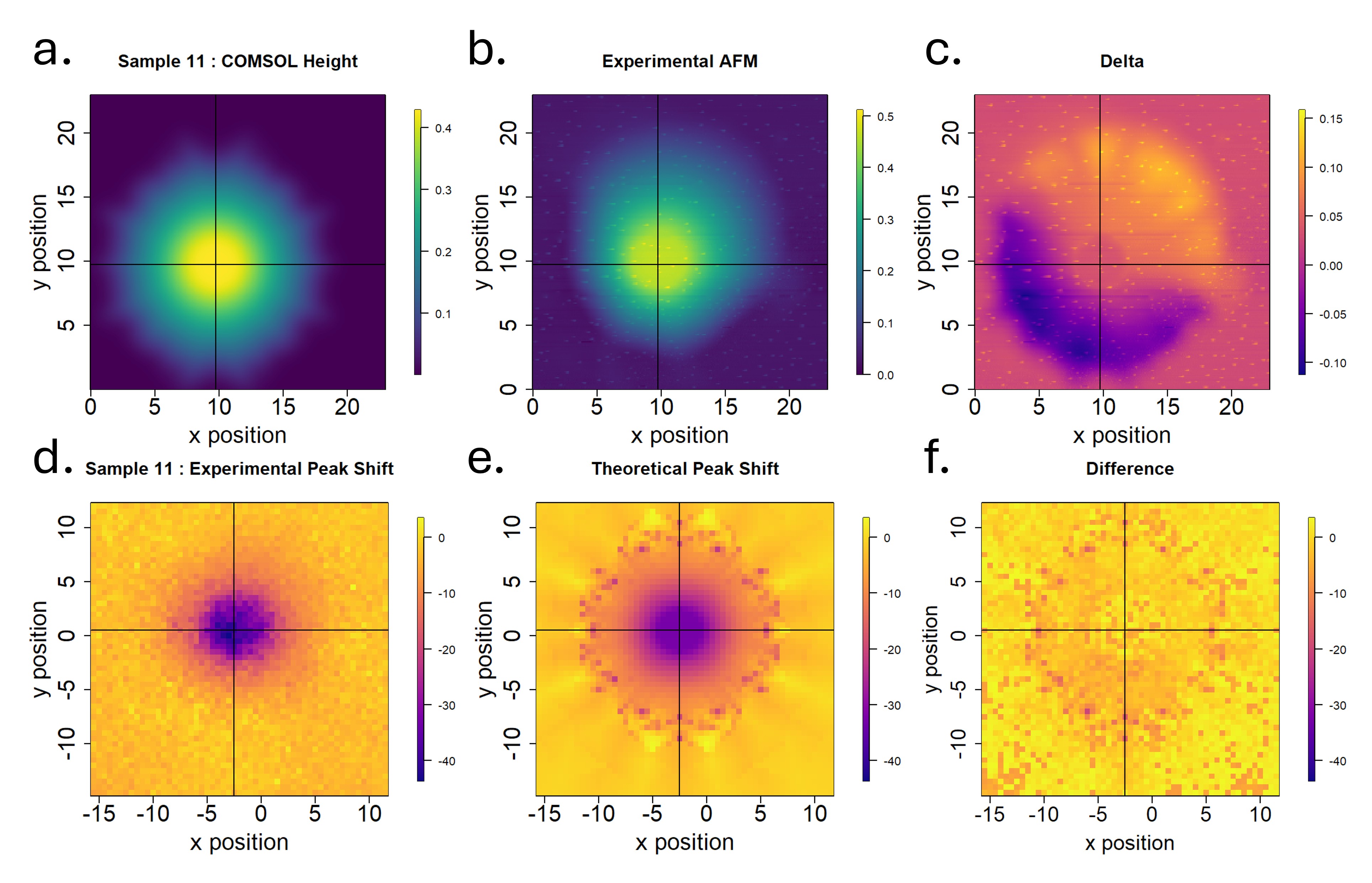}
  \caption{Sample 11. The height profile (a) simulated from COMSOL and (b) measured  by AFM, with (c) showing the difference between the experimental and predicted heights. (d) The experimental peak shift, (e) the calculated peak shift using the experimentally determined strain gauge factors, and (f) the difference between the two.}
  \label{sample_11}
\end{figure}
\begin{figure}[h]
  \includegraphics[width=0.83\linewidth]{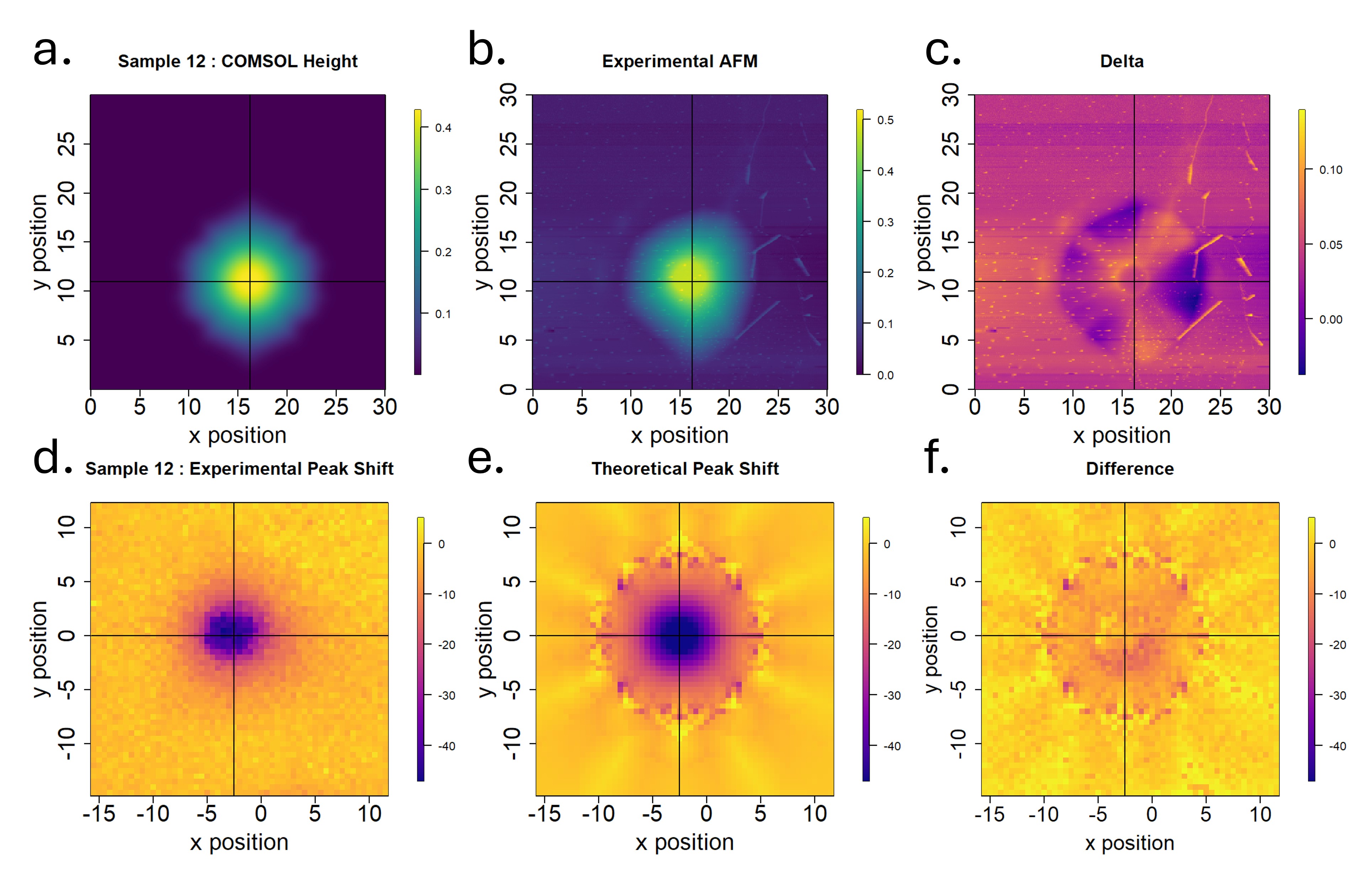}
  \caption{Sample 12. The height profile (a) simulated from COMSOL and (b) measured  by AFM, with (c) showing the difference between the experimental and predicted heights. (d) The experimental peak shift, (e) the calculated peak shift using the experimentally determined strain gauge factors, and (f) the difference between the two.}
  \label{sample_12}
\end{figure}
\begin{figure}[h]
  \includegraphics[width=0.83\linewidth]{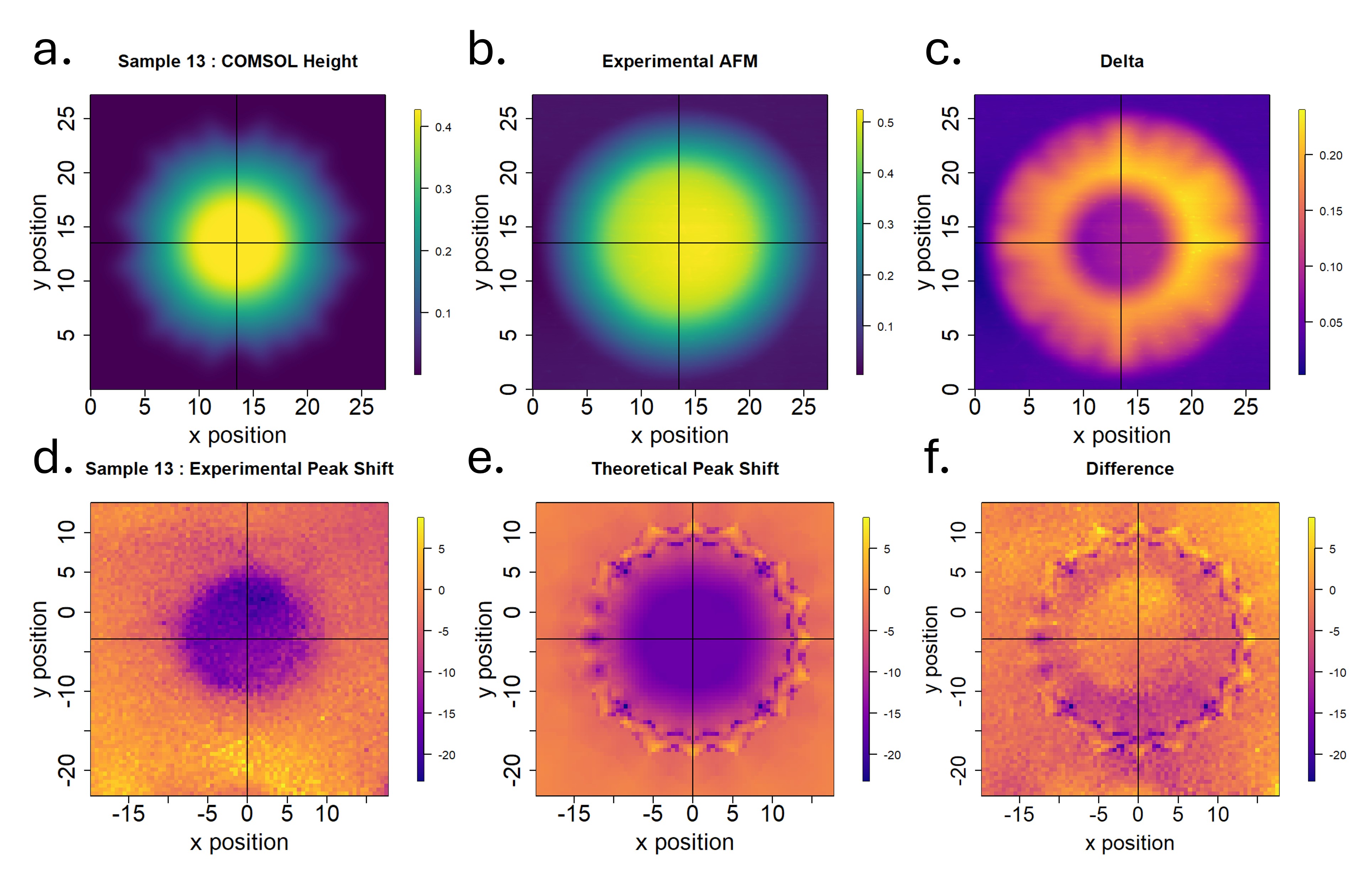}
  \caption{Sample 13. The height profile (a) simulated from COMSOL and (b) measured  by AFM, with (c) showing the difference between the experimental and predicted heights. (d) The experimental peak shift, (e) the calculated peak shift using the experimentally determined strain gauge factors, and (f) the difference between the two.}
  \label{sample_13}
\end{figure}
\begin{figure}[h]
  \includegraphics[width=0.83\linewidth]{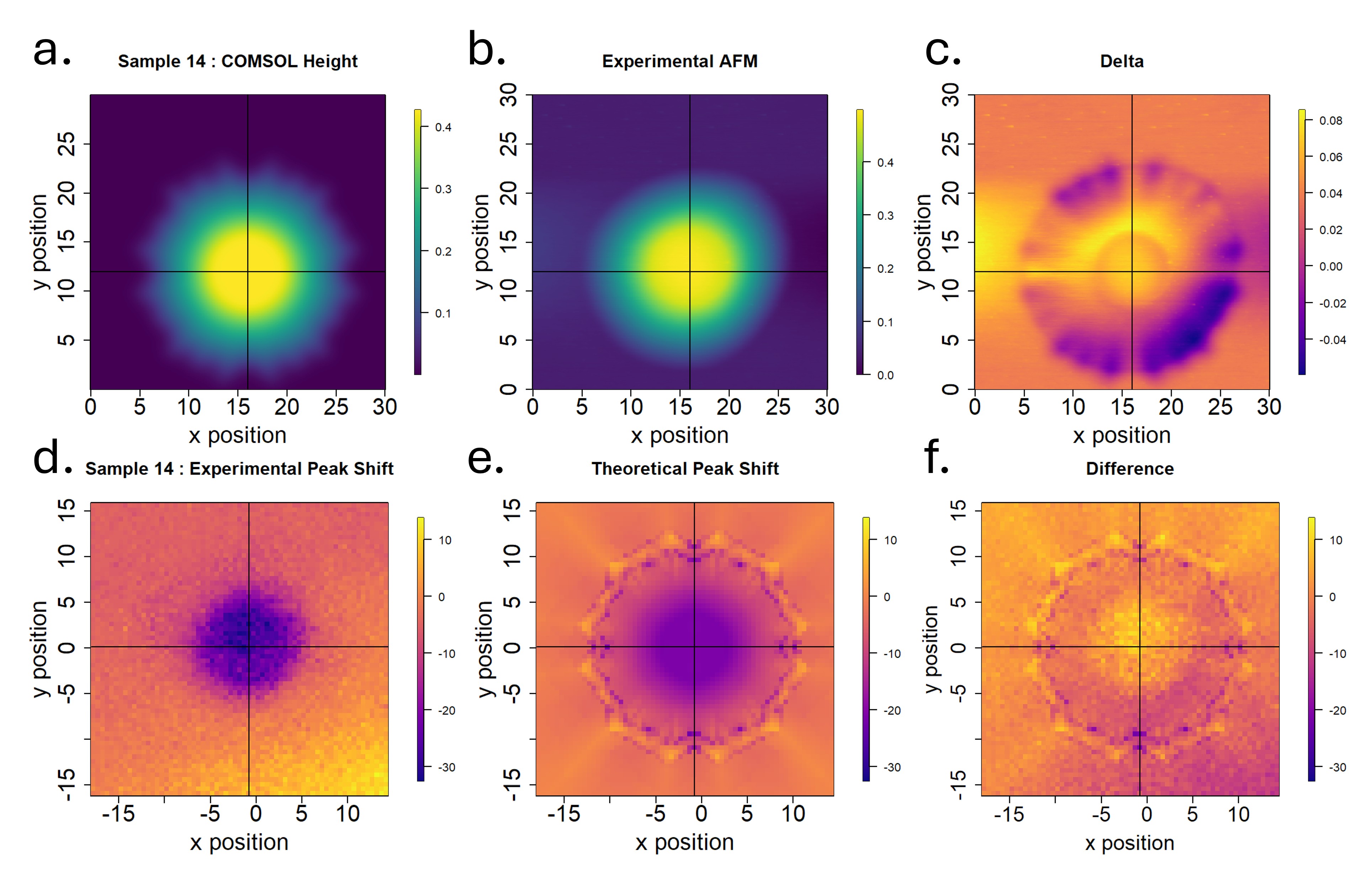}
  \caption{Sample 14. The height profile (a) simulated from COMSOL and (b) measured  by AFM, with (c) showing the difference between the experimental and predicted heights. (d) The experimental peak shift, (e) the calculated peak shift using the experimentally determined strain gauge factors, and (f) the difference between the two.}
  \label{sample_14}
\end{figure}
\begin{figure}[h]
  \includegraphics[width=0.83\linewidth]{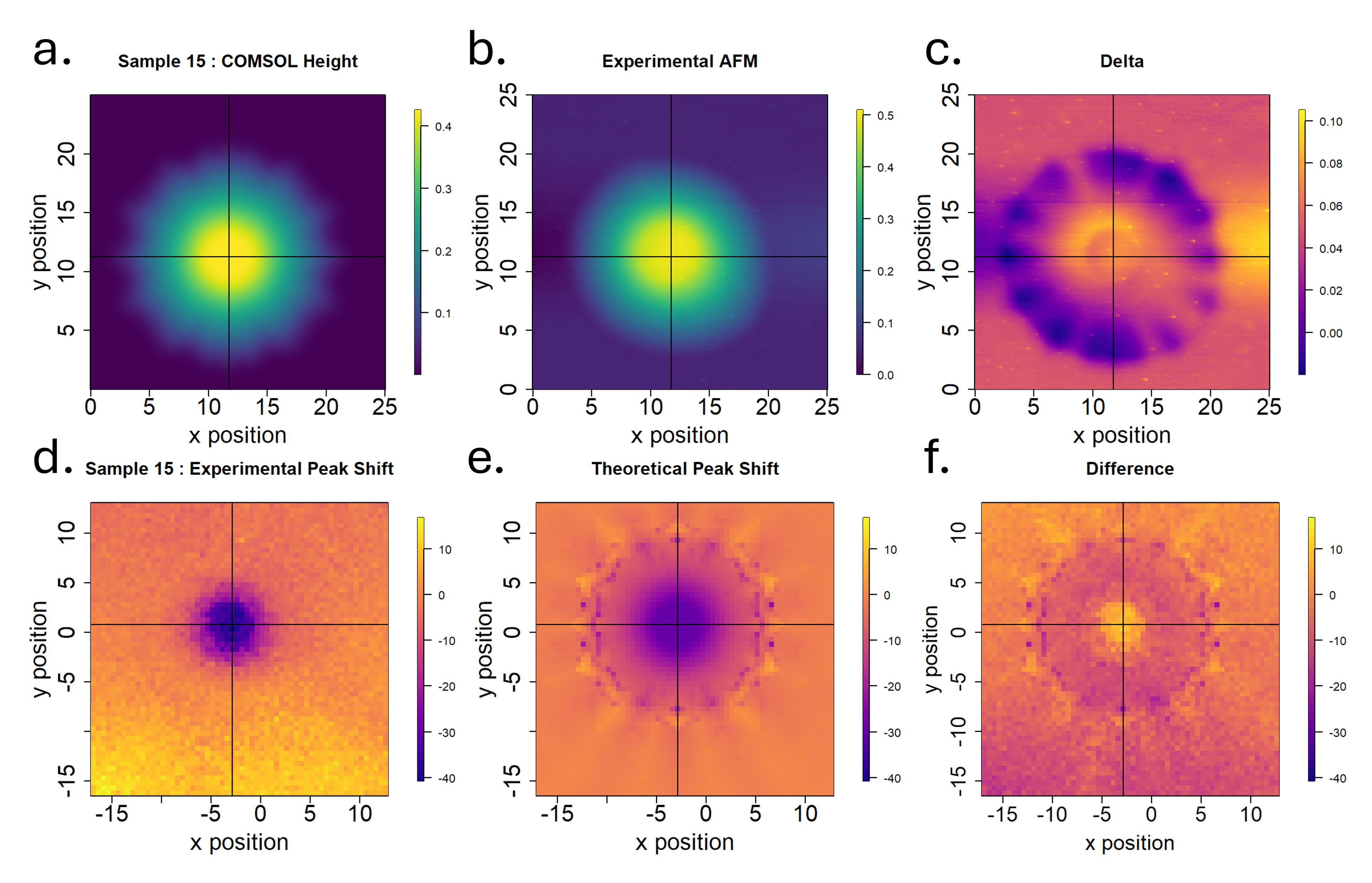}
  \caption{Sample 15. The height profile (a) simulated from COMSOL and (b) measured  by AFM, with (c) showing the difference between the experimental and predicted heights. (d) The experimental peak shift, (e) the calculated peak shift using the experimentally determined strain gauge factors, and (f) the difference between the two.}
  \label{sample_15}
\end{figure}
\begin{figure}[h]
  \includegraphics[width=0.83\linewidth]{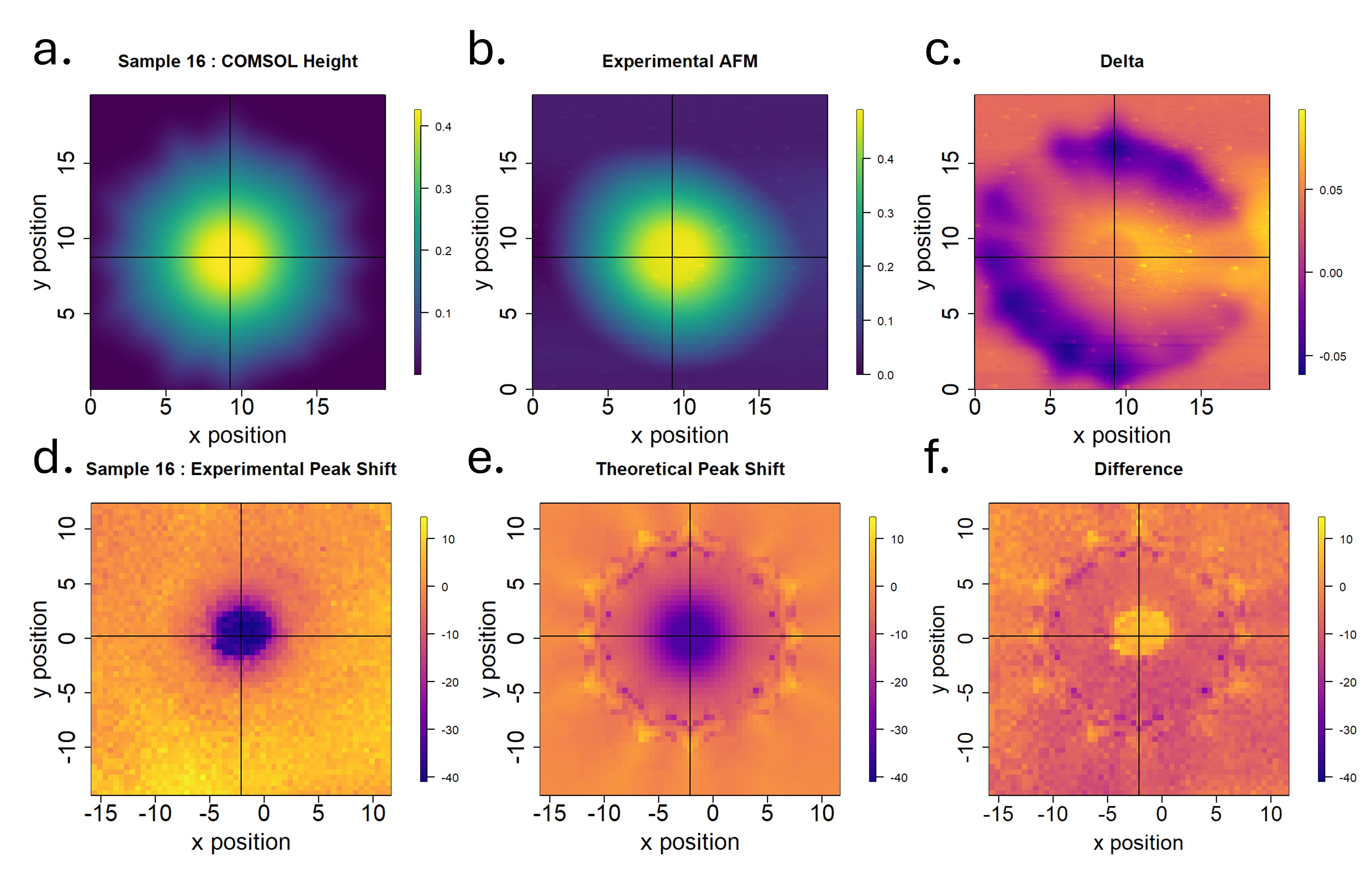}
  \caption{Sample 16. The height profile (a) simulated from COMSOL and (b) measured  by AFM, with (c) showing the difference between the experimental and predicted heights. (d) The experimental peak shift, (e) the calculated peak shift using the experimentally determined strain gauge factors, and (f) the difference between the two.}
  \label{sample_16}
\end{figure}
\begin{figure}[h]
  \includegraphics[width=0.83\linewidth]{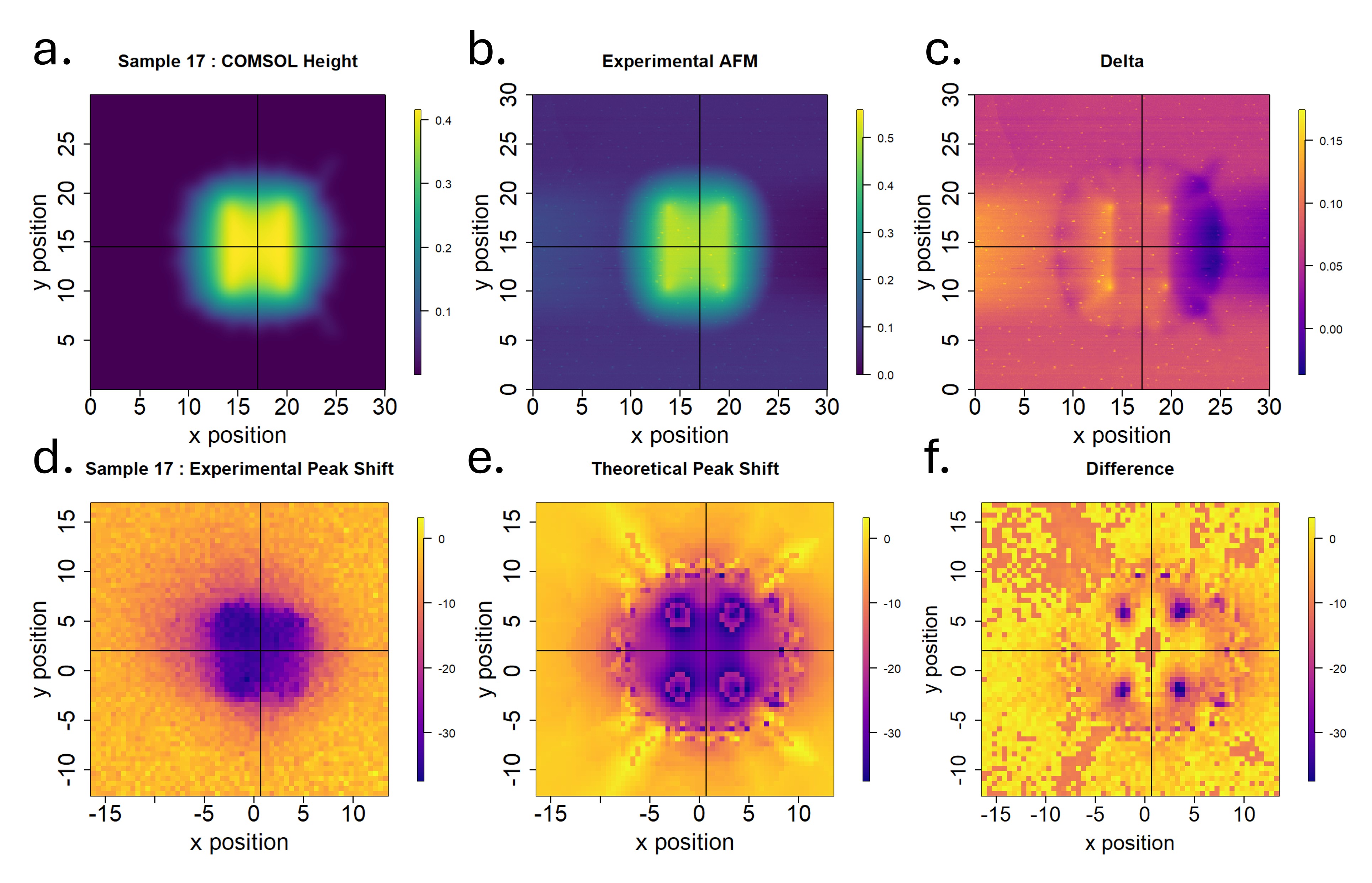}
  \caption{Sample 17. The height profile (a) simulated from COMSOL and (b) measured  by AFM, with (c) showing the difference between the experimental and predicted heights. (d) The experimental peak shift, (e) the calculated peak shift using the experimentally determined strain gauge factors, and (f) the difference between the two.}
  \label{sample_17}
\end{figure}
\begin{figure}[h]
  \includegraphics[width=0.83\linewidth]{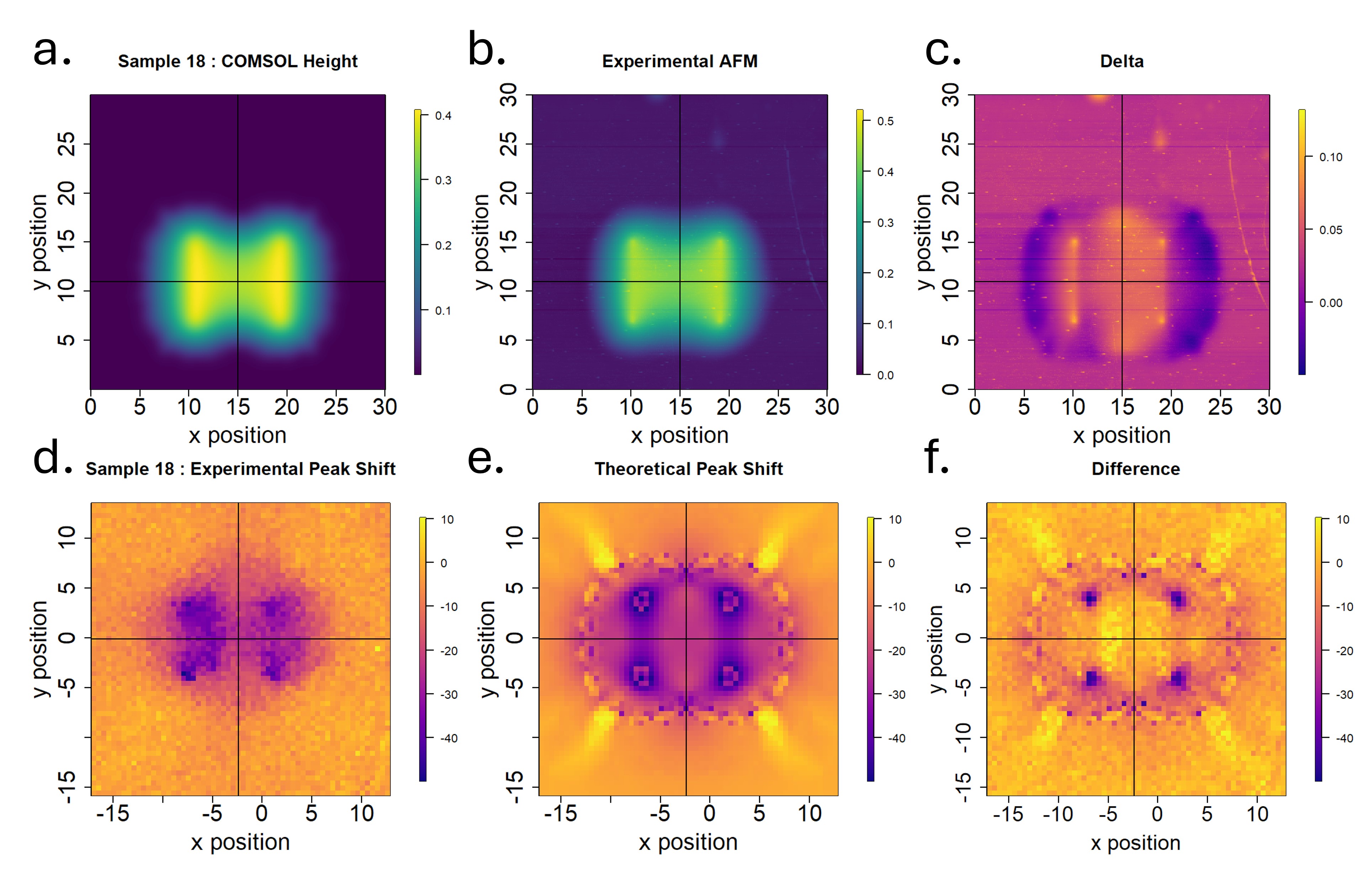}
  \caption{Sample 18. The height profile (a) simulated from COMSOL and (b) measured  by AFM, with (c) showing the difference between the experimental and predicted heights. (d) The experimental peak shift, (e) the calculated peak shift using the experimentally determined strain gauge factors, and (f) the difference between the two.}
  \label{sample_18}
\end{figure}
\begin{figure}[h]
  \includegraphics[width=0.83\linewidth]{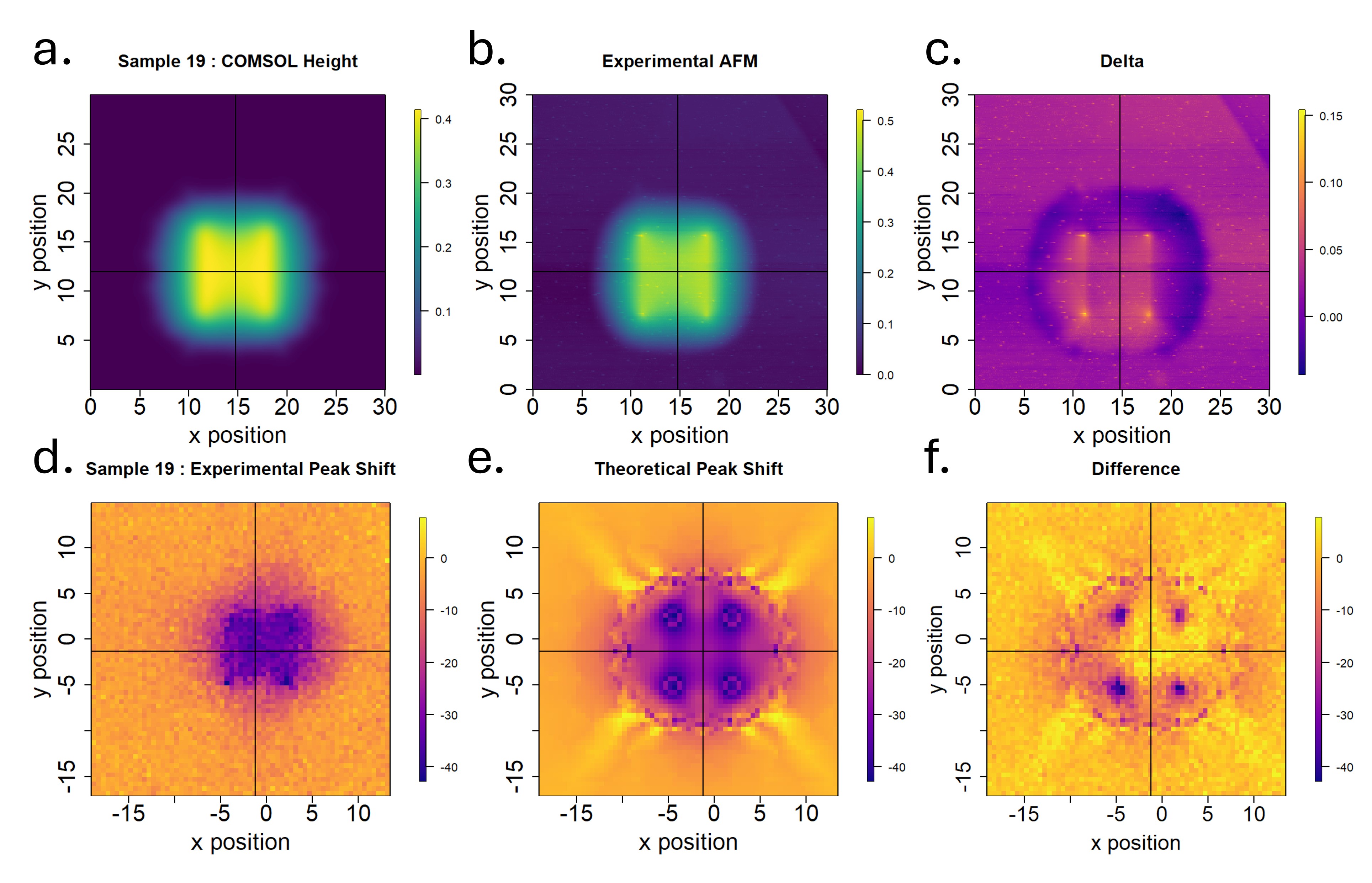}
  \caption{Sample 19. The height profile (a) simulated from COMSOL and (b) measured  by AFM, with (c) showing the difference between the experimental and predicted heights. (d) The experimental peak shift, (e) the calculated peak shift using the experimentally determined strain gauge factors, and (f) the difference between the two.}
  \label{sample_19}
\end{figure}
\begin{figure}[h]
  \includegraphics[width=0.83\linewidth]{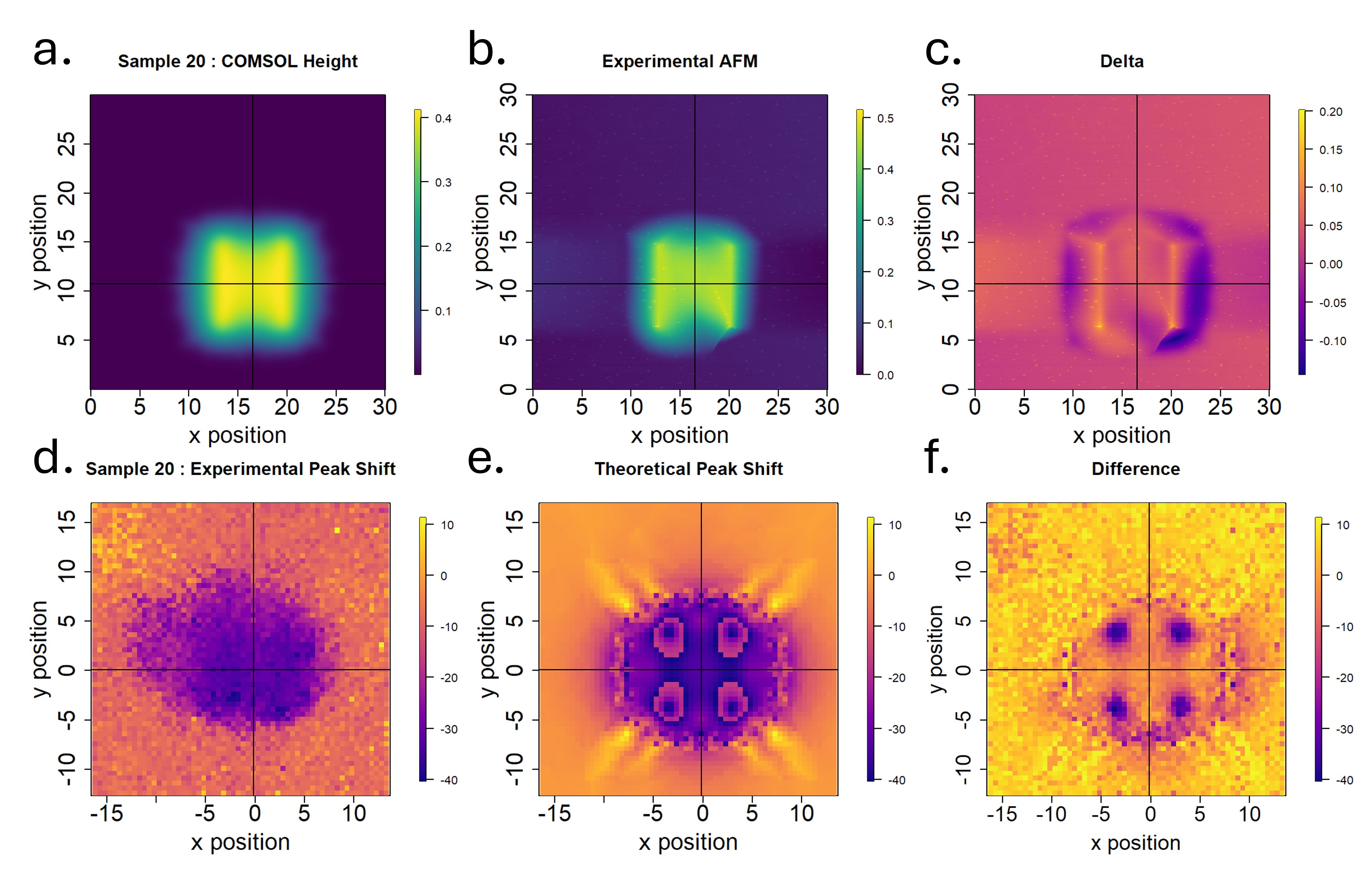}
  \caption{Sample 20. The height profile (a) simulated from COMSOL and (b) measured  by AFM, with (c) showing the difference between the experimental and predicted heights. (d) The experimental peak shift, (e) the calculated peak shift using the experimentally determined strain gauge factors, and (f) the difference between the two.}
  \label{sample_20}
\end{figure}

\begin{figure}[h]
  \includegraphics[width=0.83\linewidth]{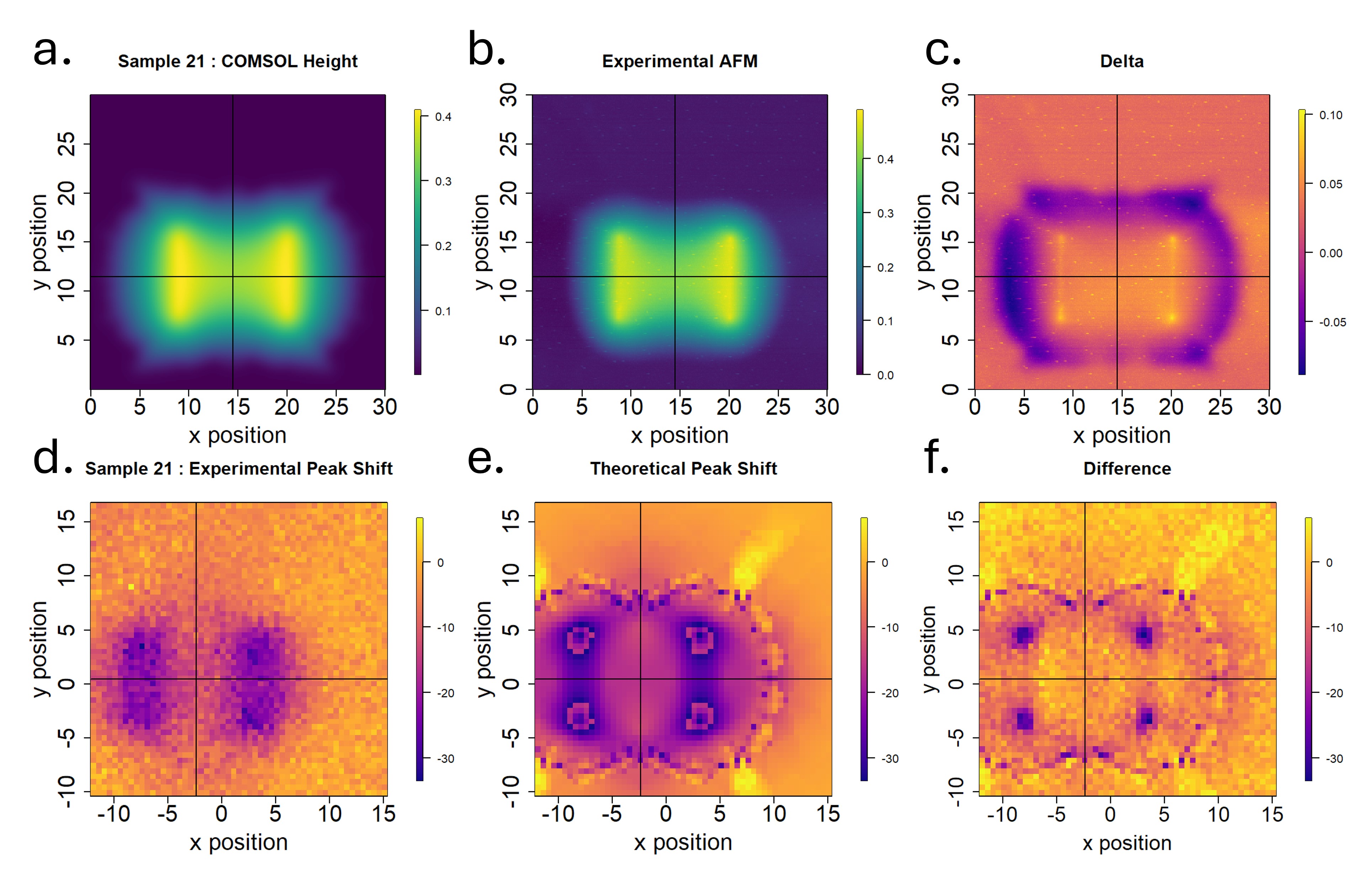}
  \caption{Sample 21. The height profile (a) simulated from COMSOL and (b) measured  by AFM, with (c) showing the difference between the experimental and predicted heights. (d) The experimental peak shift, (e) the calculated peak shift using the experimentally determined strain gauge factors, and (f) the difference between the two.}
  \label{sample_21}
\end{figure}
\begin{figure}[h]
  \includegraphics[width=0.83\linewidth]{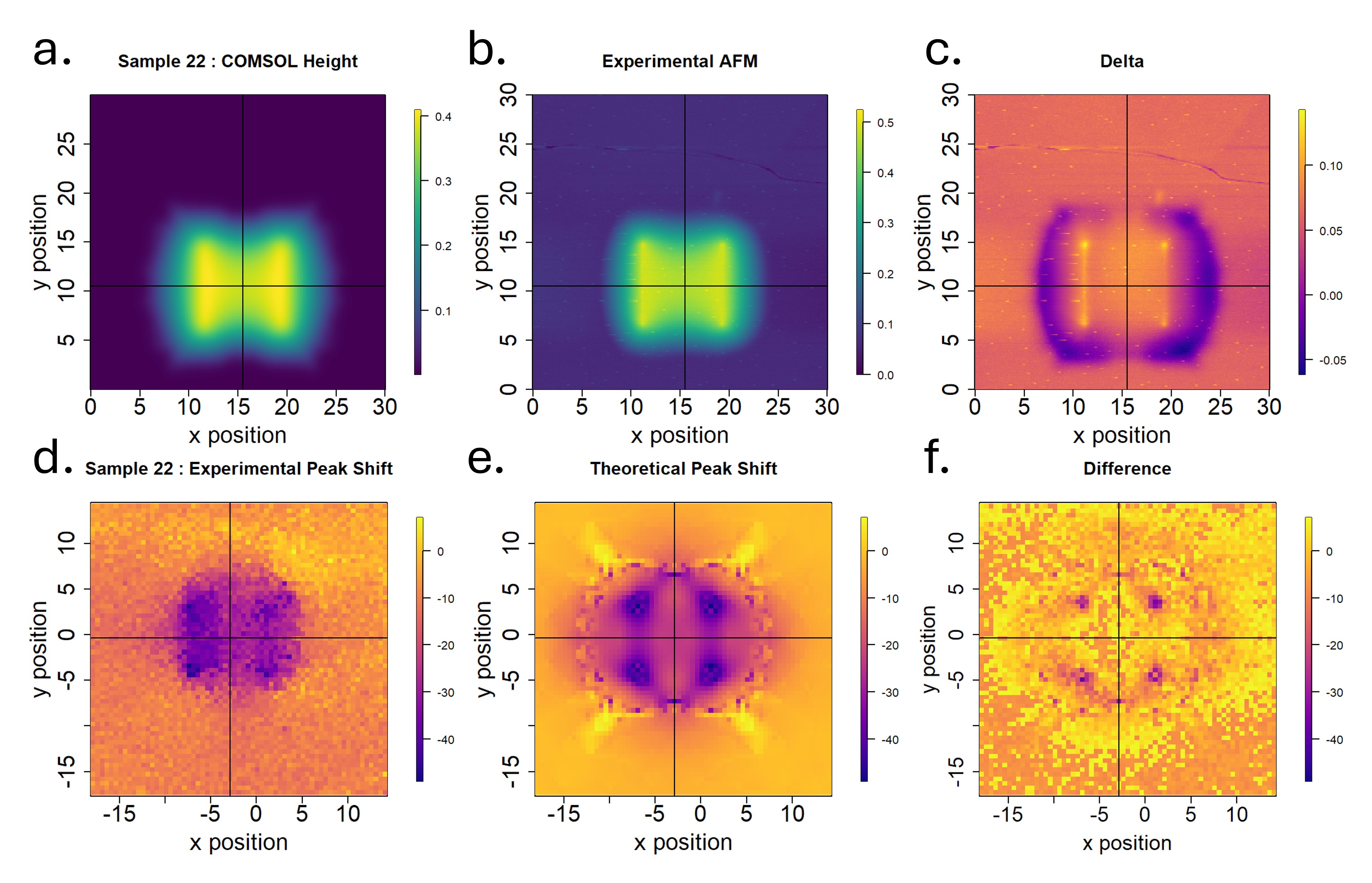}
  \caption{Sample 22. The height profile (a) simulated from COMSOL and (b) measured  by AFM, with (c) showing the difference between the experimental and predicted heights. (d) The experimental peak shift, (e) the calculated peak shift using the experimentally determined strain gauge factors, and (f) the difference between the two.}
  \label{sample_22}
\end{figure}
\begin{figure}[h]
  \includegraphics[width=0.83\linewidth]{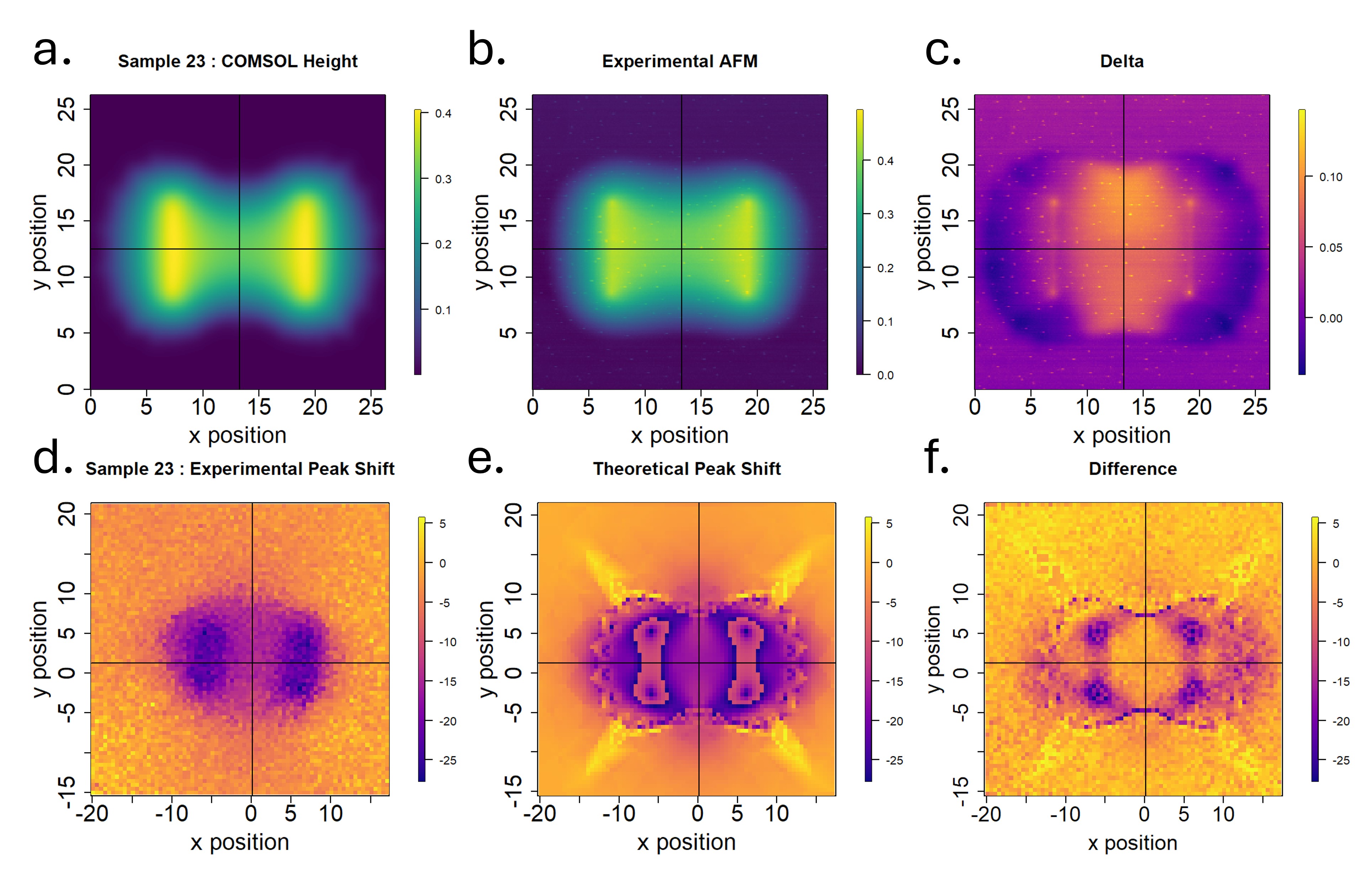}
  \caption{Sample 23. The height profile (a) simulated from COMSOL and (b) measured  by AFM, with (c) showing the difference between the experimental and predicted heights. (d) The experimental peak shift, (e) the calculated peak shift using the experimentally determined strain gauge factors, and (f) the difference between the two.}
  \label{sample_23}
\end{figure}

\clearpage
\newpage
\bibliography {ref}